\documentclass[12pt]{article}

\usepackage{amsthm}

\theoremstyle{definition}
\newtheorem{assumption}{Assumption}

\newtheorem{theorem}{Theorem}

\newtheorem{proposition}{Proposition}
\newtheorem{example}{Example}
\newtheorem{remark}{Remark}
\newtheorem*{remark*}{Remark}
\usepackage{amsmath, caption, mathrsfs, yfonts,subfig,graphicx}
\usepackage[top=1in, bottom=1in, left=1.1in, right=1.1in]{geometry}
\usepackage{array}
\newcolumntype{L}[1]{>{\raggedright\let\newline\\\arraybackslash\hspace{0pt}}m{#1}}
\newcolumntype{R}[1]{>{\raggedleft\let\newline\\\arraybackslash\hspace{0pt}}m{#1}}
\newcolumntype{C}[1]{>{\centering\let\newline\\\arraybackslash\hspace{0pt}}m{#1}}

\usepackage{graphicx,psfrag,epsf,mathtools}
\usepackage{multirow}
\usepackage{amsthm}
\usepackage{natbib}
\usepackage{dsfont, amsfonts,amssymb}
\usepackage{url} 
\usepackage{booktabs, bm, paralist,tikz,longtable,microtype,rotating,setspace} 
\usepackage[pdftex,colorlinks=true]{hyperref}
\definecolor{darkblue}{rgb}{0,0,.6}
\hypersetup{citecolor=darkblue,linkcolor=darkblue,urlcolor=darkblue}

\usepackage[noend]{algpseudocode}
\usepackage{algorithm}
\usepackage{chemformula}
\usepackage{appendix}
\usepackage{titling}

\usepackage{xr}
\newcommand{\blind}{1} 

\newcommand{\norm}[1]{\left\lVert#1\right\rVert}
\newcommand{\abs}[1]{\left|#1\right|}
\newcommand*\diff{\mathop{}\!\mathrm{d}}

\newcommand{\LL}{\mathcal{L}}
\newcommand{\X}{\mathcal{X}}

\newcommand{\inp}[2]{\langle #1\, , #2 \rangle}

\usepackage{afterpage}

\graphicspath{{plots/}}
\DeclareMathOperator*{\argmin}{\arg\!\min}

\newsavebox\CBox
\def\textBF#1{\sbox\CBox{#1}\resizebox{\wd\CBox}{\ht\CBox}{\textbf{#1}}}

\def\T{{ \mathrm{\scriptscriptstyle T} }}

\date{}

\makeatletter
\newcommand{\distas}[1]{\mathbin{\overset{#1}{Cern\z@\sim}}}%
\newsavebox{\mybox}\newsavebox{\mysim}
\newcommand{\distras}[1]{%
  \savebox{\mybox}{\hbox{Cern3pt$\scriptstyle#1$Cern3pt}}%
  \savebox{\mysim}{\hbox{$\sim$}}%
  \mathbin{\overset{#1}{Cern\z@\resizebox{\wd\mybox}{\ht\mysim}{$\sim$}}}%
}
\makeatother

\begin{document}

\def\spacingset#1{\renewcommand{\baselinestretch}%
{#1}\small\normalsize} \spacingset{1}


\if1\blind
{
  \title{Feature Extraction for Functional Time Series: Theory and Application to NIR Spectroscopy Data}
  \author{Yang Yang$^{a,}$\thanks{Corresponding author. Email address: yang.yang3@monash.edu} \\
  {\small $^a$Department of Econometrics and Business Statistics,} \\
  {\small Monash University, Melbourne, VIC 3145, Australia} \\
\\
 Yanrong Yang$^b$ \\
{\small $^b$Research School of Finance, Actuarial Studies and Statistics,} \\
{\small Australian National University, Canberra, ACT 2601, Australia} \\  
\\
Han Lin Shang$^c$ \\
{\small $^c$Department of Actuarial Studies and Business Analytics,} \\
{\small Macquarie University, Sydney, NSW 2109, Australia }}
\maketitle
} \fi

\if0\blind
{
  \bigskip
  \bigskip
  \bigskip
  \begin{center}
    {\LARGE Feature Extraction for Functional Time Series: Theory and Application to NIR Spectroscopy Data}
\end{center}
  \medskip
} \fi

\bigskip
 
\begin{abstract}
We propose a novel method to extract global and local features of functional time series. The global features concerning the dominant modes of variation over the entire function domain, and local features of function variations over particular short intervals within function domain, are both important in functional data analysis. Functional principal component analysis (FPCA), though a key feature extraction tool, only focus on capturing the dominant global features, neglecting highly localized features. We introduce a FPCA-BTW method that initially extracts global features of functional data via FPCA, and then extracts local features by block thresholding of wavelet (BTW) coefficients. Using Monte Carlo simulations, along with an empirical application on near-infrared spectroscopy data of wood panels, we illustrate that the proposed method outperforms competing methods including FPCA and sparse FPCA in the estimation functional processes. Moreover, extracted local features inheriting serial dependence of the original functional time series contribute to more accurate forecasts. Finally, we develop asymptotic properties of FPCA-BTW estimators, discovering the interaction between convergence rates of global and local features.
\end{abstract}

\noindent%
\vfill

\newpage
\doublespacing

\section{Introduction} \label{sec:1}

\begin{figure}[!hb]
\centering
\subfloat[Observed NIR spectra.]
{\includegraphics[width = 2.7in]{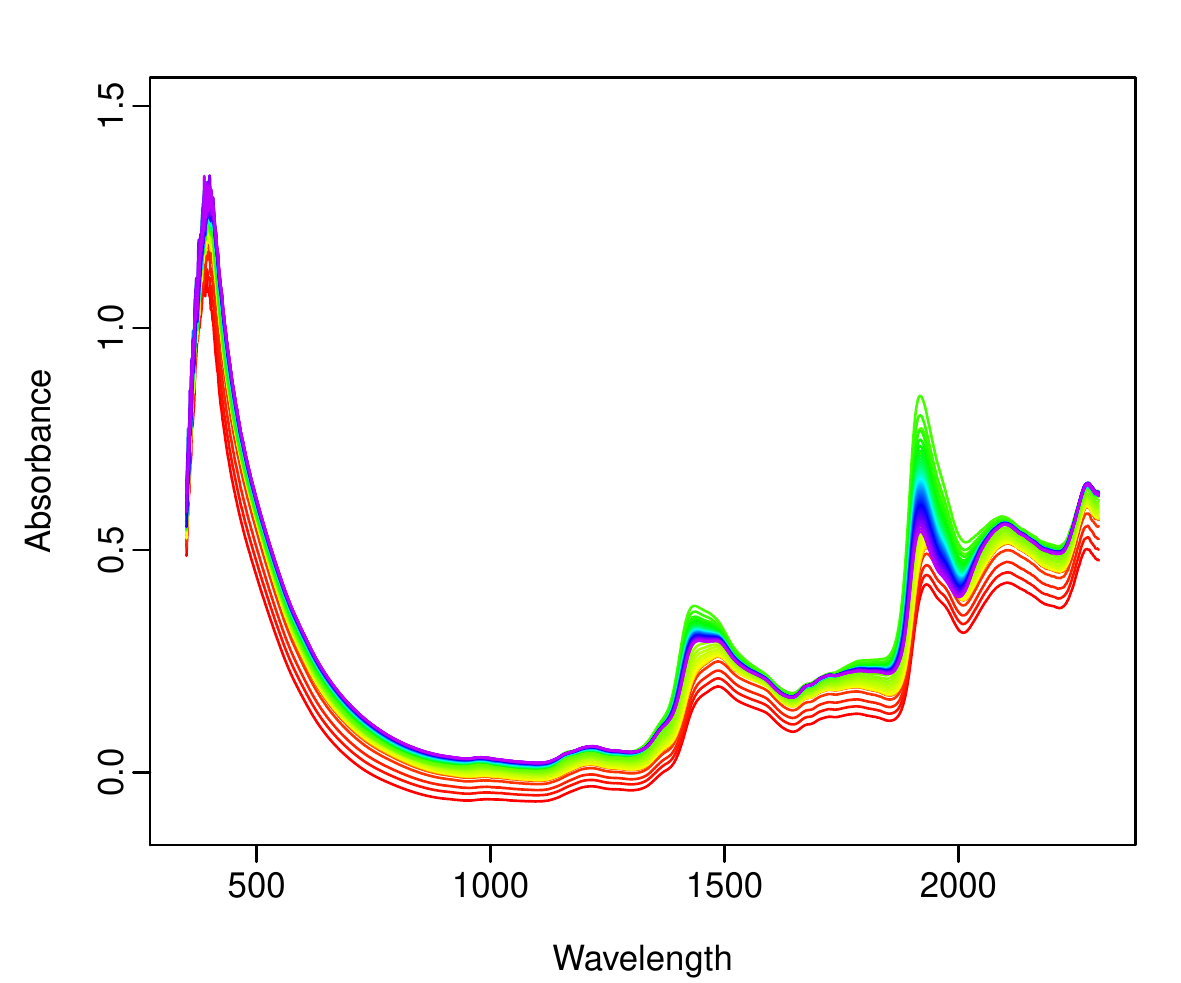}\label{fig:1a}} \quad
\subfloat[Smoothed NIR spectra.]
{\includegraphics[width = 2.7in]{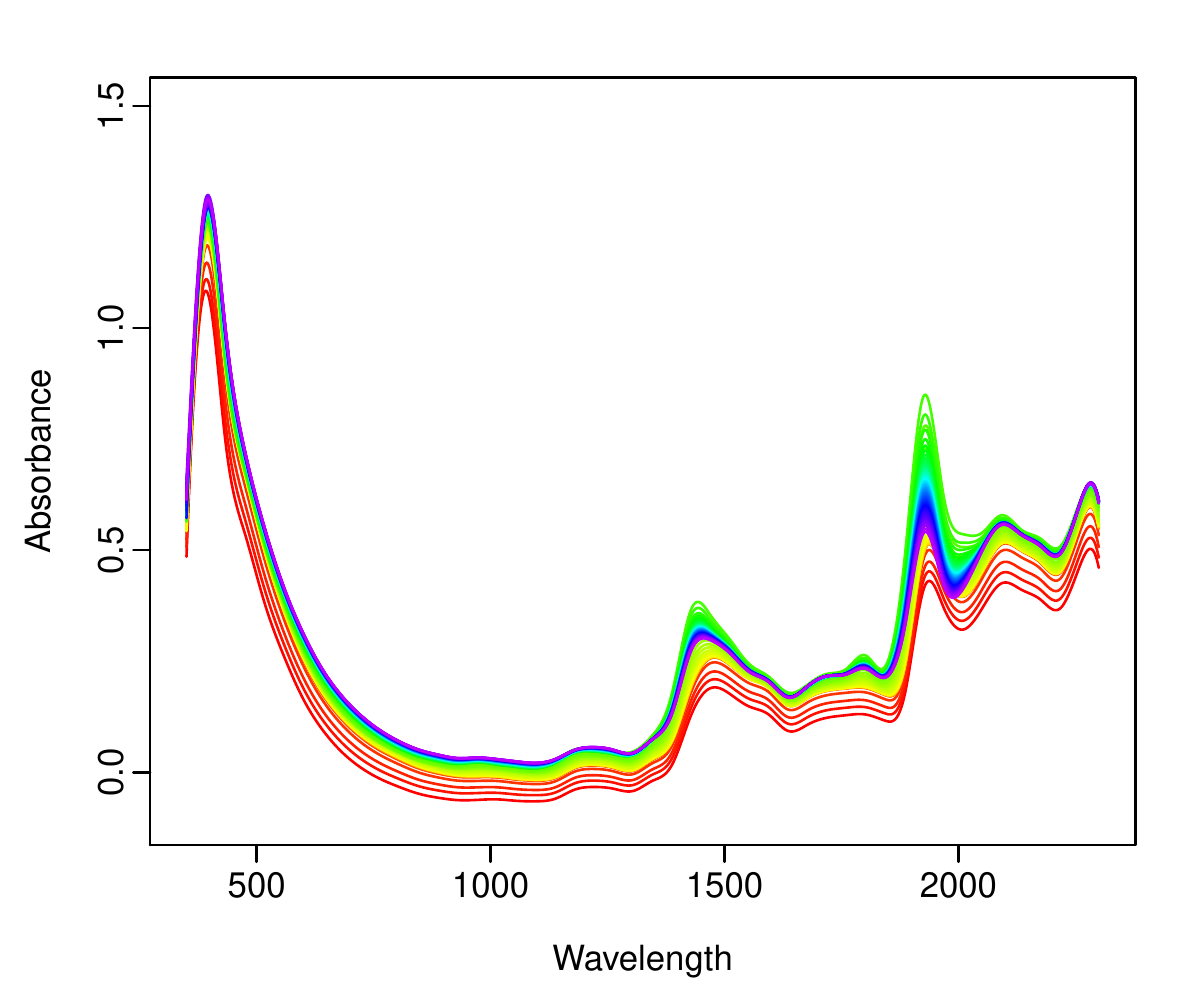}\label{fig:1d}} \\
\subfloat[Sample long-run covariance.]
{\includegraphics[width = 2.7in]{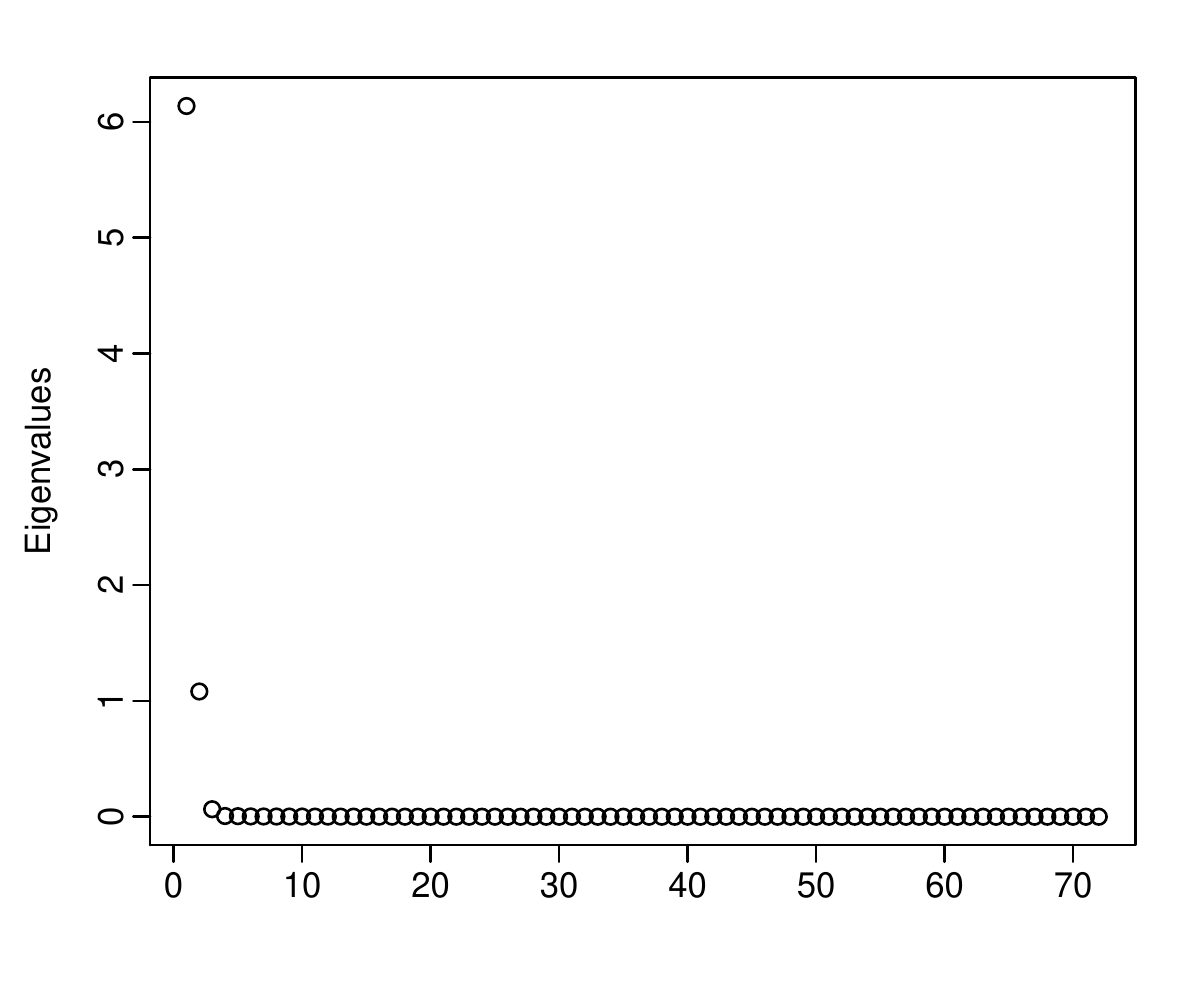}\label{fig:1e}} \quad
\subfloat[Dynamic FPCA residuals.]
{\includegraphics[width = 2.7in]{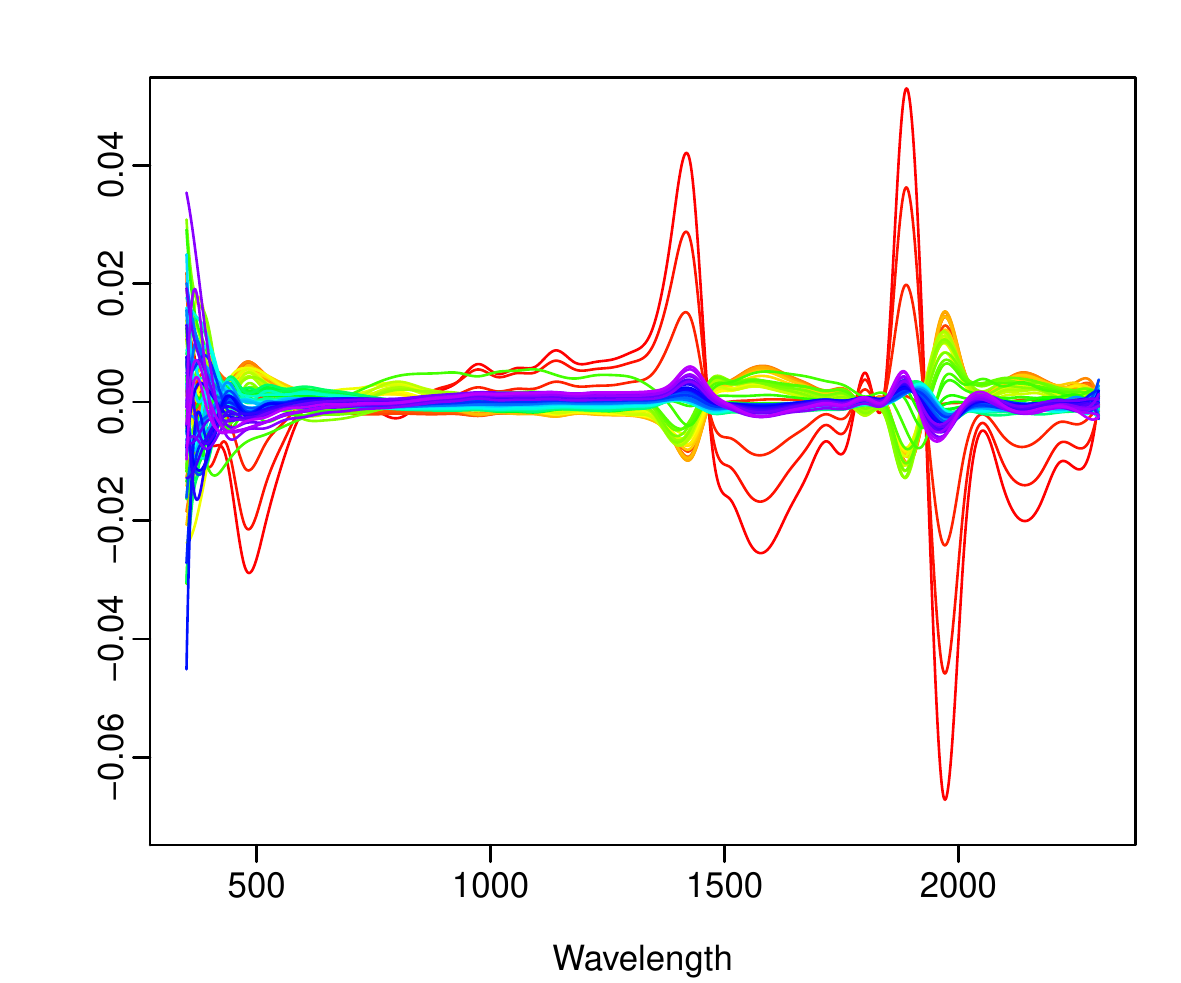}\label{fig:1f}} 
\caption{NIR absorption spectra of wood panels and residual functions after extraction of the first two dynamic functional components associated with largest empirical eigenvalues of sample long-run covariance function. Using rainbow plots, curves from the distant past are shown in red, and the most recent curves are in violet.}\label{fig:1}
\end{figure}

The rapid improvements in automated data acquisition technology allow researchers to access functional data more frequently. 
Functional data sequentially recorded over time are often considered as finite realizations of a functional stochastic process $\{\X_t(u)\}_{t\in\mathbb{Z}}$, where the time parameter $t$ is discrete, and the parameter $u$ is a continuum bounded within a finite interval domain $[a,b]$. Observations $\{\X_t(u)\}_{t\in\mathbb{Z}}$ are commonly referred to as functional time series. Functional time series can arise when a continuous-time record is separated into natural consecutive time intervals. Examples include daily concentration curves of particulate matter with an aerodynamic diameter of less than 10~$\mu m$ \citep[e.g.,][]{HK15} and monthly sea surface temperature in the ``Ni\~{n}o region'' \citep[e.g.,][]{SH11}. Alternatively, functional time series can arise when observations that are continuous functions in nature are repeatedly sampled in a period. For example, Figure~\ref{fig:1a} displays near-infrared (NIR) spectra recorded in monitoring glue curing process of wood panels in 72 experimental trials. The curves in the plot are ordered chronologically according to the colors of the rainbow \citep{HS10}.

Functional time series methods and theory have witnessed an upsurge in literature contributions in the past two decades \citep[see, e.g.,][]{Bosq00,Bosq08,KR13,ANH15,HS09,KK16,KKW16,LRS20}. Most existing functional time series modeling methods, including those in the references cited above, rely on functional principal component analysis (FPCA) to project the intrinsically infinite-dimensional functional objects onto directions of a small number of leading functional principal components. FPCA extracts only the dominant modes of variation of a functional object over its entire domain, with captured information referred to as the ``main features'' of the considered process. However, the ``minor'' components neglected by FPCA often have highly localized features possessing information on functional variations over particular short intervals within the function domain. A relatively recent dynamic FPCA introduced by \cite{HK15} employs long-run covariance to include serial dependence of the data, but suffers the same problem of loss of local features in dimension reduction. The problem of FPCA inadequately extracting local features is illustrated in Figure~\ref{fig:1}. The eigenanalysis results shown in Figure~\ref{fig:1e} indicate that the first two leading dynamic functional components explain most functional variation of smoothed NIR spectra (see Section~\ref{sec:6} for details of smoothing). Removing the empirical functional principal components from observations, residual functions of dynamic FPCA still contain sharp features around 1300~nm and 1900~nm of wavelength, as shown in Figure~\ref{fig:1f}. Thus, local features are important for the estimation of functional time series, typically in the study of NIR spectroscopy data that possess multiple significant local features.

Based on molecular overtones and combination vibrations of the investigated molecule, NIR spectroscopy generates complex absorption spectra over a region of the electromagnetic spectroscopy. Since many chemical compounds are known to have characteristic absorption bands over certain spectrum regions between 780--2500 nm, to determine composition materials of an object requires studying particular wavelength ranges (i.e., narrow bands with extreme absorption intensity) of the observed NIR spectrum together instead of examining absorptions one frequency at a time \citep{BC07}. Thus, a computational method that can extract ``local features'' covering multiple frequencies of absorption spectrum is important for NIR spectroscopy analysis in practice. Use the wood panel NIR spectrum illustrated in Figure~\ref{fig:1} above as an example. The observed local features between 350--2300~nm linked to composition materials, namely, the wood substrate, curing resin, and moisture content \citep{CS+18}. Subtle changes in experimental conditions such as temperature and pressure lead to variations of absorption bands over a series of trials. Hence, extracting and modeling local features are essential for monitoring the glue curing process of wood panels. Moreover, local features inheriting serial dependence of the original NIR curves can be used to make forecasts for future experiments. In this paper, we aim at developing a methodology for recovering local features that are ignored by FPCA and for using these extracted local features to make more accurate estimations and forecasts for functional time series.

Most existing feature extraction methods attempt to capture local features of functional data by either restricting function domain \citep[see, e.g.,][]{HH16,GCND14} or introducing sparseness penalty parameters \citep[see, e.g.,][]{HSB09, AW19} during dimension reduction. However, truncating function domains to specific intervals to enhance local feature extraction requires well aligned curves with most local features occurring in the same region. Thus, truncating methods are not suitable for analysis of NIR spectroscopy data that generally focus on identifying non-overlapping absorption spikes in observed spectra. In contrast, sparse FPCA methods impose sparsity penalties in regularized eigendecomposition to identify basis functions with local features. However, a single penalty parameter in practice is not sufficient to accommodate for local features of various magnitudes at different scales. As a result, solving optimization problems to identify the optimal penalty parameter can be tricky: a small penalty results in a significant amount of observation noise falsely identified as local features, while a large penalty fails to preserve peak heights of high-magnitude local features. 

Unlike the feature extraction methods mentioned above, \cite{JL09} considered extracting principal components of high-dimensional data in wavelet domains. The wavelet bases are considered to be natural for uncovering sparse local features in the signal for the following four reasons. First, wavelet transform is a spatially varying decomposition that adapts its effective ``window width'' to magnitudes of local oscillations in FPCA residual functions. As a result, wavelet-based algorithms can accurately estimate local features at various scales. Second, orthonormal bases of compactly supported wavelets are particularly good at estimating sharp, highly localized features. This character of wavelet transform allows effective detection of local features associated with chemicals that have very narrow absorption bands (i.e., short intervals of wavelength frequencies) but high intensities (i.e., large absorbance coefficients) in NIR spectroscopy data \citep{BC07}. Third, the wavelet transform is computationally efficient. For a given orthonormal wavelet basis, feature extraction can be completed in one step of matrix multiplication known as the ``discrete wavelet transform'' (for further detail on discrete wavelet transform, see \cite{Strang89,D92}). Fourth and the most important, many types of functional forms encountered in practice, including NIR absorption spectrum, can be sparsely and uniquely represented by a series of wavelet coefficients. Thus, wavelet transform allows a parsimonious representation of local features using only a relatively small number of estimated coefficients. 

We propose a two-step algorithm that captures global and local features of functional time series sequentially. Initially, dynamic FPCA is applied to extract global features from the smoothed functional time series. Residuals of dynamic FPCA are then transformed into wavelet domains and block thresholding of wavelet (BTW) coefficients are conducted. Advantages of the FPCA-BTW method over sparse FPCA methods in relation to local feature extraction are demonstrated using simulated data in Section~\ref{sec:5.1}, and via an empirical application in Section~\ref{sec:6}. It should be noticed that neither conducting the BTW alone, or conducting the BTW before dynamic FPCA, would effectively capture most global and local features of functional time series in a parsimonious set of estimated wavelet coefficients: First, wavelet approximations requires a fairly large number of coefficients (e.g., $2^{11}$ for the wood panel spectra, and the number of coefficients would increase if more spectrum frequencies are considered) to summarize all global and local features of a continuous function consisting of non-zero signals over its entire domain. More details of wavelet approximations will be presented in Section~\ref{sec:2.3} later. Second, implementing BTW leads to a trade-off between preserving the overall smoothness and attaining to fine details of the true signal \citep[see, page 942, Figure 1 in][for a depiction of this trade-off]{AF01}. As a result, in practice many local features need to be sacrificed to minimize estimation errors measured by an $L^2$ norm for functional time series. In contrast, after conducting FPCA in the initial step of  our proposed FPCA-BTW method isolates significant local features in the format of sparse ``spikes'' over short segments of a function that contains no signal but noise elsewhere. Then, performing the BTW in the second step yields only a small number of non-zero estimated wavelet coefficients containing information on local features as the thresholding algorithm reduces the remaining least important coefficients to zero.

To the best of our knowledge, there is no precedent research focusing on improving FPCA estimation performance via adequately extracting local features contained in ``minor'' functional components. The principal orthogonal complement thresholding method of \cite{FLM13} for the estimation of a high dimensional covariance with a conditional sparsity structure is closely analogous to our work as both methods attempt to produce improved estimation performance for processes consisting of finite common global features and sparse local features.

The rest of the paper is organized as follows. In Section~\ref{sec:2}, we provide necessary background on FPCA and wavelet approximation, before introducing the FPCA-BTW feature extraction method. Implementation details of the proposed method in estimation and forecasting of functional time series are given in Section~\ref{sec:3}. Section~\ref{sec:4} presents asymptotic properties of FPCA-BTW estimators. In Section~\ref{sec:5}, we use Monte Carlo simulations to illustrate finite sample performances of FPCA-BTW estimators regarding estimation and forecasting of functional time series. Section~\ref{sec:6} presents real data applications on NIR spectroscopy data of wood panels. Finally, Section~\ref{sec:7} concludes the paper and provides some discussion and directions for future research.

\section{Methodology}\label{sec:2}

\subsection{Notations}\label{sec:2.1}

We start by fixing the notations used in this paper. Let $\{\X_t(u) \}_{ t\in \mathbb{Z}}$ denote random functions defined on a rich enough probability space $(\Omega, \mathcal{A}, P)$. Observations $\{\X_t(u)\}_{t\in \mathbb{Z}}$ are elements of the Hilbert space $H = L^2([0, 1])$ equipped with the inner product $\inp{x}{y} = \int_0^1 x(u) y(u)du $. Each $\X_t$ is a square integrable function satisfying $\norm{\X_t}^2  = \int_{0}^{1}\X_t^2(u)du < \infty $, where the standard norm on $L^2([0,1])$ is defined as $\norm{x} = \langle x,x\rangle^{1/2}$. Define a notation $\X \in L^p_H(\Omega, \mathcal{A}, P)$ such that, for some $p >0$,  $E\norm{\X}^p < \infty$.

We consider functional time series $\{\X_t(u)\}_{t \in \mathbb{Z}}$ with a general representation given by
\begin{equation}
\X_t(u)  = \mu(u) + \sum_{k=1}^{K}\beta_{t,k} \phi_k(u) + Z_t(u) + \varepsilon_t(u), \qquad u \in [0,1],\label{eq_1}
\end{equation}
where $\mu(u) = E[\X(u)]$ is the mean function; $\{\phi_k(u)\}_{k=1}^K$ are real-valued orthogonal functions with $K$ a fixed positive integer; a set of pairwise uncorrelated real numbers $\{\beta_{t,k}\}_{k=1}^K = \{\beta_{t,1}, \ldots, \beta_{t,K}\}$ satisfy that $var(\beta_{t,i},\beta_{t,j}) = 0$ for any $i \neq j$; $\{Z_t(u)\}_{t \in \mathbb{Z}}$ is a set of functions uncorrelated with $\{\phi_k(u)\}_{k=1}^K$; $\{\varepsilon_t(u)\}_{t \in \mathbb{Z}}$ is $H$-white noise with $E\left\{\varepsilon_t(u)\right\}=0$. (See Chapter~3 of \cite{Bosq00} for further detail about strong white noise function in Hilbert space.) The $\sum_{k=1}^{K}\beta_{t,k} \phi_k(u)$ in~\eqref{eq_1} containing dominant modes of variation of $\{\X_t(u)\}_{t \in \mathbb{Z}}$ are referred to as ``global features'', whereas $\{Z_t(u)\}_{t\in\mathbb{Z}}$ with sparse localized spikes over the function domain $[0,1]$ are referred to as ``local features''. We assume that all eigenvalues of long-run covariance of local features are bounded, and the first $K$ eigenvalues of long-run covariance function of global features decrease at the rate of $O(1)$. Extraction of global features and local features from functional time series $\{\X_t(u) \}_{ t\in \mathbb{Z}}$ are introduced in Sections~\ref{sec:2.2} and~\ref{sec:2.3}, respectively.

\subsection{Extraction of global features}\label{sec:2.2}
A weakly stationary functional time series $\{\X_t(u)\}_{t \in \mathbb{Z}}$ satisfies that, for all $t \in \mathbb{Z}$, 
\begin{inparaenum}[(a)]
	\item $\X_t(u) \in L^2([0,1])$,
	\item $E[\X_t(u)] = E[\X_0(u)] = \mu(u)$, and
	\item for all $\ell \in \mathbb{Z}$ and $u,s\in [0,1]$, 
	\begin{equation}
		c_{\ell} (u,s) = \text{cov}[\X_0(u), \X_{\ell}(s)] =  \text{cov}[\X_{t+\ell}(u), \X_{t}(s)],
		\label{eq_2}
	\end{equation}
  with $\text{cov}[\X(u),\X(s)] = E[\{\X(u) - \mu(u)\}\{\X(s) - \mu(s)\}]$.
\end{inparaenum}
$C_{\ell}$ induces an operator $L^2([0,1]) \rightarrow L^2([0,1])$ given by
\begin{equation*}
  C_{\ell}(x)(u)  = \int_{0}^{1} C_{\ell}(u,s)x(s) \diff s, \qquad x \in L^2([0,1]), \quad u, s \in [0,1].
\end{equation*}
When $\ell = 0$, the autocovariance operator $C_{\ell}$ has a special case of covariance operator $C_0$ defined by $C_0(x) = \int_{0}^{1} C_0(u,s)x(s) \diff s$ for $x \in L^2([0,1])$ and $u, s \in [0,1]$.

In practice, $\{\X_t(u)\}_{t \in \mathbb{Z}}$ often consists of serially correlated observed trajectories. To incorporate serial dependence carried by lagged observations, recent studies \citep[see, e.g.,][]{RS17, Shang19} suggest computing a long-run covariance function $C(u,s)$  as
\begin{equation}
C(u,s) = \sum_{\ell = -\infty}^{\infty} c_{\ell} (u,s), \qquad u,s \in [0,1].
\label{eq_3}
\end{equation}
A long-run covariance operator $C$ is then defined as
\begin{equation*}
C(x)(u) = \int_{0}^1 C(u,s) x(s) \diff s \qquad x \in L^2([0,1]), \quad u, s \in [0,1].
\end{equation*} 
The symmetric positive-definite Hilbert-Schmidt operator $C$ admits a decomposition as
\begin{equation*}
  C(x) = \sum_{k=1}^{\infty}\lambda_k \inp{\phi_k}{x}\phi_k,  \qquad x \in L^2([0,1]),
\end{equation*}
where $\{\lambda_k\}_{ k \in \mathbb{Z}^+}$ are the nonincreasing eigenvalues, and $\{\phi_k \}_{ k \in \mathbb{Z}^+}$ the corresponding orthonormal eigenfunctions such that $C(\phi_k) = \lambda_k \phi_k$, and $\inp{\phi_i}{\phi_j} = 1$ iff $i=j$. The Karhunen--Lo\`{e}ve expansion of a stochastic process $\X_t(u)$ is then given by
\begin{equation*}
  \X_t(u) = \mu(u) + \sum_{k=1}^{\infty}{\beta}_{t,k}\phi_k(u),
\end{equation*}
where the $k$th functional component score $\beta_{t,k}$ is a projection of $\overline{\X}_t(u) =  \X_t(u) - \mu(u)$ in the direction of the $k$th eigenfunction $\phi_k(u)$, that is, $\beta_{k} = \inp{\overline{\X}_t}{\phi_k} $. 

According to~\eqref{eq_1}, the main features of the infinite-dimensional $\{\X_t(u)\}_{t \in \mathbb{Z}}$ can be summarized by its first $K$ leading components as
\begin{equation}
  \X_t(u) = \mu(u) + \sum_{k=1}^{K}{\beta}_{t,k}\phi_k(u) + e_t(u),
  \label{eq_4}
\end{equation}
where $\{e_t(u)\}_{t \in \mathbb{Z}}$ are error functions after truncation. According to Theorem 2 of \cite{HK15}, the linear combination of $\sum_{k=1}^{K}{\beta}_{t,k}\phi_k(u)$ obtained by dynamic FPCA satisfies that, for any other orthonormal basis $\{\varphi_k\}_{k \in \mathbb{Z}^+ }$ of Hilbert space $H$,
\begin{equation}
  E \left[ \norm{\overline{\X}_t - \sum_{k=1}^{K}\beta_{t,k}\phi_k }^2 \right]  \leq E \left[ \norm{\overline{\X}_t - \sum_{k=1}^{K}\inp{\X_t}{\varphi_k}\varphi_k}^2 \right].
  \label{eq_5}
\end{equation}

In rare cases, functional time series $\{\X_t(u)\}_{t \in \mathbb{Z}}$ may possess weak serial dependence. The significance of serial dependence can be determined according to the hypothesis test of \cite{HRW16}. Functional observations are treated as independent if $c_{\ell}$ of~\eqref{eq_2} at all lags apart from $\ell = 0$ are tested to be negligible. A process that decomposes the covariance operator $C_0$ to extract global features is often referred to as static FPCA to distinguish it from dynamic FPCA. In the remaining of this paper, we present feature extraction results obtained by dynamic FPCA and include feature extraction results associated with static FPCA in the Supplementary document. The aim of this paper is to demonstrate the proposed local feature extraction method can be applied to improve performances of static FPCA and dynamic FPCA, instead of comparing performances of the two versions of FPCA.

It can be seen from~\eqref{eq_5} that dynamic FPCA can find an optimal representation of global features of $\{\X_t(u)\}_{t \in \mathbb{Z}}$, but ignores most local features $\{Z_t(u)\}_{t \in \mathbb{Z}}$. Our proposed two-step feature extraction method will continue to capture any remaining local features from FPCA residuals, as described in Section~\ref{sec:2.3}.

\subsection{Extraction of local features}\label{sec:2.3}
To extract sharp and highly localized features from FPCA residuals, we consider an orthonormal system of wavelet functions. Wavelet functions combine compact support with various degrees of smoothness, which enables the extraction of signals at a variety of different scales. It has been tested that wavelets can effectively isolate signals from noisy functions in statistical applications \citep[see, e.g.,][]{Antoniadis07,Ogden2012}. Most recent wavelet applications in statistics adopt the approach of \cite{D92} to define two related and specially selected orthonormal parent wavelet functions: the scaling function $\psi$ and the mother wavelet $\Psi$. Wavelets can then be generated by dilation and translation as
\begin{equation*}
	\psi_{j,p} = 2^{j/2} \psi(2^{j} t-p), \quad \Psi_{j,p} = 2^{j/2} \Psi(2^jt-p), \qquad (j \in \mathbb{Z}^{+}, \quad p = 1, \ldots, 2^j),
\end{equation*}
where the index $j$ represents resolution level in wavelet decomposition. This wavelet system produces wavelet functions forming an orthonormal wavelet basis in $L^2([0,1])$. With a primary decomposition level $j_0 \geq 0$, local features $\{Z_t(u)\}_{t \in \mathbb{Z}}$ admit a decomposition given by
\begin{equation}
  Z_t(u) = \sum_{p=1}^{2^{j_0} } D'_{j_0,p,t} \psi_{j_0,p}(u) + \sum_{j={j_0}}^{\infty} \sum_{p=1}^{2^j} D_{j,p,t} \Psi_{j,p}(u),
  \label{eq_6}
\end{equation}
where wavelet coefficients are defined as
\begin{equation*}
  D'_{j_0,p,t} = \int_{0}^1Z_t(u) \psi_{j_0,p}(u)\diff u, \quad D_{j,p,t} = \int_{0}^1Z_t(u) \Psi_{j,p}(u)\diff u.
\end{equation*}
``Approximations'' and ``details'' of $Z_t(u)$ are stored in wavelet coefficients $D'_{j_0,p,t}$ and $D_{j,p,t}$, respectively \citep{Mallat99}.

According to~\eqref{eq_1}, residual functions $\{e_t(u)\}_{t \in \mathbb{Z}}$ consist of highly localized features $Z_t(u)$ and random noise $\varepsilon_t(u)$ given by
\begin{equation*}
  e_t(u) = Z_t(u) + \varepsilon_t(u).
\end{equation*}
The wavelet transform of $e_t(u)$ can be expressed as
\begin{equation*}
   e_t(u) = \sum_{p=1}^{2^{j_0} } \widetilde{D}'_{j_0,p,t} \psi_{j_0,p}(u) + \sum_{j={j_0}}^{\infty} \sum_{p=1}^{2^j} \widetilde{D}_{j,p,t} \Psi_{j,p}(u),
 \end{equation*} 
where the empirical wavelet coefficients $\widetilde{D}'_{j_0,p,t}$ and $\widetilde{D}_{j,p,t}$ are given by
\begin{equation*}
  \widetilde{D}'_{j_0,p,t} = \int_{0}^1e_t(u) \psi_{j_0,p}(u)\diff u, \quad \widetilde{D}_{j,p,t} = \int_{0}^1e_t(u) \Psi_{j,p}(u)\diff u.
\end{equation*}
Wavelet coefficients related to detailed structure of $e_t(u)$ and $Z_t(u)$ thus satisfy that, for any $t \in \mathbb{Z}$,
\begin{equation}
 \widetilde{D}_{j,p,t} = D_{j,p,t} + \epsilon_{j,p,t}, \qquad (j \in \mathbb{Z}^{+}, \quad p = 1, \ldots, 2^j),
 \label{eq_7}
\end{equation}
where $\epsilon_{j,p,t} =  \int_{0}^1\varepsilon_t(u)\Psi_{j,p}(u)\diff u$ represents a wavelet transform of contamination noise. Since local features $\{Z_t(u)\}_{t \in \mathbb{Z}}$ are sparse, a vector of wavelet coefficients $D_t = \{D_{j_0,1,t}, \ldots, \allowbreak D_{j_0,2^{j_0},t}, \ldots\}$ contains many zeros. Extracting local features is then equivalent to determining non-zero wavelet coefficients $D_{j,p,t}$. From a statistical modeling perspective, the denoising problem of~\eqref{eq_7} has been commonly approached by shrinking the empirical wavelet coefficients $\{\widetilde{D}_{j,p,t}\}_{j \in \mathbb{Z}^{+} , p = 1, \ldots, 2^j}$ one by one \citep[see, e.g.,][]{DJ94,AF01}. However, local features of functional data often occur over short intervals within the function domain that correspond to several consecutive wavelet coefficients at fine resolution levels. To determine chemical content of an object by NIR spectroscopy, simultaneously considering the non-zero wavelet coefficients corresponding to certain distinctive absorption bands of known chemical compounds provides more accurate composition results than examining absorption value at any single frequency. For example, local features depicting extreme absorption bands of approximately 1900~nm, shown in Figure~\ref{fig:1f}, are summarized into 21 consecutive empirical wavelet coefficients at the resolution level $j=11$. Thus, to enhance extraction of local features, adjacent wavelet coefficients should be modeled together as a group. For this purpose, we adopt a block thresholding approach of \cite{Cai02} to make simultaneous selection of empirical wavelet coefficients in groups as follows. At each resolution level $j$, divide the empirical wavelet coefficients $\widetilde{D}_{j,p,t}$ into non-overlapping blocks of length $L$. Denote indices of the coefficients in the $a$th block at level $j$ by $j_a$, i.e., $j_a = \{(j,p): (a-1)L+1 \leq p \leq aL\}$. Let $S^2_{j_a} = \sum_{p\in j_a} \widetilde{D}_{j,p,t}^2$ denote the sum of squares of the empirical coefficients in the block. A block is significant if its $S^2_{j_a}$ is larger than a threshold $T_w = \lambda^{\star} L \sigma^2/2^J$, where $\lambda^{\star}$ is a threshold constant and $\sigma$ is the noise level. Retaining significant wavelet coefficients while discarding the remaining negligible coefficients leads to a local feature estimator as
\begin{equation}
\widehat{Z}_t(u) = \sum_{k=0}^{2^{j_0} } \widetilde{D}'_{j_0,p,t} \psi_{j_0,p}(u) + \sum_{j={j_0}}^{J-1} \sum_{a} \left(\sum_{p \in j_a} \widetilde{D}_{j,p,t} \Psi_{j,p}(u) \mathds{1}(S^2_{j_a} > T_w)\right),
\label{eq_8}
\end{equation}
where $a$ varies for different resolution levels and $\mathds{1}(\cdot)$ represents the binary indicator function.

In~\eqref{eq_8}, the block length $L$ and the threshold constant $T_w$ together control global and local adaptivity of the estimator $\widehat{Z}_t(u)$. A global adaptive estimator adjusts to the overall regularity of the target function, and a locally adaptive estimator focuses on optimally adapting to subtle and highly localized features along the curve. The optimal selection of parameters $L$ and $T_w$, together with other implementation details about the FPCA-BTW feature extraction method, are described in Section~\ref{sec:3}. 

\section{Implementation details}\label{sec:3}

\subsection{Long-run covariance estimation}\label{sec:3.1}
We first present technical details of extracting global features of a finite sample functional time series. To consider serial dependence of stationary functional observations $\{\X_t(u)\}_{t=1}^T$, we compute the empirical long-run covariance function as
\begin{equation}
  \widehat{C}_{h, q}(u,s) = \sum_{\ell = -T}^{T} W_q \left( \frac{\ell}{h} \right) \widehat{c}_{\ell}(u,s),
\label{eq_9}
\end{equation}
where $W_{q}$ is a symmetric weight function with bounded support of order $q$, and $h$ is a bandwidth parameter; the estimator of $c_{\ell}(u,s)$ is defined in the form of
\begin{equation*}
\widehat{c}_{\ell}(u,s) = \begin{cases}
\frac{1}{T} \sum_{j=1}^{T-\ell} \left[\X_j(u) - \widehat{\mu}(u) \right] \left[\X_{j+\ell}(s) - \widehat{\mu}(s) \right], & \ell \geq 0; \\
\frac{1}{T} \sum_{j=1-\ell}^{T} \left[\X_j(u) - \widehat{\mu}(u) \right] \left[\X_{j+\ell}(s) - \widehat{\mu}(s) \right], & \ell < 0.
\end{cases}
\end{equation*}
The optimal bandwidth parameter $h$ is selected via the ``plug-in'' algorithm proposed in \cite{RS17}. More details about estimating the corresponding $\widehat{C}_{\widehat{h}_\text{opt}}(u,s)$ are provided in Appendix~\ref{app_b1} in the Supplementary document. The empirical long-run covariance operator is then given by
\begin{equation*}
  \widehat{C}(x)(u)=\int_{0}^{1} \widehat{C}_{\widehat{h}_\text{opt}}(u,s) x(s) \diff s, \qquad x \in  L^2([0,1]).
\end{equation*}

Performing eigendecomposition on the empirical long-run covariance operator yields
\begin{equation*}
   \widehat{C}(x) = \sum_{k=1}^{\infty}\widehat{\lambda}_k \inp{\widehat{\phi}_k}{x}\widehat{\phi}_k,  \qquad x \in L^2([0,1]),
\end{equation*}
where $\{\widehat{\phi}_k\}_{k\in\mathbb{Z^{+}}}$ are the empirical eigenfunctions, and $\{\widehat{\lambda}_k\}_{k\in\mathbb{Z^{+}}}$ are associated eigenvalues.
To facilitate dimension reduction, the dimension of global features $\widehat{K}$ need to be empirically determined. Existing functional time series methods generally select $\widehat{K}$ by requiring that retained functional components should explain a certain level of the total variance, approximately 85\% \citep[see, e.g.,][]{Chiou12, HK15, Shang19}. However, this criterion of cumulative percentage of explained variation has the disadvantage of incorrectly selecting too many components as global features when fast-diverging eigenvalues are present in FPCA analysis. To precisely extract global features, following \cite{LRS20}, the value of $K$ is determined as the integer minimizing ratios of two adjacent empirical eigenvalues given by
\begin{equation}
\widehat{K} = \argmin_{1\leq k \leq k_{max} }\left\{ \frac{\widehat{\lambda}_{k+1}}{\widehat{\lambda}_k} \times \mathds{1}\left(\frac{\widehat{\lambda}_{k}}{\widehat{\lambda}_1} \geq \tau \right) +  \mathds{1}\left(\frac{\widehat{\lambda}_{k}}{\widehat{\lambda}_1} < \tau \right) \right\},
\label{eq_10}
\end{equation}
where $k_{max}$ is a prespecified positive integer, $\tau$ is a prespecified small positive number, and $\mathds{1}(\cdot)$ is the indicator function. When without priori information about a possible maximum of $K$, it is unproblematic to choose a relatively large $k_{max}$, e.g., $k_{max} = \#\{k| \widehat{\lambda}_k \geq \sum_{k=1}^{T}\widehat{\lambda}_k/T, k\geq 1 \}$ \citep{AH13}. Given that the small empirical eigenvalues $\widehat{\lambda}_k$ for some $K<k<k_{max}$ are likely to be practically zero, we adopt the threshold constant $\tau = 1/\ln({\max{\{\widehat{\lambda}_1},T\}})$ to ensure consistency of $\widehat{K}$.

As described in Section~\ref{sec:2.2}, it is possible to have nearly independent $\{\X_t(u)\}_{t=1}^T$ in practice. The sample covariance operator of independent observations is computed as
\begin{equation*}
  \widehat{C}_0(x)(u)=\int_{0}^{1} \widehat{c}_0(u,s) x(s) \diff s, \qquad x \in  L^2([0,1]),
\end{equation*}
where $\widehat{c}_0(u,s) = \frac{1}{T}\sum_{t=1}^{T}[\X_t(u)-\widehat{\mu}(u)][\X_t(s)-\widehat{\mu}(s)]$ is the empirical covariance kernel with the empirical mean function $\widehat{\mu}(u) = \frac{1}{T}\sum_{t=1}^{T}\X_t(u)$. For such data static FPCA is applied to extract global features from $\widehat{C}_0(x)(u)$ using the same criterion as~\eqref{eq_10}. 

With FPCA results, functional time series can be estimated by
\begin{equation*}
	\widehat{\X}_t(u) = \widehat{\mu}(u) + \sum_{k=1}^{\widehat{K}}\widehat{\beta}_{t,k} \widehat{\phi}_k(u), 
\end{equation*}
where $\sum_{k=1}^{K}\widehat{\beta}_{t,k} \widehat{\phi}_k(u)$ represents the extracted global features, with the empirical principal component scores defined by $\widehat{\beta}_{t,k} = \int_{0}^1 \left[\X_t(u) - \widehat{\mu}(u)\right] \widehat{\phi}_k(u) \diff u$. Removing the estimated mean function and the extracted global features from functional observations leaves residual functions given by
\begin{equation*}
	\widehat{e}_t(u) = \X_t(u) - \widehat{\mu}(u) - \sum_{k=1}^{\widehat{K}}\widehat{\beta}_{t,k} \widehat{\phi}_k(u).
\end{equation*}
In Section~\ref{sec:3.2}, we present details of recovering local features from $\{\widehat{e}_t(u)\}_{t=1}^{T}$ through block thresholding of wavelet coefficients.

\subsection{Estimation of wavelet coefficients}\label{sec:3.2}
The continuous wavelet transform formalized by \cite{GM84} can be implemented in computer software such as \textbf{R} \citep{Team15} to extract local features from FPCA residuals $\{\widehat{e}_t(u)\}_{t=1}^T$. However, in practice, we most likely only observe discretized values $\{\X_t(u_i)\}_{i=1}^{n_t}$, with $n_t$ denoting the number of grid points in the $t$-th curve. Removing global features evaluated at each grid point leaves discrete residuals $\{\widehat{e}_t(u_i)\}_{i=1}^{n_t}$. When equally spaced grids satisfy $n_t = 2^J$ for $t = 1, \ldots, T$, wavelet transform of $\{\widehat{e}_t(u)\}_{t=1}^T$ can be performed in $O(2^J)$ operations \citep{Mallat89}. 

In situations when functional observations have nondyadic, varying or unequally spaced grid points, the non-linear regularized Sobolev interpolator of \cite{AF01} is adopted to perform the wavelet transform. Local feature extraction can then be completed in the following steps. First, select an orthonormal wavelet family to obtain an orthogonal DWT base matrix ${W}$ with dimension $N \times N$, where $N = 2^J \geq \max(n_1,\ldots,n_T)$ is a dyadic integer. There are many discrete wavelet families available in the literature. We follow \cite{ZOR12} and consider the Daubechies least asymmetric wavelets with 10 vanishing moments in the analysis of NIR spectroscopy data. Denote ${A}$ as a matrix of dimension $n_t \times N$ whose  $i$th row corresponds to the row of the matrix ${W}^{\T}$. We interpolate the vector $\widehat{{e}}_t = [\widehat{e}_t(u_1), \ldots, \widehat{e}_t(u_{n_t})]^{\T}$ as
\begin{equation}
  \widetilde{{D}}_t =  {A}^{\T}\widehat{{e}}_t,
  \label{eq_11}
\end{equation}
where $\widetilde{{D}}_t = [\widetilde{D}'_{j_0,1,t}, \ldots, \widetilde{D}'_{j_0,2^{j_0},t}, \widetilde{D}_{j_0,1,t}, \ldots, \widetilde{D}_{j_0,2^{j_0},t}, \ldots, \widetilde{D}_{J-1,1,t}, \ldots, \widetilde{D}_{J-1,2^{J-1},t}]^{\T}$ is a vector of size $N$ \citep{AF01}. The optimal parameters for block thresholding are then selected according to \cite{Cai02}. Specifically, for the block size $L$ and the threshold constant $\lambda^{\star}$ in~\eqref{eq_8}, $L = 2^{\lfloor \log_2 (\ln(2^J)) \rfloor}$ and $\lambda^{\star} = 4.5052$ are chosen. Noise level of residual functions are estimated by taking the median absolute deviation (MAD) as
\begin{equation*}
  \widehat{\sigma} = \frac{\text{MAD}\{\widetilde{D}_{J-1,p,t}/v_{J-1,p,t}^{1/2}: v_{J-1,p,t} > 0.0001 \}}{0.6745},
\end{equation*}
where $\{\widetilde{D}_{J-1,p,t}\}_{p=1,\ldots,2^{J-1}}$ are the empirical wavelet coefficients at the resolution level $J-1$, and $\{v_{J-1,p,t}\}_{p=1,\ldots,2^{J-1}}$ are diagonal elements of the matrix ${V} = {A}^{\T}{A}$ \citep{AF01}. Next, the first-round block thresholding is implemented according to~\eqref{eq_8}, and intermediate results are denoted as $\widetilde{{D}}^{\ast}_t$. Subtracting the inverse transform of $\widetilde{{D}}^{\ast}_t$ from discrete residuals gives
\begin{equation*}
  \widehat{{e}}^{\ast}_t = \widehat{{e}}_t - {A}\widetilde{{D}}^{\ast}_t.
\end{equation*}
The second round empirical wavelet coefficients are then computed as
\begin{equation*}
   \widetilde{{D}}^{\dagger}_t = \widetilde{{D}}^{\ast}_t + {A}^{\T} \widehat{{e}}^{\ast}_t.
\end{equation*}
Finally, performing block thresholding again on $\widetilde{{D}}^{\dagger}_t$ yields the final BTW coefficients $\widehat{{D}}_t = [\widehat{D}'_{j_0,1,t}, \ldots, \widehat{D}'_{j_0,2^{j_0},t}, \widehat{D}_{j_0,1,t}, \ldots, \widehat{D}_{j_0,2^{j_0},t}, \ldots,  \widehat{D}_{J-1,1,t}, \ldots, \widehat{D}_{J-1,2^{J-1},t}]^{\T}$ with many zero entries reflecting the sparseness of the local features. Note that we keep the ``approximation'' wavelet coefficients unchanged as $\widehat{D}'_{j_0,p,t} = \widetilde{D}'_{j_0,p,t}$ for all $p = 1, \ldots, 2^{j_0}$. According to \cite{Solo2001}, implementing the estimation method of \cite{AF01} through a two-round block thresholding process simplifies computation. Applying the above procedure to each discrete residual function $\widehat{e}_t$ leads to a sparse $N \times T$ matrix of BTW coefficients $\widehat{{D}} = [\widehat{{D}}_{1}, \ldots, \widehat{{D}}_{T}]$. The extracted local features are then given by a product ${A} \widehat{{D}}$. 

Using the extracted global and local features, we can make improved estimation of the considered functional process and its covariance structure, and produce more accurate forecasts. We demonstrate applications of the proposed feature extraction method using simulated samples in Section~\ref{sec:5} and real NIR spectroscopy data in Section~\ref{sec:6}. Additional technical details about long-run covariance estimation and applications of the FPCA-BTW method are provided in Appendix~\ref{app_b2} in the supplementary document.

\section{Asymptotic properties}\label{sec:4}

Before presenting assumptions and asymptotic results of long-run covariance based FPCA-BTW estimators, we introduce some notations. Let $\LL = \LL(H,H)$ be the space of bounded linear operators from $H$ to $H$. We define the operator norm $\norm{A}_{\LL} = \sup_{\norm{x}\leq1} \norm{A(x)}$ for $A \in \LL$.
The operator $A$ is compact if there exists two orthonormal bases $\{\nu_k\}$ and $\{v_k\}$, and a real sequence $\{\lambda_k\}$ converging to zero, such that 
\begin{equation*}
  A(x) = \sum_{k=1}^{\infty} \lambda_k \inp{x}{\nu_k}v_k, \qquad x \in H.
\end{equation*}
A compact operator is said to be a Hilbert-Schmidt operator if $\sum_{k=1}^{\infty}\lambda_k^2 < \infty$. We denote the Hilbert-Schmidt norm by $\norm{A}_{\mathcal{S}}$. For any Hilbert-Schmidt operator $A$, one can show that $\norm{A}_{\mathcal{S}}^2 = \sum_{k\geq1}\lambda_k^2$ and $\norm{A}_{\LL} \leq \norm{A}_{\mathcal{S}}$ \citep[][Chapter 2]{HK12}. 

\begin{assumption}
	Functions $\{\X_t(u), u\in[0,1]\}_{t\in\mathbb{Z}}$ are $L^4\operatorname{-}m\operatorname{-}$approximable, taking values in $L^2([0,1])$, satisfying the following conditions:
	\begin{enumerate}[(i)]
		\item  $X_t$ admits the representation $\X_t = f(\delta_t, \delta_{t-1}, \delta_{t-2} \ldots, \delta_{t-m+1}, \delta_{t-m}, \delta_{t-m-1}, \ldots)$ with $\delta_i$ i.i.d. elements taking values in a measurable space $S$ and a measurable function $f: \mathcal{S}^{\infty} \rightarrow H$. \label{cd1}
		\item $E\norm{\X_0}^{4+d} < \infty$ for some $d >0$, and \label{cd2}
		\item $\{\X_t(u), u\in[0,1]\}_{t \in \mathbb{Z}}$ can be approximated by $m$-dependent sequences
		\begin{equation*}
		\X_t^{(m)} = f(\delta_t, \delta_{t-1}, \delta_{t-2} , \ldots, \delta_{t-m+1}, \delta_{t,t-m}^{(m)}, \delta_{t,t-m-1}^{(m)}, \ldots),
		\end{equation*}
		where $\{\delta_{t,i}^{(m)}\}$ are independent copies of sequence $\{\delta_{t}\}_{-\infty<t<\infty}$ defined on the same measurable space $S$ such that $\sum_{m=1}^{\infty} \upsilon_4(\X_t - \X_t^{(m)}) < \infty$ with $ \upsilon_4(\X_t - \X_t^{(m)}) = \left\lbrace E\norm{\X_t - \X_t^{(m)}}^4 \right\rbrace^{1/4}$.
		\label{cd3}
	\end{enumerate}
	\label{asp1}
\end{assumption}

\begin{remark}
	Assumption~\ref{asp1} follows the dependence concept for functional time series introduced in \cite{HK10}. This assumption is often considered as equivalent conditions to the classic mixing conditions in function spaces \citep[see, e.g.,][]{BHR16,HRW16,RS17}. Condition~\eqref{cd3} specifies the level of dependence that is allowed within process $\{\X_t(u)\}_{t\in\mathbb{Z}}$ in relation to how well it can be approximated by finite $m$-dependent processes. Condition~\eqref{cd3} can also be satisfied when $\upsilon_4(\X_t - \X_t^{(m)}) = O(m^{-\rho})$ for some $\rho > 4$. Roughly speaking, the $\X_t^{(m)}$ defined by the coupling construction in Condition~\eqref{cd3} can be determined by the first $m$ elements $\delta_t, \delta_{t-1}, \ldots, \delta_{t-m+1}$. When the measurable space $S$ coincides with $H$, the sequence $\{\widetilde{\X}_t^{(m)}\}$ given by
	\begin{equation*}
		\widetilde{\X}_t^{(m)} = f(\delta_t, \delta_{t-1}, \delta_{t-2} , \ldots, \delta_{t-m+1}, 0,0,\ldots)
	\end{equation*}
	is also strictly stationary and $m$-dependent, satisfying $\sum_{m=1}^{\infty}\upsilon_4(\X_t - \widetilde{\X}_t^{(m)}) < \infty$.
  \end{remark}

\begin{assumption}
	The kernel function $W_q(\cdot)$ in~\eqref{eq_9} satisfies the following standard conditions:
	\begin{align}
		&W_q(0) = 1, \quad W_q(u) \leq 1, \quad W_q(u) = W_q(-u), \quad W_q(u) = 0 \text{ if } \abs{u} > g \text{ for some constant } g >0, \nonumber \\
		& \text{and } W_q(u) \text{ is Lipschitz continuous on } [-g, g].
		\label{cd2.1}
	\end{align}
	  There exists a $q>0$ such that 
	  \begin{equation*}
	    0 < \lim_{u\to0}\frac{W_q(u)-1}{\abs{u}^{q}} = \mathcal{W}_q < \infty, 
	  \end{equation*}
		and there exists $q' > q$ such that  
		\begin{equation*}
		\sum_{\ell =-\infty}^{\infty} \left|\ell\right|^{q'} \norm{c_{\ell}} < \infty,
		\end{equation*}
		where $c_{\ell}$ is the lag-$\ell$ autocovariance function defined in~\eqref{eq_2}.
	  \label{asp2}
\end{assumption}

\begin{remark}
	Assumption~\ref{asp2} limits the growing rate of $W_q(u)$ at $u=0$, with $q$ referred to as the characteristic exponent of the kernel function by \cite{Parzen57}. The smoother the kernel $W_q(u)$ at zero, the larger the value of $q$ for which $\mathcal{W}_q$ is finite. This assumption has been widely adopted in studies on limit behaviors of the long-run covariance estimator \citep[e.g.,][]{BHR16,RS17}. 
\end{remark}

The conditions in Assumptions~\ref{asp1} and~\ref{asp2} can be easily verified for most stationary time series models based on independent innovations. In the following example we illustrate the applicability of Assumptions~\ref{asp1} and~\ref{asp2} using a standard functional linear process \citep{Bosq00}.
\begin{example}
	(Functional autoregressive process). Suppose $\Phi \in \mathcal{L}$  satisfies $\norm{\Phi}_{\mathcal{L}} < 1$. Let $\{\varepsilon_t\}_{t \in \mathbb{Z}}$ be a sequence of i.i.d. random elements of mean zero taking values in $L^2([0,1])$ satisfying $E\norm{\varepsilon_0}^2 < \infty$. There exists a unique stationary sequence of random process $\{\X_t(u), u \in [0,1]\}_{t \in \mathbb{Z}}$ taking the form
	\begin{equation*}
		\X_t(u) = \Phi(\X_{t-1})(u) + \varepsilon_t(u),
	\end{equation*}
	which is referred to as functional autoregressive process of order one (\textsc{FAR(1)}). The \textsc{FAR(1)} process admits the expansion $\X_t(u) = \sum_{j=0}^{\infty}\Phi^j(\varepsilon_{t-j})(u)$ where $\Phi^j$ is the $j$th iterate of the operator $\Phi$. According to Condition~\eqref{cd3}, we can define an approximation $\X_t^{(m)}(u) = \sum_{j=0}^{m-1}\Phi^j(\varepsilon_{t-j})(u) + \sum_{j=m}^{\infty}\Phi^j(\varepsilon_{t-j}^{(m)})(u)$. The approximation error can then be expressed as $\X_t(u) - \X_t^{(m)}(u) =\sum_{j=m}^{\infty}(\Phi^j(\varepsilon_{t-j})(u)-\Phi^j(\varepsilon_{t-j}^{(m)})(u)) $. Using Cauchy--Schwarz inequality, it can be verified that every operator $A \in \mathcal{L}$ satisfies $\upsilon_4(A(\X)) \leq \norm{A}_{\mathcal{L}}\upsilon_4(\X)$. Then, it follows that $\upsilon_4(\X_t - \X_t^{(m)})  \leq 2\sum_{j=m}^{\infty}\norm{\Phi}^j_{\mathcal{L}}\upsilon_4(\X_t - \X_t^{(m)}) = O(\norm{\Phi}^m_{\mathcal{L}}\upsilon_4(\varepsilon_0) )$. This shows that for the F\textsc{AR(1)} process, Assumption~\ref{asp1} holds as long as $\norm{\varepsilon_0}$ has moments up to order $4+d$ for some $d>0$. In addition, Lemma 3.2 of \cite{Bosq00} indicates that $E\norm{c_{\ell}}^2 \leq \norm{\Phi}^{\ell}_{\mathcal{L}} E\norm{\X_0}^2 $. Assumption~\ref{asp2} then holds since we have assumed that $E\norm{\X_0}^2\leq  \sum_{j=0}^{\infty} \norm{\Phi}^j_{\mathcal{L}} E\norm{\varepsilon_0}^2 < \infty $.
\end{example}

To ensure the consistency of the long-run covariance estimator $\widehat{C}_{h,q}$ in \eqref{eq_9}, we impose the following condition on the bandwidth parameter $h$.

\begin{assumption}
The bandwidth parameter $h$ of long-run covariance estimator in~\eqref{eq_9} satisfies
	\begin{equation*}
	  h = h(T) \to \infty \textnormal{ and } \frac{h(T)}{T} \to 0, \textnormal{ as } T\to \infty.
	\end{equation*}
We use the optimal value of $h$ selected according to the plug-in bandwidth selection procedure of \cite{RS17} in developing asymptotic results and conducting simulation and empirical studies in this paper. 
	\label{asp3}
\end{assumption}

\begin{remark}
    Assumption~\ref{asp3} is a weaker and more standard condition compared to the one used in the Theorem 4.2 of \cite{HK10}, that is, $h^2/T \rightarrow 0$. Details of the plug-in algorithm are provided in Appendix~\ref{app_b1} in the supplementary document.
\end{remark}

\begin{assumption} 
	The eigenvalues of the long-run covariance operator $C$ are finite, positive, and distinctive, i.e., $ \infty > \lambda_1 > \lambda_2 > \ldots$. There exists a positive integer $K$ such that 
	\begin{equation}
		  \frac{\sum_{k=K+1}^{\infty} \lambda_k}{\sum_{k=1}^{K} \lambda_k} =  o(1). \label{asp4_eq}
	\end{equation}
	\label{asp4}
\end{assumption}

\begin{remark}
	Distinctive eigenvalues of covariance operators are commonly adopted in the literature to ensure identification of eigenfunctions \citep[see, e.g.,][]{HK10,HK15}. Assumption~\ref{asp4} requires that the sum of the ``insignificant'' eigenvalues $\{\lambda_{K+1}, \lambda_{K+2} \ldots \}$ tend to zero sufficiently rapidly. Thus, the $K$-dimensional global feature contains ``most information" of $\X_t(u)$ \citep[see e.g.,][]{HV06,BYZ10}. Roughly speaking, Assumption~\ref{asp4} requires that the first $K$ eigenvalues $\{\lambda_1, \ldots \lambda_{K}\}$ have greater orders than the remaining eigenvalues in the sense of \eqref{asp4_eq}. For example, denoting $a/b \rightarrow 1$ by ``$a \sim b$'', \cite{LRS20} proposed that eigenvalues of long-run covariance function satisfying the conditions
  \begin{inparaenum}[(a)]
  \item $\lambda_k \sim \rho_k T^{3-2\alpha_k}$ for $k = 1, \ldots, K$ with coefficients $\infty > \rho_1 \geq \rho_2 \geq \ldots \rho_K > 0$ and $1/2 < \alpha_1 < \alpha_2 < \ldots < \alpha_K < 1$, and
  \item $\sum_{k = K+1}^{\infty} \lambda_k = O(T)$.
  \end{inparaenum}
  Given that $T^{3-2\alpha_1} > T^{3-2\alpha_2} > \ldots T^{3-2\alpha_K} > T$ for a fixed $K$, the sum of $\sum_{k=1}^{K} \lambda_k$ has an order of $T^{3-2\alpha_1}$. It can then be readily seen that $\sum_{k=K+1}^{\infty} \lambda_k / \sum_{k=1}^{K} \lambda_k = O(T)/ T^{3-2\alpha_1} = o(1)$ as $T \rightarrow \infty$. Hence, Assumption~\ref{asp4} is satisfied with non-zero ``insignificant'' eigenvalues $\{\lambda_{K+1}, \lambda_{K+2} \ldots \}$. We are going to identify $K$ and estimate the dynamic space $\mathcal{M}$ spanned by the (deterministic) eigenfunctions $\phi_1(u), \ldots, \phi_K(u)$.
\end{remark}

\begin{assumption}
    The dynamic FPC scores $\{\beta_{t,k}\}$ are uncorrelated across $k$ at all different lags, i.e., $\text{cov}(\beta_{t,i}, \beta_{t+h,j})$ with $i \neq j$, $i, j = 1, \cdots, K$, and $h \in \mathbb{Z}$. \label{asp4_2}
\end{assumption}

\begin{remark}
  Assumption~\ref{asp4_2} specifies the uncorrelatedness of dynamic FPC scores, which is considered as one of the important properties of dynamic FPC scores \citep[see, e.g., page 329, proposition 3(b) in][]{HK15}.
\end{remark}

\begin{assumption}
	The empirical eigenfunctions are in the same direction of the true eigenfunction, i.e., $\inp{\phi_k}{\widehat{\phi}_k} > 0$.
	\label{asp5}
\end{assumption}

\begin{remark}
  Under Assumption~\ref{asp4}, the empirical eigenfunctions $\widehat{\phi}_k$ recovered are in the same direction, or in the opposite direction, with the true eigenfunction $\phi_k$, i.e., $\text{sign}(\inp{\widehat{\phi}_k}{\phi_k}) = \pm 1$. With Assumption~\ref{asp5}, the derivations of equations and proofs are simplified. Note that Assumption~\ref{asp5} is optional for conducting the Karhunen--Lo\`{e}ve expansion of a stochastic process $\X(u)$ given that $\inp{\X}{\phi}\phi$ and $\inp{\X}{-\phi}(-\phi)$ are identical.
\end{remark}

\begin{assumption}
	Let $n$ denote the number of observations on each curve. Let $N = 2^J \geq n$ be a dyadic integer. As $N,n \rightarrow \infty$, we assume that $(n\log^an)^{-1}N$ tends to a constant for some $a>0$. Let $G_n$ be the empirical distribution function of the grid points $\{u_1, \ldots, u_n\}$. Suppose that there exists a distribution $G(u)$ with density $g(u)$, which is bounded away from 0 and infinity such that  
	\begin{equation*}
		G_n(u) \rightarrow G(u) \text{ for all } t \in [0,1] \text{ as } n \rightarrow \infty.
	\end{equation*}
	Further, $g(u)$ has the $\alpha$th bounded derivative. 
	\label{asp6}
\end{assumption}

\begin{remark}
	Assumption~\ref{asp6} specifies technical conditions ensuring the estimator of~\eqref{eq_11} is closely approximate to the true signal over the Besov space $B^{\alpha}_{P, Q}$ (see Appendix~\ref{app_a1}). The same assumption was adopted in \cite{AF01} in the development of their Theorem 6. Functional data $\{\X_t(u_1),\ldots, \X_t(u_n)\}$ measured at dense grids $\{u_1,\ldots,u_n\}$ can easily satisfy Assumption~\ref{asp6}. Since the global feature extraction is conducted before the local feature extraction, selections of $N$ and $n$ have no impact on convergence of global feature estimators.
\end{remark}

\begin{proposition}
	Under Assumptions~\ref{asp1} to~\ref{asp5}, as $T \rightarrow \infty$ there is
  \begin{equation*}
   	\text{Pr}(\widehat{K} = K) \rightarrow 1,
  \end{equation*} 
  where $\widehat{K}$ is determined by~\eqref{eq_10}.
  \label{pr1}
\end{proposition}

\begin{remark}
  The estimation approach of~\eqref{eq_10} has one similarity with the ``scree plot'' method of \cite{Chiou12}: the estimated dimension of functional principal component is chosen to be the point at which the ordered eigenvalues drop substantially. Similar decision rules are often used to estimate the number of factors for high-dimensional factor models; see \cite{LY12, LYB12}, and \cite{AH13}. For functional time series with short memory, \cite{BYZ10} adopted an estimator similar to~\eqref{eq_10} in analysis of the lagged autocovariance operator $C_{\ell}$ for $\ell \neq 0$ of the $K$-dimensional functions satisfying $\lambda_{K} = 0$ and $\lambda_{K+1} = 0$. Most recently, \cite{LRS20} used  an estimator similar to~\eqref{eq_10} to identify the dimension of the dominant subspace in the long memory functional time series. We fill in the literature gap by using the estimator of~\eqref{eq_10} when estimating the dimension of long-run covariance operator for short memory functional time series. 
	\label{rmk6}
\end{remark}

We are now ready to present consistency properties of global and local feature estimators in the following theorems.

\begin{theorem}
	Denote the $k$th empirical functional component by $\widehat{\beta}_{t,k}$ and its associated score by $\widehat{\phi}_k$. Under Assumptions~\ref{asp1} to~\ref{asp5}, as $T \rightarrow \infty$,
	\begin{equation*}
		\norm{\sum_{k=1}^{K}\beta_{t,k}\phi_k(u) - \sum_{k=1}^{\widehat{K}}\widehat{\beta}_{t,k}\widehat{\phi}_k(u)} = O_P\left(T^{-2/5}\right).
	\end{equation*}
	\label{thm1}
\end{theorem}

\begin{remark}
	The convergence rate of global feature estimators depends on the weight function $W_q$ and the bandwidth $h$ in~\eqref{eq_9}. We use a flat-top weight function with quadratic spectral kernel (more details see Appendix \ref{app_b1}) that has been considered by \cite{Andrews91}, together with the optimal bandwidth selected according to the plug-in method of \cite{RS17}. The order of $T^{-2/5}$ associated with the selected $W_q$ matches findings of \cite{PR96} when the optimal bandwidth is used. 
\end{remark}

\begin{theorem}
	Denote the number of dyadic points in wavelet transform by $N$. Under Assumptions~\ref{asp1} to~\ref{asp6}, as $N,T\rightarrow \infty$, and for some $\alpha > 0$,
	\begin{equation*}
		\norm{Z_t(u) - \widetilde{Z}_t(u)} = O_P(N^{-\alpha/(1+2\alpha)} +  T^{-2/5}),
	\end{equation*}
	where $Z_t(u)$ and $\widetilde{Z}_t(u)$ are defined in~\eqref{eq_6} and~\eqref{eq_8}, respectively.
	\label{thm2}
\end{theorem}

\begin{remark}
  Theorem~\ref{thm1} states the convergence rate for FPCA-based global feature estimators when the optimal bandwidth selected by the plug-in algorithm of \cite{RS17} is used. Here, $\alpha$ indicates the degree of smoothness of the true signal of local features in a Besov ball $B^{\alpha}_{P, Q}$ (see \ref{lm7} in Appendix~\ref{app_a1} for the definition of $B^{\alpha}_{P, Q}$). Loosely speaking, the true signal in the Besov space $B^{\alpha}_{P, Q}$ has $\alpha$ bounded derivatives in $L^P$ space, with finer gradation of smoothness further controlled by the parameter $Q$ \citep[see, e.g.,][for definitions and properties of Besov spaces]{Meyer92}. Given that local features are estimated after the extraction of global features, convergence of local feature estimators should depend on global feature estimators. This conjecture is confirmed by the term of $O_P\left(T^{-2/5}\right)$, i.e., the convergence rate of FPCA global feature estimators, in the derived convergence rate for BTW local feature estimators.
\end{remark}

\begin{theorem}
	Under Assumptions~\ref{asp1} to~\ref{asp6}, as $N,T\rightarrow \infty$, 
	\begin{equation*}
		\norm{X_t(u) - \widehat{\X}_t(u)} = O_P( N^{-\alpha/(1+2\alpha)} + T^{-2/5}).
	\end{equation*}
	\label{thm3}
\end{theorem}

\begin{remark}
	Theorem~\ref{thm3} indicates the estimation error for $\X_t(u)$ includes a component from the estimation of global features, and another component from the estimation of local features. As $N,T\rightarrow \infty$, both components converge to zero, and we have the total estimation error subsequently converges to zero. 
\end{remark}

\section{Monte Carlo experiments}\label{sec:5}
Finite sample performances of FPCA-BTW estimators are examined through two Monte Carlo experiments. The FPCA-BTW method is applied to make estimation of the functional process and its covariance structure, and produce out-of-sample forecasts. The data generating process for each experiment is calibrated according to a real NIR dataset. Throughout this section, the dimension of global features $K$ is a fixed integer estimated by~\eqref{eq_10}.

\subsection{Experiment 1}\label{sec:5.1}
\begin{figure}[!htbp]
\centering
\subfloat[NIR spectrum of chloroform.]
{\includegraphics[width = 2.7in]{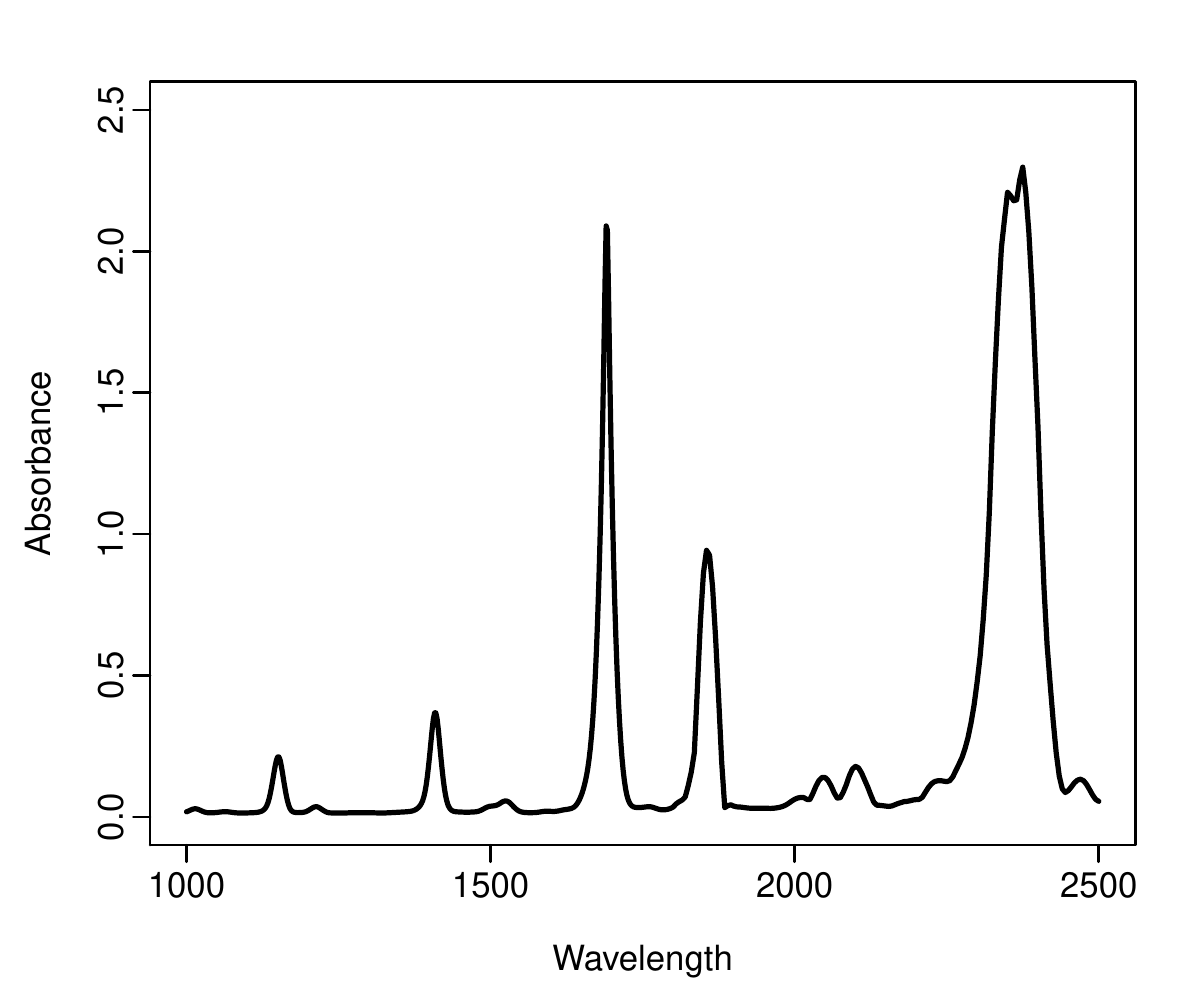}\label{fig:2a}} \qquad
\subfloat[Orthonormalized basis functions.]
{\includegraphics[width = 2.7in]{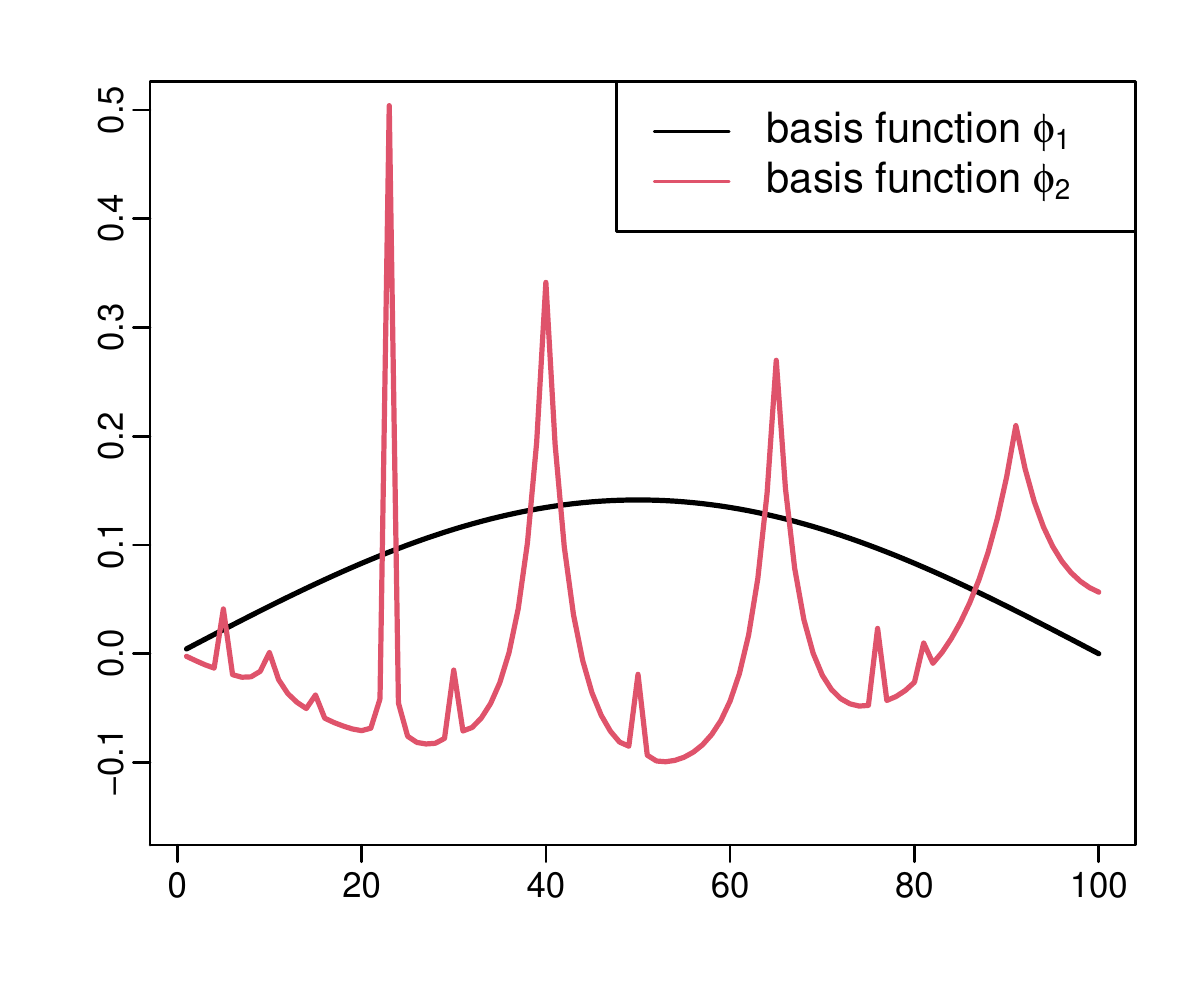}\label{fig:2b}}
\caption{Motivation data and designed basis functions for Experiment 1.}\label{fig:2}
\end{figure}

Many common chemical compounds (e.g., chlorinated alkanes) have complex NIR spectroscopy spectra consisting of mixed sharp spikes and lower peaks \citep[see, e.g.,][Figure 21.3]{BC07}. As an example, Figure~\ref{fig:2a} illustrates the NIR spectrum of chloroform with formula \ch{CHCl3} consisting of two sharp and highly localized features at approximately 1750~nm and 2400~nm, and several lower peaks scattering between 1000~nm and 1300~nm. To generate functional data imitating NIR spectroscopy spectra of such chemical compounds, we select $\phi_1(u) = \sin(\pi u)$ as a basis function for global features, and extend the ``bumps'' function of \cite{DJ94} to simulate local features. Specifically, choose a kernel function $f_{\text{kernel}}(u) = (1+\abs{u})^{-4}$ to generate $\phi_2(u) = f_{\text{Bumps}}(u) = \sum_{j=1}^{11}s_j f_{\text{kernel}}\left((u-u_j)/w_j\right) $, where $u_j$, $w_j$ and $s_j$ are location, bandwidth and scaling parameters, respectively. 

Figure~\ref{fig:2b} presents the orthonormalized basis functions used in this experiment. Coefficients $\{\beta_{t,k}: k = 1,2\}_{t=1}^T$ are generated from autoregressive models of order 1 (\textsc{AR(1)}) of the form $\beta_{t,k} = \theta_k \beta_{t-1,k} + \omega_{t,k}$. Select $\theta_1 = 0.8$ and $\omega_{t,1} \sim N(0,4)$ for $\{\beta_{t,1}\}_{t=1}^T$, while choosing $\theta_2 = 0.2$ and $\omega_{t,2} \sim N(0,0.01)$ for $\{\beta_{t,2}\}_{t=1}^T$. Combining the generated global and local features gives the true simulated process as $\X_t^{\text{TRUE}} = \beta_{t,1}\phi_1(u) + Z_t(u) = \beta_{t,1}\phi_1(u) + \beta_{t,2}\phi_2(u)$. Generate independent noise as $\varepsilon_t(u) = 0.01B_t(u)$ with $B_t(u)$ i.i.d. standard Brownian motion $\{B_t(u)\}_{t=1}^T$. Finally, functional time series $\{\X_t(u)\}_{t=1}^T$ is calculated as $\X_t(u) = \X_t^{\text{TRUE}} + \varepsilon_t(u)$ for $u \in [0,1]$.
 
For each sample size $T \in \{25, 50, 100\}$, dynamic FPCA is applied to extract global features, with obtained results denoted by $\widehat{g}_t^{FPCA}(u)$. The BTW method, together with competing methods including the unified sparse and functional PCA (SFPCA) method of \cite{AW19} and the two-way FPCA (TWFPCA) method of \cite{HSB09}, are applied to extract local features $\widehat{Z}_t(u)$ from FPCA residuals. SFPCA and TWFPCA are implemented with a grid search parameter selection approach provided by the \textit{MoMA} package \citep{MoMA} in \textbf{R} \citep{Team15}. Estimation accuracy is assessed by relative squared error (RSE) defined in a simple Riemann sum as
\begin{equation*}
  \text{RSE} = \sum_{t=1}^{T} \frac{\norm{\X^{\text{TRUE}}_t - \widehat{g}_t^{\text{FPCA}} - \widehat{Z}_t}^2}{\norm{\X^{\text{TRUE}}_t - \widehat{g}_t^{\text{FPCA}}}^2} = \sum_{t=1}^{T}\sum_{i=1}^{100} \frac{\abs{\X^{\text{TRUE}}_t(u_i) - \widehat{g}_t^{\text{FPCA}}(u_i) - \widehat{Z}_t(u_i)}^2}{\abs{\X^{\text{TRUE}}_t(u_i) - \widehat{g}_t^{\text{FPCA}}(u_i)}^2},
\end{equation*}
where $i = \{1, \ldots, 100\}$ denote equally spaced discrete realizations over $[0,1]$. Given that the denominator of RSE corresponds to the reconstruction accuracy of the FPCA estimator, any estimation method with $\text{RSE} < 1$ has a more accurate estimation performance than the conventional FPCA method. Moreover, the numerator of RSE is proportional to mean squared estimation error 
defined by $T^{-1}\sum_{t=1}^{T}\norm{\X_t(u)-\widehat{\X}_t(u)}^2$. Thus, small RSE indicates an efficient local feature extraction method.

\begin{table}[!htbp]
	\centering
	\small
  	\tabcolsep 0.2in	
	\caption{\small Mean RSE and running time of various local feature extraction methods (standard errors in parentheses). The bold entries highlighting the best performing method for each setting.} 
	\label{table_1}
	{\renewcommand{\arraystretch}{1}%
	\begin{tabular}{@{\extracolsep{5pt}}L{2cm}C{1cm}C{2.5cm}C{2.5cm}C{2.5cm}@{}}
		\toprule
		 Sample size & & SFPCA & TWFPCA & BTW \\
		\midrule 
		\multirow{2}{*}{$T=25$} & RSE & 0.687 (0.077) & 0.749 (0.058) & \textBF{0.663} (0.079) \\
		& Time	& 15.462 (0.603) & 21.085 (2.872) & \textBF{0.154} (0.104) \\
		 \cline{1-5}
		 \multirow{2}{*}{$T=50$} & RSE & 0.659 (0.070) & 0.737 (0.047) & \textBF{0.639} (0.067) \\ 
		 & Time & 34.288 (1.289) & 20.206 (3.686) & \textBF{0.143} (0.023) \\ 
		 \cline{1-5}
		 \multirow{2}{*}{$T=100$} & RSE & 0.649 (0.052) & 0.731 (0.034) & \textBF{0.629} (0.052) \\ 
		 & Time & 89.555 (2.579) &  21.089 (2.881) & \textBF{0.244} (0.041) \\ 
		\bottomrule
	\end{tabular}
	}
\end{table}

Table~\ref{table_1} presents RSE averaged over 100 replications for three considered local feature extraction methods, together with computation time (in seconds) for a single iteration in \textbf{R} \citep{Team15} on an AMD Ryzen Threadripper 1950X CPU at 3.40GHz. 
It can be seen that the BTW local feature extraction method consistently outperforms competing methods in estimation accuracy and computation efficiency. All three methods report RSE significantly less than 1, indicating that extracting local features after FPCA dramatically improves estimation accuracy. We note the existence of a greedy ``coordinate-wise'' Bayesian Information Criterion (\textsc{BIC}) optimization scheme by \cite{AW19} that can significantly reduce computation time for the SFPCA and TWFPCA methods. However, in this experiment the \textsc{BIC} optimization approach produces RSEs around 1, suggesting that inappropriate penalty parameters are being selected. We present RSEs in relation to static FPCA in Appendix~\ref{app_b2} in the Supplementary document.

\subsection{Experiment 2}\label{sec:5.2}
Significant spikes of local features of functional time series are often visible in surface plots of long-run covariance functions. A good example of such data is spectroscopy of absorbance on samples of ground pork recorded on a Tecator infrared spectrometer in the region 850 to 1050~nm \citep{Thodberg96}. Following \cite{FV06}, we apply dynamic FPCA to 77 Tecator NIR spectroscopy spectra corresponding to samples with large fat content. Fitting the leading empirical scores associated with the leading functional components to an \textsc{AR(1)} model returns an estimated coefficient $0.2487$. Using analysis results of the Tecator data, we calibrate the data generating process for Experiment 2 by choosing $\phi_1(u) = \frac{1}{\sqrt{2\pi}}\exp{\{-\frac{u^2}{2}\}}$ and generating $\{\beta_{t,1}\}_{t=1}^T$ from $\beta_{t,1} = 0.2487 \beta_{t-1,1} + \omega_{t,1}$, where $\omega_{t,1} \sim N(0,1)$. To amplify bumps in covariances, local features are generated as
\begin{equation*}
  Z_t(u) = \begin{cases}
    0.5 Z_{t-1}(u) + 0.1B^{\ast}_t(u), &  0.25 \leq u < 0.5 \\
    0, &   \text{elsewhere} \\
  \end{cases},
\end{equation*}
where i.i.d. Brownian motion innovations $\{B^{\ast}_t(u), u \in [0.25, 0.5]\}_{t=1}^T$ satisfy $B^{\ast}_t(0.25) = 0$. Finally, independent noise is generated as $\varepsilon_t(u) = \sqrt{0.001}B_t(u)$ with $B_t(u)$ i.i.d. standard Brownian motion $\{B_t(u), u \in [0,1]\}_{t=1}^T$. Functional time series is computed as $\X_t(u) = \beta_{t}\phi_1(u) + Z_t(u) + \varepsilon_t(u)$ for $u \in [0,1]$.

For each simulated time series $\{\X_t(u)\}_{t=1}^T$, we apply the FPCA-BTW method to extract global and local features, and use the extracted features to reconstruct long-run covariance functions. The dynamic FPCA estimators are considered as comparison benchmarks. Estimation accuracy for covariance is assessed according to relative error (RE) given by
\begin{equation*}
  \text{RE}  =  \sqrt{\sum_{i=1}^{40} \sum_{j=1}^{40} \frac{\left|C(u_{i},s_{j}) - \widehat{C}(u_i,s_{j})\right| ^2}{\left| C(u_{i},s_{j})\right| ^2} },
\end{equation*}
where $C(u,s)$ is the theoretical long-run covariance function, and $\widehat{C}(u,s)$ is the reconstructed estimator using extracted features; $i,j = \{1,\ldots, 40\}$ denote equally spaced grid points over $[0,1]$. 

\begin{figure}[!htbp]
	\centering
	\includegraphics[width = 4.5in]{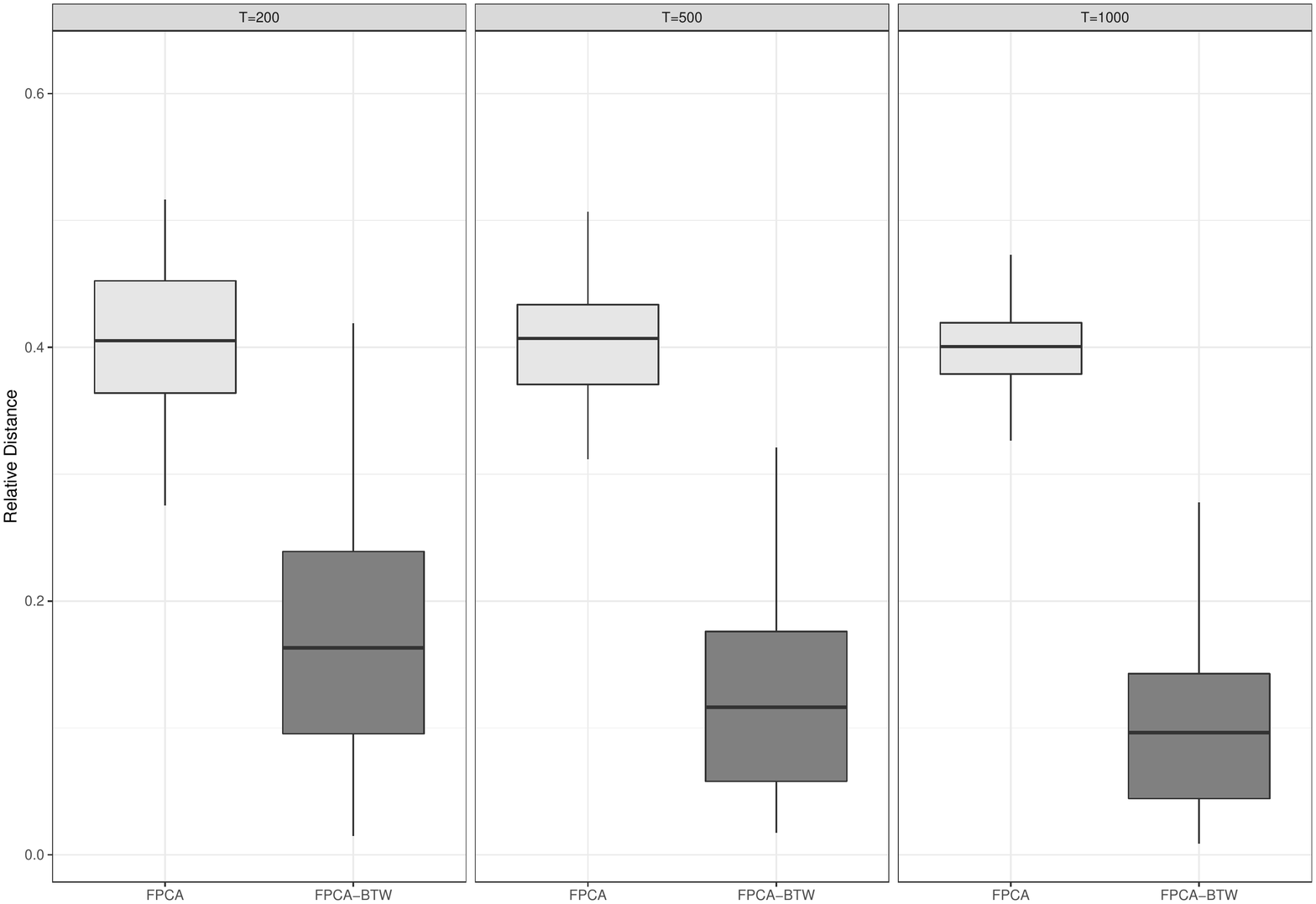}
	\caption{Relative errors of long-run covariance estimators.}\label{fig:3}
\end{figure}

For each $T \in \{200, 500, 1000\}$, we replicate the experiment 100 times. Throughout the experiment, the empirical dimension of global features is determined to be $\widehat{K} = 1$ by~\eqref{eq_10}. Figure~\ref{fig:3} shows that the FPCA-BTW method produces smaller reconstruction errors than the FPCA method. Hence, the extracted local features are tested to improve long-run covariance estimation accuracy. Finally, it can be easily observed that both FPCA and FPCA-BTW methods report smaller estimation errors when sample sizes increase. 

\begin{figure}[!htbp]
 	\centering
 	\subfloat[\label{fig:4d}]{\includegraphics[width = 1.8in]{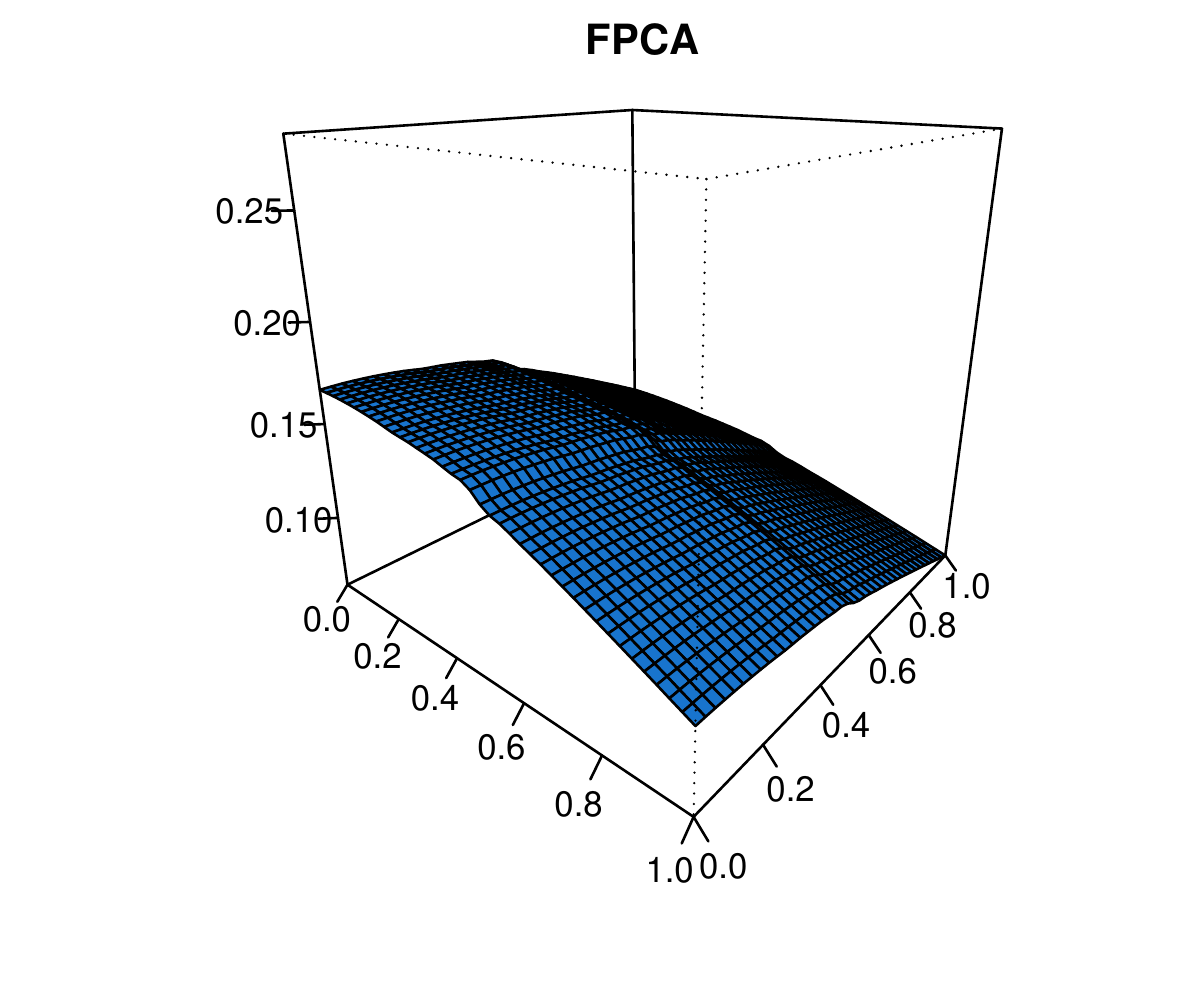}}
	\subfloat[\label{fig:4e}]{\includegraphics[width = 1.8in]{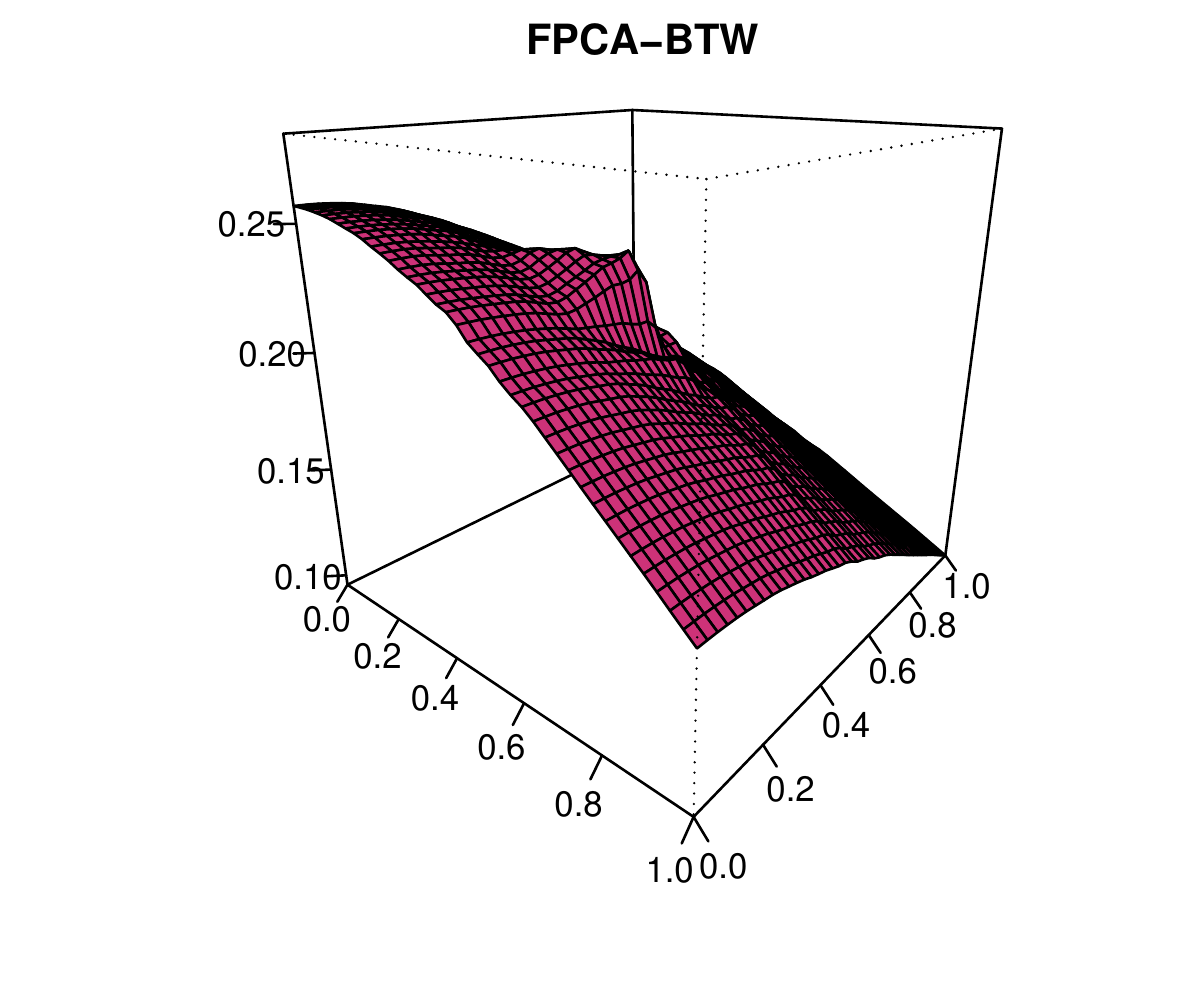}}
	\subfloat[\label{fig:4f}]{\includegraphics[width = 1.8in]{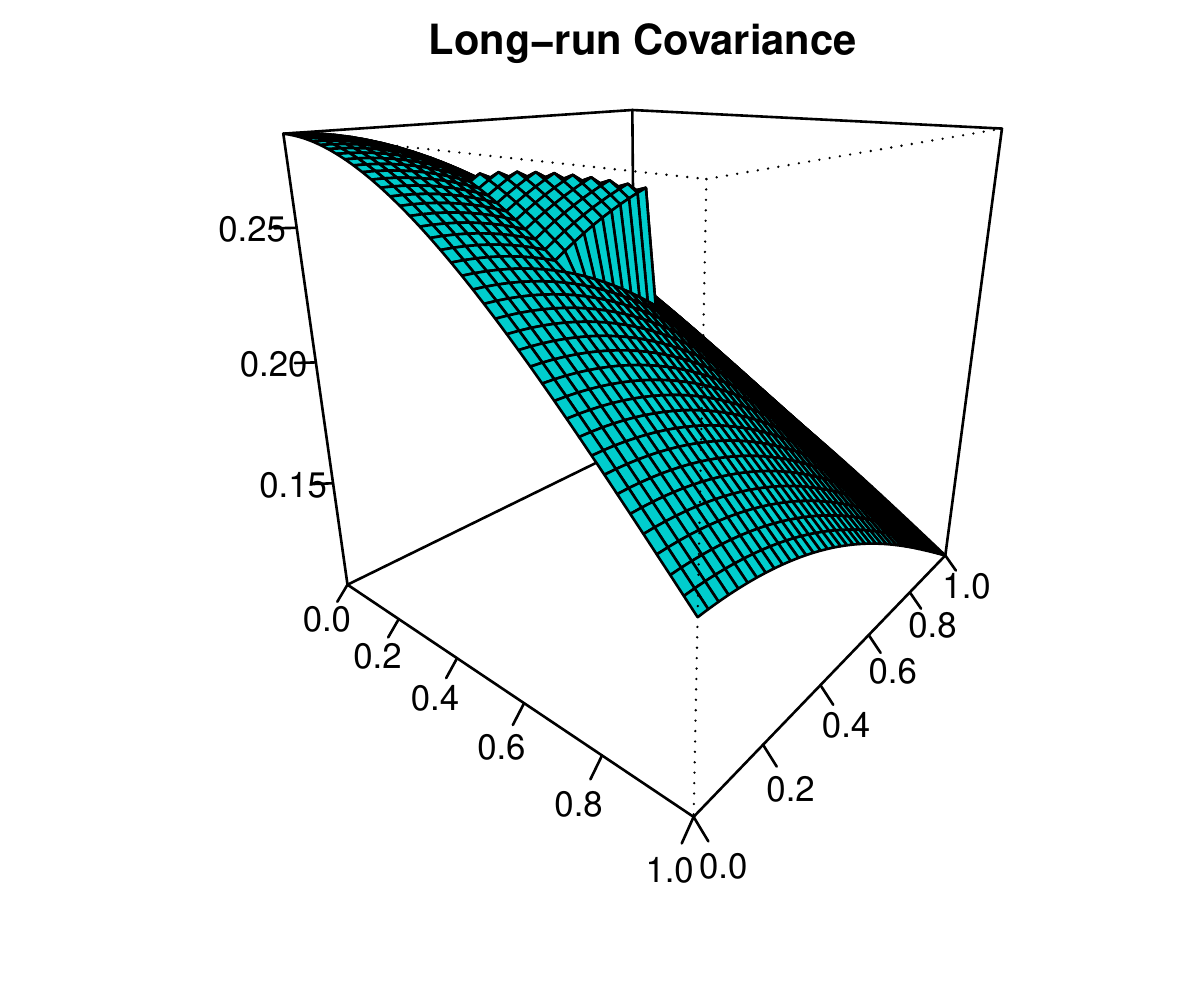}} 
 \caption{Surface plots of the mean long-run covariance estimators over 100 simulations obtained by FPCA (blue) and FPCA-BTW (red) for $T = 200$, along with the true theoretical functions (cyan).}\label{fig:4}
 \end{figure}

Figure~\ref{fig:4} visualizes the advantage of FPCA-BTW estimators in long-run covariance estimation when sample size $T=200$. Figure~\ref{fig:4f} presents the theoretical long-run covariance function that has a ``pyramid-shaped bump" corresponding to local features $Z_t(u)$. Estimators depicted by Figure~\ref{fig:4d} fail to capture the ``bump'' of local features. In contrast, FPCA-BTW estimators successfully recover most information about local features in the presence of intentionally added noise. This experiment shows that local features are essential for the estimation of the long-run covariance function of functional time series.

Monte Carlo experiments introduced in Section~\ref{sec:5} prove that FPCA-BTW produces the best feature extraction performance among considered methods. We also design experiments to show that local features extracted by BTW help to improve point forecast accuracy, with details included in Appendix~\ref{app_b3} in the Supplementary document. In the next section, advantages of the FPCA-BTW method at feature extraction and forecasting functional time series are demonstrated using the empirical wood panel NIR spectroscopy data.

\section{Empirical application}\label{sec:6}

The wood panel NIR spectroscopy data illustrated in Figure~\ref{fig:1} consists of spectra of absorbance (in negative base ten logarithm of the transmittance) recorded at wavelengths from 350 to 2500~nm in 1~nm intervals in a series of 72 experimental trials. Removing observations from 2301 to 2500~nm because of considerable noise gives $n = 1951$ discrete realizations on each curve. Figure~\ref{fig:1a} indicates that raw spectra curves are contaminated by observational noise. Denoting the observed NIR absorbance values at wavelength $i$ in the $t$th curve as $Y_{t}(u_i)$, the data can be expressed as
\begin{equation*}
  Y_{t}(u_i) = \X_t(u_i) + \varepsilon_t(u_i), \qquad i = 1, \ldots, 1951, \quad t = 1, \ldots, 72,
\end{equation*}
where $\X_t(u)$ is the true underlying smooth process, and $\varepsilon_t(u)$ is a random noise function. The smoothed functions displayed in Figure~\ref{fig:1d} are obtained by minimizing the penalized residual sum of squares (PENSSE) given by
\begin{equation*}
  \text{PENSSE} = \sum_{t=1}^{72}\sum_{i=1}^{1951} [\X_t(u_i) - Y_t(u_i)]^2 + \lambda_{BS}\sum_{t=1}^{72} \int_{0}^{1} \frac{d^2 \X_t(u)}{du^2}du, \qquad t = 1, \ldots, 72, \quad u \in [0,1],
\end{equation*}
where $\lambda_{BS}$ is a B-spline smoothing parameter selected by the \textit{fda.usc} package in \textbf{R} \citep{Team15} through  generalized cross-validations. The functional KPSS test of \cite{HKR14} confirms that $\{\X_t(u)\}_{t=1}^{72}$ is stationary at the 5\% significance level with a \textit{p}-value of 0.053. 

\subsection{Feature extraction performance}\label{sec:6.1}
\begin{figure}[!htb]
\centering
\includegraphics[width = 3.5in]{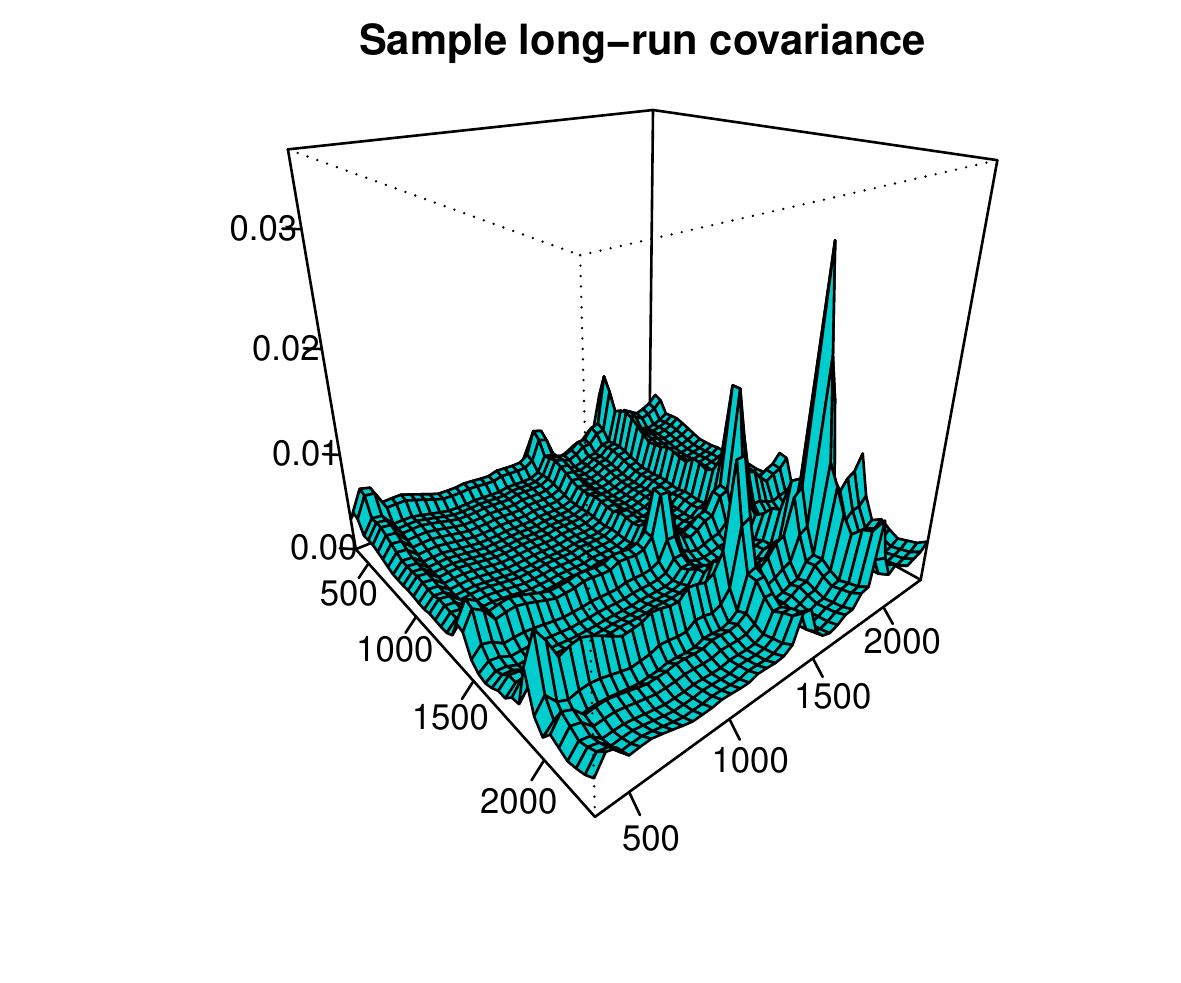}
\caption{Sample long-run covariance function of smoothed wood panel NIR spectroscopy spectra.}\label{fig:5}
\end{figure}

We compare feature extraction performances of the FPCA-BTW method with competing sparse FPCA methods. The sample long-run covariance function for $\{\X_t(u)\}_{t=1}^{72}$ is computed following the procedures described in Section~\ref{sec:3.1}, and is presented in Figure~\ref{fig:5}. It can be seen that sharp spikes mainly occur between 1300 and 1900 nm, indicating that autocovariance functions at non-zero of lags $\{\X_t(u)\}_{t=1}^{72}$ also possess information exclusive to local features. 

Decomposing the sample long-run covariance operator reports the dimension of global features as $\widehat{K} = 2$ by~\eqref{eq_10}, and this result is confirmed by the empirical eigenvalues presented in Figure~\ref{fig:1e}. Using only global features we compute a long-run covariance estimator $\widehat{C}^{\text{FPCA}}(u,s)$. Next, we apply the BTW method, and also the SFPCA and TWFPCA methods mentioned in Section~\ref{sec:5.1}, to recover local features from dynamic FPCA residuals. With the extracted local features, we compute another estimator $\widehat{C}^{\text{local}}(u,s)$. The true process of NIR absorbance is not observable in practice. To assess effectiveness of various local feature extraction methods, a sample relative error  is defined as
\begin{equation*}
  \text{Sample relative error}  =  \sqrt{\sum_{i=1}^{500} \sum_{j=1}^{500} \frac{\left|C^{\text{sample}}(u_{i},s_{j})- \widehat{C}^{\text{FPCA}}(u_i,s_{j}) - \widehat{C}^{\text{local}}(u_i,s_{j})\right| ^2}{\left| C^{\text{sample}}(u_{i},s_{j}) - \widehat{C}^{\text{FPCA}}(u_i,s_{j})\right| ^2} },
\end{equation*}
where $i$ and $j$ denote equally spaced grid points over $[0,1]$. To accelerate the computation, we pick $n = 500$ equally spaced grids on each $\X_t(u)$ and get sample relative errors of $0.255$, $0.386$ and $7.866e-5$ for the sparse FPCA method \citep{AW19}, the two-way FPCA method \citep{HSB09} and the BTW method, respectively. A sample RE close to 0 indicates that nearly all relevant information of the sample long-run covariance has been utilized in modeling the functional time series $\{\X_t(u)\}_{t=1}^{72}$. The obtained sample relative errors indicate that BTW is an optimal method for recovering sharp and highly localized features for functional data.

\subsection{Forecasting performance} \label{sec:6.2}
The forecasting performance of various global and local feature extraction methods are compared. First, the smoothed functions $\{\X_t(u)\}_{t=1}^{72}$ are divided into a training set $\{\X_1(u), \ldots, \X_{62}(u)\}$ and a testing set $\{\widehat{\X}_{63}(u), \ldots, \widehat{\X}_{72}(u) \}$.  We apply the FPCA-BTW method to the training set, and use the obtained global and local features to make out-of-sample forecasts. Adopting the expanding window approach of \cite{ZW06}, in total we produce ten one-step-ahead forecasts, nine two-step-ahead forecasts, and so on, up to one 10-step-ahead forecast. Point forecasts obtained without considering local featires under the same expanding window setting serve as comparison benchmarks in this application.

To accelerate computation, we pick $n = 500$ equally spaced grids on each $\X_t(u)$, and compute the mean absolute forecast error (MAFE) and the root mean squared forecast error (RMSFE) as
\begin{align*}
\text{MAFE}(h) &= \frac{1}{500\times (11-h)}\sum^{10}_{\varsigma = h}\sum^{500}_{i=1}|\X_{62+\varsigma}(u_i) - \widehat{\X}_{62+\varsigma|62+\varsigma-h}(u_i)|, \\
\text{RMSFE}(h) &= \sqrt{\frac{1}{500\times (11-h)}\sum^{10}_{\varsigma = h}\sum^{500}_{i=1}\left \lbrace \X_{62+\varsigma}(u_i) - \widehat{\X}_{62+\varsigma|62+\varsigma-h}(u_i)\right \rbrace^2 },
\end{align*}
where $\X_{62+\varsigma}(u_i)$ represents the actual holdout sample at the $i$th wavelength of the $\varsigma$th curve, and $\widehat{\X}_{62+\varsigma}(u_i)$ is the corresponding point forecasts. Averaging over ten forecast horizons, we obtain summary statistics given by
\begin{align*}
\text{Median (MAFE)} & = \frac{1}{2}[\text{MAFE}(h=5) + \text{MAFE}(h=6)], \\
\text{Mean (RMSFE)} & = \frac{1}{10}\sum^{10}_{h=1}\text{RMSFE}(h).
\end{align*}
The median statistic is suitable for handling the absolute error MAFE while the mean statistic is good at handling the squared error RMSFE \citep{Gneiting11}.

Point forecast evaluation results are reported in Tables~\ref{table_4}. The forecasts constructed using only global features are shown in the columns with the heading ``None'', with the remaining columns reporting forecasts produced with global and local features extracted by various methods. It can be easily seen that forecasts produced with local features are consistently more accurate. This result highlights the importance of incorporating local features in forecasting NIR spectroscopy spectra time series. Further, it can be seen that BTW consistently outperforms the competing methods in recovering local features relevant to forecasting. Thus, we recommend FPCA-BTW method in modeling and forecasting functional time series in practice. In addition, a comparison with point forecast evaluation results shown in Appendix~\ref{app_b2} indicates that dynamic FPCA produces more accurate point forecasts than static FPCA for the NIR spectroscopy data. This finding indicates that incorporating serial dependence carried by lagged NIR spectroscopy observations improves point forecast accuracy. 

\begin{table}[!htbp]
	\centering
	\small
	\tabcolsep 0.09in
	\caption{\small Mean MAFEs and RMSFEs of point forecasts averaged over 100 replications. The bold entries highlight the feature extraction method with higher forecast accuracy.} 
	{\renewcommand{\arraystretch}{1}%
\begin{tabular}{lccccccccc}
		\toprule
		& \multicolumn{4}{c}{MAFE} && \multicolumn{4}{c}{RMSFE}\\\cline{2-5}\cline{7-10}
		$h$ & None & BTW & SFPCA & TWFPCA && None & BTW & SFPCA & TWFPCA \\
		\midrule
		1 & 0.482 & \textBF{0.430} & 0.450 & 0.450 && 0.870 & \textBF{0.837} & 0.841 & 0.846 \\
		2 & 0.502 & \textBF{0.449} & 0.473 & 0.475 && 0.882 & \textBF{0.841} & 0.852 & 0.857 \\ 
		3 & 0.528 & \textBF{0.475} & 0.498 & 0.502 && 0.910 & \textBF{0.872} & 0.878 & 0.884 \\
		4 & 0.537 & \textBF{0.486} & 0.511 & 0.516 && 0.918 & \textBF{0.870} & 0.884 & 0.891 \\ 
		5 & 0.543 & \textBF{0.491} & 0.513 & 0.518 && 0.939 & \textBF{0.891} & 0.902 & 0.910 \\ 
		6 & 0.580 & \textBF{0.533} & 0.556 & 0.566 && 0.988 & \textBF{0.938} & 0.951 & 0.962 \\
		7 & 0.598 & \textBF{0.552} & 0.575 & 0.592 && 1.016 & \textBF{0.959} & 0.976 & 0.994 \\ 
		8 & 0.645 & \textBF{0.596} & 0.627 & 0.650 && 1.041 & \textBF{0.972} & 1.003 & 1.030 \\ 
		9 & 0.704 & \textBF{0.646} & 0.685 & 0.717 && 1.115 & \textBF{1.044} & 1.072 & 1.105 \\
		10& 0.593 & \textBF{0.531} & 0.548 & 0.577 && 1.144 & \textBF{1.082} & 1.109 & 1.133 \\
		Mean & 0.571 & \textBF{0.519} & 0.544 & 0.556 && 0.982 & \textBF{0.931} & 0.942 & 0.961 \\ 
		Median & 0.561 & \textBF{0.511} & 0.530 & 0.542 && 0.963 & \textBF{0.915} & 0.927 & 0.936 \\
		\bottomrule
	\end{tabular}
  }\label{table_4}
\end{table}

\begin{figure}[htbp]
	\centering
	\subfloat{\includegraphics[height = 3.5in]{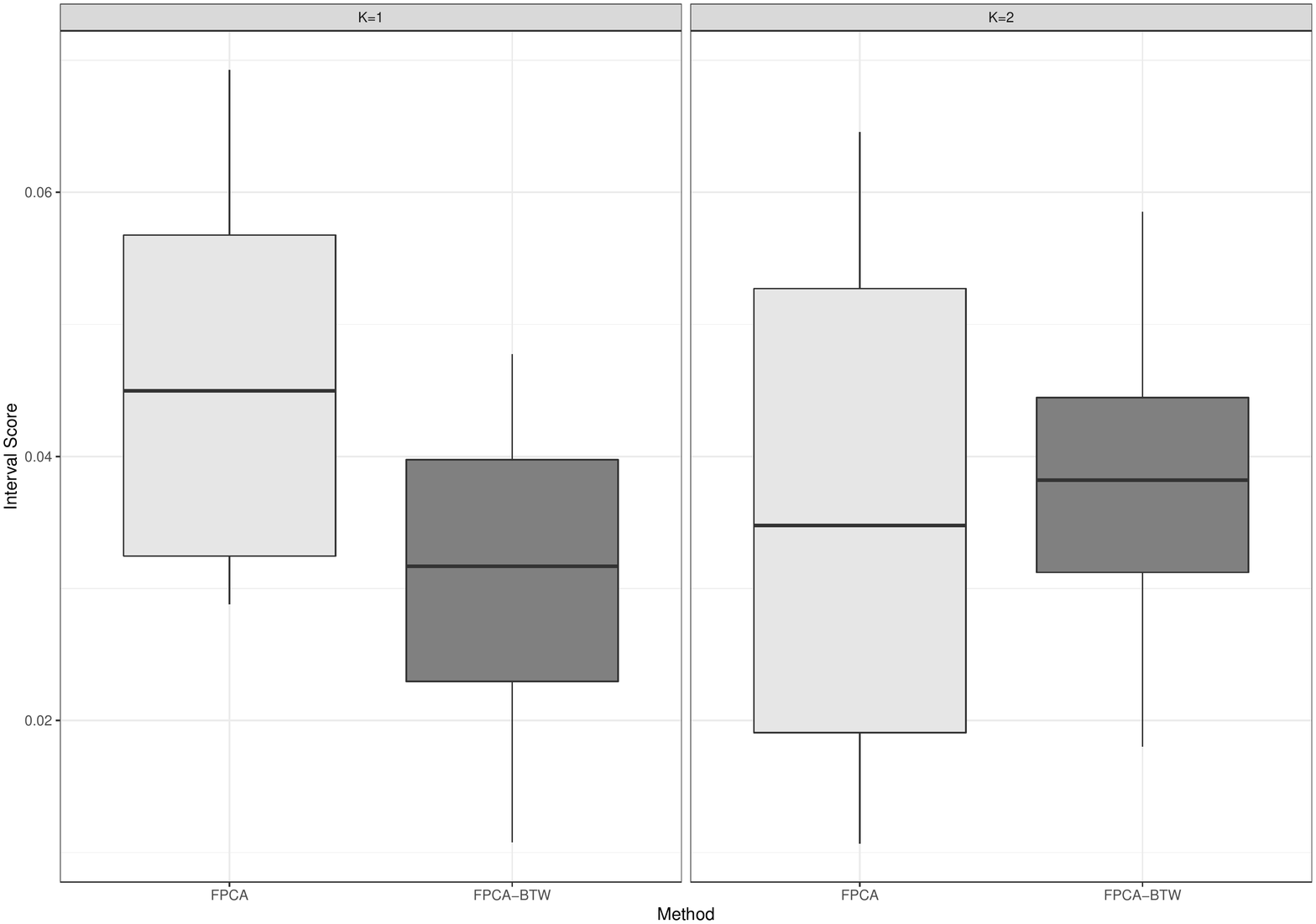}} 
	\caption{Scores of pointwise interval forecasts produced by FPCA method and FPCA-BTW method.}
	\label{fig:6}
\end{figure}

Table~\ref{table_4} shows that the FPCA-BTW method produces the most accurate point forecasts. Therefore, we do not further consider other competing feature extraction methods. To access the forecast uncertainty of FPCA-BTW method, we adapt the approach of \cite{ANH15} and compute pointwise prediction intervals at the $100(1-a)\%$ nominal coverage probability. Technical details of interval forecasts are provided in Appendix~\ref{app_b3}. Pointwise predictions intervals are evaluated using the interval score of \cite{GR07} given by
\begin{align*} 
S_{a} \left[ \widehat{\X}_{T+h}^{\text{lb}}(u_i), \widehat{\X}_{T+h}^{\text{ub}}(u_i); \X_{T+h}(u_i)  \right] & =  \left[ \widehat{\X}_{T+h}^{\text{ub}}(u_i) -  \widehat{\X}_{T+h}^{\text{lb}}(u_i)  \right] \\
& +  \frac{2}{a}\left[ \widehat{\X}_{T+h}^{\text{lb}}(u_i) - \X_{T+h}(u_i)  \right] \mathds{1}\left\lbrace \X_{T+h}(u_i) < \widehat{\X}_{T+h}^{\text{lb}}(u_i)  \right\rbrace\\
&  + \frac{2}{a}\left[  \X_{T+h}(u_i) -  \widehat{\X}_{T+h}^{\text{ub}}(u_i)  \right] \mathds{1} \left\lbrace \X_{T+h}(u_i) >  \widehat{\X}_{T+h}^{\text{ub}}(u_i)   \right\rbrace,
\end{align*}
where $\widehat{\X}_{T+h}^{\text{lb}}(u_i)$ and $\widehat{\X}_{T+h}^{\text{ub}}(u_i)$ denote lower and upper bounds of a symmetric $100(1-a)\%$ prediction interval, and the level of significance is customarily selected as $a = 0.2$. To accelerate computation, we again pick $n = 500$ equally spaced grids on each $\X_t(u)$. Averaging over different points in a curve and different forecast horizons, the mean interval score is defined as
\begin{equation*}
\bar{S}_{\alpha}(h) = \frac{1}{500\times (11-h)}\sum^{10}_{\varsigma = h}\sum^{500}_{i=1}S_{a} \left[ \widehat{\X}_{T+h}^{\text{lb}}(u_i), \widehat{\X}_{T+h}^{\text{ub}}(u_i); \X_{T+h}(u_i)  \right], 
\end{equation*}
where $ S_{a}\left[ \widehat{\X}_{T+h}^{\text{lb}}(u_i), \widehat{\X}_{T+h}^{\text{ub}}(u_i); \X_{T+h}(u_i)  \right]$ denotes the interval score at the $\varsigma$th curve in the testing set. The interval scores summarized in Figure~\ref{fig:6} confirm that incorporating the local features produces more accurate interval forecasts. 
Moreover, extracting local features after retaining only one empirical functional component gives decent interval forecasts, which further highlights that the BTW method can effectively recover nearly all relevant information of functional data.

\section{Conclusion}\label{sec:7}

We propose a novel feature extraction method for functional time series. The proposed FPCA-BTW method improves the feature extraction performance of FPCA by recovering sharp, highly localized features from dimension reduction residuals. Local features extracted by BTW possess information of functional variations over particular short intervals within function domain, contributing to improved estimation results and more accurate forecasts. Theoretical properties of FPCA-BTW method are developed. Superior estimation and forecasting performances of FPCA-BTW estimators in finite samples are verified by Monte Carlo experiments and an empirical application to wood panel NIR spectroscopy data. 

There are several ways in which the present paper can be further extended. First, this paper employed \citeauthor{Cai02}'s (\citeyear{Cai02}) parametric blockwise threshold approach to select wavelet coefficients. To make the proposed FPCA-BTW a nonparametric feature extraction method, the block size and threshold level at different resolution levels need to be selected based on characteristics of observations. A possible extension of the current method is adopting the data-driven block thresholding approach of \cite{Cai09} to enhance extraction of local features. Moreover, this paper considered forecasting functional time series with extracted linear features. Non-linear extensions of functional regression, for example the continuously additive model of \cite{MWY13}, provide enhanced flexibility and structural stability. Another possible extension of the FPCA-BTW method may consider functional additive model \citep{MY08} as the main dimension reduction tool. Since the inspirational work on functional manifold models by \cite{DG05}, functional manifold models have witnessed increasing contributions in methodology and applications \citep[see, e.g.,][]{LY19}.

\section*{Acknowledgment}

The authors thank Professor Jiguo Cao from Simon Fraser University for providing us with the lumber data set.

\bigskip
\begin{center}
{\large\bf SUPPLEMENTARY MATERIAL}
\end{center}

\begin{description}

\item[Supplementary document:] Document containing detailed proofs of the theoretical results and additional technical details of implementing FPCA-BTW method. 
\end{description}

\clearpage 

\newpage
\begin{singlespacing}
\bibliographystyle{agsm}
\bibliography{datadriven}
\end{singlespacing}
\end{document}


\def\spacingset#1{\renewcommand{\baselinestretch}%
{#1}\small\normalsize} \spacingset{1}




\title{Supplementary to ``Feature Extraction for Functional Time Series: Theory and Application to NIR Spectroscopy Data''}
\maketitle

\begin{appendices}
\doublespacing

Appendix~\ref{appendixA} provides detailed proofs of theoretical results stated in the main document. Appendix~\ref{appendixB} present additional technical details of estimation long-run covariance function as well as applications of the FPCA-BTW feature extraction method.

\section{}\label{appendixA}

In Appendix~\ref{app_a1}, we provide proofs of consistency of FPCA global feature estimators. In Appendix~\ref{app_a2}, we present proofs of consistency of BTW local feature estimators. Finally, Appendix~\ref{app_a3} provides proof of consistency of FPCA-BTW estimators for functional time series. The preliminary lemmas, together with additional notations facilitating development, are presented before proofs of the main results throughout Appendix~\ref{appendixA}.

\subsection{Consistency of global feature estimators}\label{app_a1}

\begin{lemma}
  Assume that $\{\X_t\}_{t\in\mathbb{Z}}$ is $L^4\operatorname{-}m\operatorname{-}$approximable. Then 
  \begin{enumerate}[(i)]
    \item the autocovariance operator $C_{\ell}$ defined by the kernel function $c_{\ell}(u,s)$ in~\eqref{eq_2} satisfies
    \begin{equation*}
    \sum_{\ell \in \mathbb{Z}}\norm{C_{\ell}}_{\mathcal{S}} < \infty.
  \end{equation*} \label{lemma1.1}
    \item The sample mean function $\widehat{\mu}(u) = T^{-1}\sum_{t=1}^{T}\X_t(u)$ satisfies $E\norm{\widehat{\mu}(u)-\mu(u)}^2 = O(1/T)$. \label{lemma1.2}
  \end{enumerate}
    \label{lemma1}
\end{lemma}

\begin{proof}[Proof of Lemma~\ref{lemma1}]  
  For $L^4\operatorname{-}m\operatorname{-}$approximable processes $\{\X_t\}_{t\in\mathbb{Z}}$, the proof of~\eqref{lemma1.1} follows the Proposition 6 in \cite{HK15}, and the proof~\eqref{lemma1.2} follows the Theorem 4.1 in \cite{HK12b}.
\end{proof}

Since estimation of the long-run covariance function $\widehat{C}_{h,q}(u,s)$ does not depend on its mean function $E[\X_t(u)] = E[\X_0(u)] = \mu(u)$, in the following derivations we can assume $E[\X_0(u)] = 0$ without losing generality.

\begin{lemma} 
  Under Assumptions~\ref{asp1} to~\ref{asp3}, the estimated long-run covariance operator 
  \begin{equation*}
    \widehat{C}_{h,q}(x)(u) = \int_{0}^{1}\widehat{C}_{h,q}(u,s)x(s)\diff s, \qquad x \in L^2([0,1])
  \end{equation*}
  and the true long-run covariance operator
  \begin{equation*}
    C(x)(u)  = \int_{0}^{1}C(u,s)x(s) \diff s, \qquad x \in L^2([0,1])
  \end{equation*}
  satisfy 
  \begin{equation*}
    \norm{\widehat{C}_{h,q} - C}_{\mathcal{S}} = o_P(1) .
  \end{equation*}
  \label{lm1}
\end{lemma}

\begin{proof}[Proof of Lemma~\ref{lm1}]  
  Under Assumption~\ref{asp1}, by Theorem 2.3 in \cite{BHR16} we have
\begin{align}
  E\norm{\widehat{C}_{h,q} - C  }^2_{\mathcal{S}} & = \frac{h}{T} \left(\iint C^2(u,s)\diff u \diff s + \left(\int_{0}^{1} C(u,u)\diff u\right)^2 \right) \int_{-\infty}^{\infty} W_q^2(x) \diff x \nonumber  \\
  & \qquad + h^{-2q} \norm{\mathcal{W}_q \sum_{\ell = -\infty}^{\infty} |\ell|^{q}\gamma_{\ell}(u,s)}^2  + o(\frac{h}{T} + h^{-2q}) \nonumber \\
  & =: \frac{h}{T} \mathfrak{c}_1 +  h^{-2q} \mathfrak{c}_2 +  o\left(\frac{h}{T} + h^{-2q}\right),
  \label{eq1}
\end{align}
where $W_q(\cdot)$ is a symmetric kernel function and $\mathcal{W}_q = \lim_{u\to0}\frac{W_q(u)-1}{\abs{u}^{q}}$. 

First, consider the first term $\mathfrak{c}_1 h/T$ in \eqref{eq1}. Under Assumption~\ref{asp1}, the true long-run covariance function $C(u,s)$ is an element of $L^2([0,1]^2)$ satisfying $\iint C^2(u,s) \diff u \diff s < \infty$ and $\int_{0}^{1} C(u,u)\diff u < \infty $; see Appendix A.2 in \cite{HKR13}. By condition~\eqref{cd2.1} in Assumption~\ref{asp2}, the kernel function $W_q(u)$ over a bounded support $[-g,g]$ satisfies $\int_{-\infty}^{\infty} W_q^2(x) \diff x = \int_{-g}^{g} W_q^2(x) \diff x < \infty$. We then have $\mathfrak{c}_1 h/T= O\left(\frac{h}{T}\right)$. 

Next, consider the second term $h^{-2q} \mathfrak{c}_2$ in \eqref{eq1}. \cite{Andrews91} shows that the considered flat-top kernel in~\eqref{w_ft} has $q >0$ while the considered QS kernel in~\eqref{w_qs} has $q = 2$. By Assumption~\ref{asp2}, $\norm{\mathcal{W}_q \sum_{\ell = -\infty}^{\infty} |\ell|^{q}\gamma_{\ell}(u,s)}^2 = \mathcal{W}_q^2 \norm{\sum_{\ell = -\infty}^{\infty} |\ell|^{q}\gamma_{\ell}(u,s)}^2 < \infty$. We then have $h^{-2q} \mathfrak{c}_2 = O(h^{-2q})$. By Assumption~\ref{asp3}, we then have 
\begin{equation*}
  E\norm{\widehat{C}_{h,q} - C }^2_{\mathcal{S}} = O\left(\frac{h}{T} + h^{-2q}\right).
\end{equation*}
The Chebyshev's inequality and Assumption~\ref{asp3} imply that
\begin{equation*}
    \norm{\widehat{C}_{h,q}  - C }_{\mathcal{S}} = O_P\left(\left(\frac{h}{T}\right)^{\frac{1}{2}} + h^{-q}\right).
\end{equation*}
\end{proof}

\begin{lemma} (Theorem 2.2 in \cite{RS17}).
  Under Assumption~\ref{asp1}, using a flat-top weight function and a quadratic spectral kernel function in the estimation of~\eqref{eq_9} gives 
  \begin{equation*}
    \widehat{h}_{\text{opt}} = O(T^{1/5}).
  \end{equation*}
  \label{lm2}
\end{lemma}

\begin{proof}[Proof of Lemma~\ref{lm2}]  
  When a flat-top weight function and a quadratic spectral kernel function (see Section \ref{app_b1} for more details) is used in the ``plug-in'' method, by Theorem 2.2 in \cite{RS17}, the estimated optimal bandwidth satisfies
  \begin{equation*}
    \widehat{h}_{\text{opt}} = h_{\text{opt}} \left(1 + O_P\left(\frac{\log^{5/2}(T)}{\sqrt{T}} \right) \right).
  \end{equation*}
  Applying L'Hospital's rule, we can easily verify that $\lim_{T\rightarrow\infty}\frac{\log^{5/2}(T)}{\sqrt{T}} = \frac{15}{T^{1/2}\log^{1/2}(T)} = 0$. 
  Moreover, equation (2.15) in \cite{BHR16} suggests that $h_{\text{opt}} = O(T^{1/5})$,. Hence, we have $\widehat{h}_{\text{opt}} = O(T^{1/5})$.
\end{proof}

\begin{lemma} (Lemma 3.2 in~\cite{HK10}).
  Under Assumption~\ref{asp1}, for two compact operators $P, Q \in \LL$ with singular value decompositions 
  \begin{equation*}
  P(x) = \sum_{k=1}^{\infty}\xi_k \inp{x}{v_k} v_k, \qquad Q(x) = \sum_{k=1}^{\infty}\zeta_k\inp{x}{\nu_k}\nu_k,
  \end{equation*}  
  with $P(v_k) = \xi_kv_k$ and $\sum_k\xi_k^2 < \infty$. Consequently $\xi_k$ are eigenvalues of $P$ and $\nu_k$ the corresponding eigenfunctions. We also define
  \begin{equation*}
  v_k' = \widehat{s}_kv_k, \qquad \widehat{s}_k = \text{sign}(\inp{\nu_k}{v_k}).
  \end{equation*} 
  If $P$ has its eigenvalues satisfy
  \begin{equation*}
  \xi_1 > \xi_2 > \cdots> \xi_K > \xi_{K+1},
  \end{equation*}
  then
  \begin{equation*}
  \norm{\nu_k - v_k'}  \leq \frac{2\sqrt{2}}{\alpha_k} \norm{Q-P}_{\LL}, \qquad 1 \leq k \leq K,
  \end{equation*}
  where $\alpha_1 = \xi_1 - \xi_2$ and $\alpha_k = \min(\xi_{k-1} -\xi_k, \xi_k - \xi_{k+1}), \, 2 \leq k \leq K$.
  \label{lm3}
\end{lemma}

\begin{proof}[Proof of Lemma~\ref{lm3}]
  \cite{HK10} outlined the proof to their Lemma 3.2, and later provided more detailed proof of this lemma in their book \citep[page.~35]{HK12}.
\end{proof}

\begin{lemma} Under Assumptions~\ref{asp1} to~\ref{asp5}, for each $1\leq k\leq K$, the empirical eigenfunctions and the true eigenfunctions satisfy $ \norm{\widehat{\phi}_k(u) - \phi_k(u)} =  O_P\left(T^{-2/5}\right) $  as $T \rightarrow \infty$.
 \label{lm4}
\end{lemma}

\begin{proof}[Proof of Lemma~\ref{lm4}]  
  For the proof of this lemma, we replace operators $P$ and $Q$ in Lemma~\ref{lm3} by the true long-run covariance operator $C$ and the estimated long-run covariance operator $\widehat{C}$, respectively. $v_k$ and $\nu_k$ in Lemma~\ref{lm3} can then be replaced by the true eigenfunction $\phi_k$ and the empirical eigenfunction $\widehat{\phi}_k$, respectively. By Assumption~\ref{asp5}, $\phi_k$ and $\widehat{\phi}_k$ are in the same direction, and hence $\widehat{s}_k = 1$. Conditioning on $\widehat{K} = K$, we have 
  \begin{equation*}
    \norm{\widehat{\phi}_k(u) - \phi_k(u)} \leq \frac{2\sqrt{2}}{\alpha_k} \norm{\widehat{C} - C}_{\mathcal{L}} \leq \frac{2\sqrt{2}}{\alpha_k} \norm{\widehat{C} - C}_{\mathcal{S}}, \qquad 1 \leq k \leq K,
    \end{equation*}
  where $\alpha_1 = \lambda_1 - \lambda_2$ and $\alpha_k = \min(\lambda_{k-1} - \lambda_k, \lambda_k - \lambda_{k+1})$. Assumption~\ref{asp4} indicates the first $K+1$ eigenvalues $\lambda_1, \cdots, \lambda_{K+1}$ are distinctive. So $\alpha_k$ does not equal to 0. In the estimation of long-run covariance function, the $\widehat{h}_{\text{opt}}$ of~\eqref{hopt} and the quadratic spectral kernel function are used. Hence, by Lemma~\ref{lm1} and~\ref{lm2}, we have 
\begin{align*}
  \norm{\widehat{\phi}_k(u) - \phi_k(u)} & =  O_P\left(\left(\frac{\widehat{h}_{\text{opt}}}{T}\right)^{\frac{1}{2}} +\widehat{h}_{\text{opt}}^{-2}\right) \\
  & = O_P\left(T^{-2/5}+ T^{-2/5}\right) \\
  & = O_P\left(T^{-2/5}\right) 
\end{align*}
\end{proof}

\begin{lemma}
  The estimated FPC scores $\widehat{\beta}_{t,k}$ and the true FPC scores $\beta_{t,k}$ associated with the $t$\textsuperscript{th} ($t \in \mathbb{Z}$) function $\X_t(u)$ satisfy
  \begin{equation*}
    |\widehat{\beta}_{t,k} - \beta_{t,k}| =  O_P\left(T^{-2/5}\right) , \qquad 1 \leq k \leq K.
  \end{equation*}
  \label{lm5}
\end{lemma}

\begin{proof}[Proof of Lemma~\ref{lm6}]  
 For each $1 \leq k \leq K$,
  \begin{align*}
  |\widehat{\beta}_{t,k} - \beta_{t,k}| & = |\inp{\X_t(u)}{\widehat{\phi}_k(u)}-\inp{\X_t(u)}{\phi_k(u)}| \\
  & \leq |\inp{\X_t(u)}{\widehat{\phi}_k(u) - \phi_k(u)}| \\
  & \leq \norm{\widehat{\phi}_k(u) - \phi_k(u)}\norm{\X_t(u)}\quad (\text{Cauchy-Schwarz inequality}) 
  \end{align*}
  By Lemma~\ref{lm4}, 
  \begin{equation*}
  |\widehat{\beta}_{t,k} - \beta_{t,k}|  = O_P(T^{-2/5})\left( \norm{\X_t(u)} \right) 
  \end{equation*}

  It remains to show that $ \norm{\X_t(u)}  = O_P(1)$. By Assumption~\ref{asp1}, the weak stationarity assumption requires that $E[\norm{\X_t(u)}^2] = E[\norm{\X_1(u)}^2]$. Hence, for any $x > 0$, Markov's inequality implies that
  \begin{equation*}
    \text{Pr}\left(\norm{\X_t(u)} > x\right) \leq \frac{E\norm{\X_t(u)}}{x} \leq \frac{\sqrt{E[\norm{\X_1(u)}^2]}}{x} < \infty.
  \end{equation*}
\end{proof}

We define some notations before introducing the next lemma. Let $\mathscr{X}$ be a subspace (closed linear manifold) of a separable Hilbert space $H$, and let $\mathscr{Y}$ be the orthogonal complement of $\mathscr{X}$. Let $A$ be a closed linear operator on $H$. Let $X$ and $Y$ be the injections of $\mathscr{X}$ and $\mathscr{Y}$ into $H$. Then $\mathscr{X}$ is an invariant subspace of $A$ if and only if $\mathscr{X} \subset \mathcal{D}(A)$ (the domain of $A$) and
\begin{equation*}
  Y^{\ast}AX = 0.
\end{equation*}
We denote the set of all linear operators $P: \mathscr{X} \rightarrow \mathscr{Y}$ by $\mathcal{B}(\mathscr{X}, \mathscr{Y})$, and set $\mathcal{B}(\mathscr{X}) = \mathcal{B}(\mathscr{X}, \mathscr{X})$. Let $F \in \mathcal{B}(\mathscr{X})$ and $G \in \mathcal{B}(\mathscr{Y})$. Then $F$ and $G$ define an operator $T \in \mathcal{B}[\mathcal{B}(\mathscr{X}, \mathscr{Y})]$ by
\begin{equation*}
  T(P) = PF - GP, \qquad P \in \mathcal{B}(\mathscr{X}, \mathscr{Y}). 
\end{equation*}
We now define the separation of two operators $F$ and $G$ as 
\begin{equation*}
  \text{sep}(F,G) = 
  \begin{cases}
    \norm{T^{-1}}^{-1}_{\mathcal{L}}, & \text{if } 0 \notin \sigma(T) \\
    0, & \text{if } 0 \in \sigma(T)
  \end{cases},
\end{equation*}
where $\sigma(T)$ denotes the spectrum of $T$.

\begin{lemma}(Theorem 4.1 and Lemma 3.2 in \cite{Stewart71}).
  Let $A$ be a closed operator defined on a separable Hilbert space $H$ whose domain is dense in $H$. Let $\mathscr{X} \subset \mathcal{D}(A)$ (the domain of $A$) be a subspace, and let $\mathscr{Y}$ be its orthogonal complement. Let $\mathscr{Y}_A$ be the projection of $\mathcal{D}(A)$ onto $\mathscr{Y}$. Let $X$, $Y$, and $Y_A$ be the injections of $\mathscr{X}$, $\mathscr{Y}$, and $\mathscr{Y}_A$ into $H$. The adjoints of $X$ and $Y_A$,
  denoted by $X^{\ast}$ and $Y_A^{\ast}$, satisfy $X^{\ast}X = I$ (the identity on $\mathscr{X}$), $Y_A^{\ast}Y_A = I_A$ (the identity on $\mathscr{Y}_A$), $X^{\ast}Y_A = 0$, and $Y^{\ast}_AX = 0$.
  Let
  \begin{align*}
    B_{11} & = X^{\ast} A X, \qquad B_{12} = X^{\ast} A Y_A, \\
    B_{21} & = Y^{\ast}_A A X, \qquad B_{22} = Y^{\ast}_A A Y_A.
  \end{align*}
  Let a perturbation $E \in \mathcal{B}(H)$ satisfy
  \begin{equation}
    \begin{split}
    E_{11} & = X^{\ast}EX, \qquad E_{12} = X^{\ast}EY, \\
    E_{21} & = Y^{\ast}EX, \qquad E_{22} = Y^{\ast}EY.
    \end{split}
    \label{E}
  \end{equation}
  Let $\eta = \norm{B_{12}}_{\mathcal{L}} + \norm{E_{12}}_{\mathcal{L}}$, and $\theta = \text{sep}(B_{11},B_{22}) - \norm{E_{11}}_{\mathcal{L}} - \norm{E_{22}}_{\mathcal{L}}$. If $\mathscr{X}$ is an invariant subspace of $A$ and $\kappa_1 = \theta^{-2}  \eta \norm{E_{21}}_{\mathcal{L}} < 1/4$, there is a $P \in \mathcal{B}(\mathscr{X}, \mathscr{Y}_A)$ satisfying 
  \begin{equation*}
    \norm{P}_{\mathcal{L}} \leq \frac{\norm{E_{21}}_{\mathcal{L}}}{\theta} (1+\kappa_1) < 2\frac{\norm{E_{21}}_{\mathcal{L}}}{\theta}
  \end{equation*} 
  such that $\mathscr{R}(X + Y_A P)$ (the range of $X + Y_A P$) is an invariant subspace of $A+E$. Let
  \begin{equation*}
    X' = (X+Y_AP)(I + P^{\ast}P)^{-1/2},
  \end{equation*}
  and $\mathscr{X}' = \mathscr{R}(X')$. Then $\mathscr{X}' \subset \mathscr{D}(A)$ is a subspace.
  \label{lm6}
\end{lemma}

\begin{proof}[Proof of Proposition~\ref{pr1}]  
  Denote a complete set of orthonormal basis on $L^2([0,1])$ by $\{\phi_k\}_{k \in \mathbb{Z}^+}$. The long-run covariance operator $C$ and its estimator $\widehat{C}_{h,q}$ are both positive-definite Hilbert-Schmidt operators, and thus admit the decomposition
  \begin{align*}
    C(x) & = \sum_{k=1}^{\infty}\lambda_k \inp{x}{\phi_k} \phi_k, \qquad \text{and} \\
    \widehat{C}_{h,q}(x) & = \sum_{k=1}^{\infty}\widehat{\lambda}_k \inp{x}{\phi_k} \phi_k, \qquad x \in L^2([0,1]).
  \end{align*}
  We then have
  \begin{equation}
    \sum_{k=1}^{T}\widehat{\lambda}_k = \sum_{k=1}^{\infty} \inp{\phi_k}{\widehat{C}_{h,q}(\phi_k)},
    \label{th0_eq1}
  \end{equation}
  and 
  \begin{equation}
    \sum_{k=1}^{\infty} \lambda_k = \sum_{k=1}^{\infty}\inp{\phi_k}{C(\phi_k)}.
    \label{th0_eq2}
  \end{equation}

  Deducting~\eqref{th0_eq2} from~\eqref{th0_eq1} gives
  \begin{equation}
    \sum_{k=1}^{T}(\widehat{\lambda}_k- \lambda_k) = \sum_{k=1}^{\infty} \inp{\phi_k}{(\widehat{C}_{h,q}-C)(\phi_k)} + \sum_{k=T+1}^{\infty}\lambda_k.
    \label{th0_eq3}
  \end{equation}
  By definition, the long-run covariance operator $C$ can be expressed as
  \begin{align*}
    C(x) & = \sum_{\ell=-\infty}^{\infty}  E[\inp{\X_t}{x}\X_{t+\ell}] \\ 
    & = E[\inp{\X_t}{x}\X_t] + \sum_{|\ell|\geq1}^{\infty} E[\inp{\X_t}{x}\X_{t+\ell}] + \sum_{|\ell|\geq1}^{\infty} E[\inp{\X_t}{x}\X_{t-\ell}] \\
    &=: C_0(x) + \sum_{|\ell|\geq1}^{\infty}C_{\ell}(x) + \sum_{|\ell|\geq1}^{\infty} C_{-\ell}(x) . 
  \end{align*}
  It can be seen that $C$ is self adjoint because, by a direct verification,
  \begin{equation*} 
    C_{\ell}^{\ast}(x) \equiv E[\inp{\X_t}{x}\X_{t-\ell}] =  C_{-\ell}(x),
  \end{equation*}
  where the superscript $\cdot^{\ast}$ denotes the adjoint operator. Similarly, we can prove that $\widehat{C}_{h,q}$ is also self adjoint.
  In addition, using the relations $C(\phi_k) = \lambda_k \phi_k$ and 
  $\widehat{C}_{h,q}(\widehat{\phi}_k) = \widehat{\lambda}_k \widehat{\phi}_k$, we have for $k = 1, \cdots, K$,
  \begin{align*}
    \abs{\inp{\phi_k}{(\widehat{C}_{h,q}-C)(\widehat{\phi}_k)} - (\widehat{\lambda}_k - \lambda_k)} & = \abs{\inp{\phi_k}{\widehat{C}_{h,q}(\widehat{\phi}_k)} - \inp{\phi_k}{C(\widehat{\phi}_k)} - (\widehat{\lambda}_k - \lambda_k) } \\
    & = \abs{\inp{\phi_k}{\widehat{C}_{h,q}(\widehat{\phi}_k)} - \inp{C(\phi_k)}{\widehat{\phi}_k} - (\widehat{\lambda}_k - \lambda_k) } \quad (\text{self adjoint } C) \\
    & = \abs{(\widehat{\lambda}_k - \lambda_k)(\inp{\phi_k}{\widehat{\phi}_k} - 1)}  \\
    & = \abs{\widehat{\lambda}_k - \lambda_k||\inp{\phi_k}{\widehat{\phi}_k} - 1)} \\
    & = \abs{\widehat{\lambda}_k - \lambda_k||\inp{\phi_k}{\widehat{\phi}_k - \phi_k}}  \\
    & \leq \abs{\widehat{\lambda}_k - \lambda_k}\norm{\phi_k}\norm{\widehat{\phi}_k - \phi_k}  \\
    & = \abs{\widehat{\lambda}_k - \lambda_k}\norm{\widehat{\phi}_k - \phi_k}.
  \end{align*}
  By Lemma 4.2 in \cite{Bosq00}, $\sup_{k\geq 1}|\widehat{\lambda}_k - \lambda_k| \leq \norm{\widehat{C}_{h,q} - C}_{\mathcal{S}}$.
  When the optimal bandwidth $\widehat{h}_{\text{opt}}$ is used, by Lemma \ref{lm1} and Lemma \ref{lm2}, we then have
  \begin{equation}
    \widehat{\lambda}_k = \lambda_k + O_P\left(T^{-2/5}\right) , \qquad k = 1, \cdots, K.
    \label{lambda_diff}
  \end{equation}

  Now consider $k = K+1, \cdots, k_{max}$. Define the following operators
  \begin{align*}
    Q_1(x) & := \sum_{k=K+1}^{k_{max}} \inp{x}{\phi_k} \phi_k, \\
    Q_2(x) & := \sum_{k=1}^{K} \inp{x}{\phi_k} \phi_k,  \\
    \widehat{Q}_1(x) & := \sum_{k=K+1}^{k_{max}} \inp{x}{\widehat{\phi}_k} \widehat{\phi}_k, \qquad x \in L^2([0,1]).
  \end{align*}
  Then $Q_1$, $Q_2$, the long-run covariance operator $C$, and the difference of operators $\widehat{C}_{h,q} - C$ correspond to $X$, $Y_A$, $A$ and $E$ in Lemma~\ref{lm6}. Using the fact that $CQ_2 = \sum_{k=1}^{K} \lambda_k \phi_k$ and $\inp{\phi_j}{\phi_k} = 0 \,
   \forall j \neq k$, we can get
  \begin{equation*}
    \norm{B_{12}}_{\mathcal{L}} := \norm{Q_1^{\ast}CQ_2}_{\mathcal{L}} = \norm{Q_1^{\ast}(\sum_{k=1}^{K} \lambda_k \phi_k)}_{\mathcal{L}} = 0.
  \end{equation*}
  Next, by results of Lemma~\ref{lm1} and Lemma~\ref{lm2}, 
  \begin{align*}
    \norm{E_{11}}_{\mathcal{L}} & := \norm{Q_1^{\ast}(\widehat{C}_{h,q} - C)Q_1}_{\mathcal{L}} \\
    & \leq \norm{Q_1}_{\mathcal{L}}^2 \norm{\widehat{C}_{h,q} - C}_{\mathcal{S}}  \\
    & \leq  O_P(T^{-2/5}),
  \end{align*}
  where the last inequality is due to 
  \begin{align*}
    \norm{Q_1}_{\mathcal{L}}^2 & = \sup_{\norm{x} \leq 1}\left\lbrace\sum_{k=K+1}^{k_{max}}\inp{x}{\phi_k}^2 \right\rbrace \\
    & < \sup_{\norm{x}\leq 1}\left\lbrace \sum_{k=1}^{\infty} \inp{x}{\phi_k}^2 \right\rbrace \\
    & = \sup_{\norm{x}\leq 1}\norm{x}^2.
  \end{align*}
  In Lemma~\ref{lm6}, $\mathscr{Y}$ is the closure of $\mathscr{Y}_A$ and $Y$ the extension of $Y_A$ to $\mathscr{Y}$. We define a $Q_3(x) := \sum_{k=1}^{K} \inp{x}{\phi_k} \phi_k + \sum_{k={k_{max}+1}}^{\infty}\inp{x}{\phi_k} \phi_k$, with $x \in L^2([0,1])$ corresponding to $Y$ in the lemma. It can be easily seen that
  \begin{align*}
    \norm{Q_3}_{\mathcal{L}}^2 & = \norm{\sum_{k=1}^{K} \inp{x}{\phi_k} \phi_k + \sum_{k={k_{max}+1}}^{\infty}\inp{x}{\phi_k} \phi_k}_{\mathcal{L}}^2 \\
    & \leq \sup_{\norm{x}\leq 1}\left\lbrace \sum_{k=1}^{K} \inp{x}{\phi_k}^2+ \sum_{k={k_{max}+1}}^{\infty}\inp{x}{\phi_k}^2 \right\rbrace \\
    & < \sup_{\norm{x}\leq 1}\left\lbrace\sum_{k=1}^{\infty} \inp{x}{\phi_k}^2 \right\rbrace\\
    & = \sup_{\norm{x}\leq 1}\norm{x}^2.
  \end{align*}
  We then have  
  \begin{align*}
    \norm{E_{21}}_{\mathcal{L}} & := \norm{Q_3^{\ast}(\widehat{C}_{h,q} - C)Q_1}_{\mathcal{L}} \\
    & \leq \norm{Q_3^{\ast}}_{\mathcal{L}} \norm{\widehat{C}_{h,q} - C}_{\mathcal{S}} \norm{Q_1}_{\mathcal{L}} = O_P(T^{-2/5}),
  \end{align*}
  \begin{align*}
    \norm{E_{12}}_{\mathcal{L}} & := \norm{Q_1^{\ast}(\widehat{C}_{h,q}-C)Q_3}_{\mathcal{L}} \\
    & \leq \norm{Q_1}_{\mathcal{L}} \norm{\widehat{C}_{h,q} - C}_{\mathcal{S}} \norm{Q_3}_{\mathcal{L}} = O_P(T^{-2/5}),
  \end{align*}
  and
  \begin{align*}
    \norm{E_{22}}_{\mathcal{L}} & := \norm{Q_3^{\ast}(\widehat{C}_{h,q} - C)Q_3}_{\mathcal{L}} \\
    & \leq \norm{Q_3}_{\mathcal{L}}^2 \norm{\widehat{C}_{h,q} - C}_{\mathcal{S}} = O_P(T^{-2/5}).
  \end{align*}
  We have previously checked that the long-run covariance operator $C$ is self-adjoint. This corresponds to the well-known fact that if $C$ is Hermitian and $\phi$ is an approximate normalized eigenvector, then $\phi^{\ast} C \phi$ is an approximate eigenvalue. Thus we have $B_{11}$ and $B_{22} := Q_2^{\ast} C Q_2$. The separation between of $B_{11}$ and $B_{22}$ satisfies 
  \begin{align*}
    \text{sep}(B_{11}, B_{22}) & \geq \min_{\lambda_i \in \lambda(B_{11}), \lambda_j \in \lambda(B_{22})} \abs{\lambda_i - \lambda_j} \\
    & \geq \abs{\lambda_{K} - \lambda_{K+1}}  > 0
  \end{align*}
  where the last inequality due to Assumption~\ref{asp4} requiring that $\lambda_K >0$ and $\lambda_{K+1}/\lambda_{K} =o(1) $. 

  Now we readily have the condition in Lemma~\ref{lm6} satisfied such that
  \begin{align*}
    \theta^{-2}  \eta \norm{E_{21}}_{\mathcal{L}} & \leq \frac{\left(\norm{Q_1^{\ast}CQ_2}_{\mathcal{L}} + \norm{Q_1^{\ast}(\widehat{C}_{h,q} - C)Q_3}_{\mathcal{L}}\right)\norm{Q_3^{\ast}(\widehat{C}_{h,q} - C)Q_1}_{\mathcal{L}}}{\left(\abs{\lambda_{K+1} - \lambda_K} - \norm{Q_1^{\ast}(\widehat{C}_{h,q} - C)Q_1}_{\mathcal{L}} - \norm{Q_3^{\ast}(\widehat{C}_{h,q} - C)Q_3}_{\mathcal{L}} \right)^2} \\
    & \leq \frac{\left(O_P(T^{-2/5})+O_P(T^{-2/5})\right)O_P(T^{-2/5})}{\left(\abs{\lambda_{K+1} - \lambda_K} - O_P(T^{-2/5}) - O_P(T^{-2/5})\right)^2  } < \frac{1}{4}.
  \end{align*}
  By Lemma~\ref{lm6}, we can then write
  \begin{equation*}
    \widehat{Q}_1 = (Q_1 + Q_2P)(I + P^{\ast}P)^{-1/2},
  \end{equation*}
  with
  \begin{align}
    \norm{P}_{\mathcal{L}} & \leq \frac{2\norm{E_{21}}_{\mathcal{L}}}{\text{sep}(Q_1^{\ast}CQ_1,Q_2^{\ast}CQ_2) - \norm{E_{11}}_{\mathcal{L}} - \norm{E_{22}}_{\mathcal{L}}} \nonumber \\
    & \leq \frac{2\norm{Q_3^{\ast}(\widehat{C}_{h,q} - C)Q_1}_{\mathcal{L}}}{\abs{\lambda_{K+1} - \lambda_K} - \norm{Q_1^{\ast}(\widehat{C}_{h,q} - C)Q_1}_{\mathcal{L}} - \norm{Q_3^{\ast}(\widehat{C}_{h,q} - C)Q_3}_{\mathcal{L}}} \\
    & \leq \frac{2\times O_P(T^{-2/5})}{\abs{\lambda_{K+1} - \lambda_K} - O_P(T^{-2/5}) - O_P(T^{-2/5})} = O_P(T^{-2/5}).
    \label{P}
  \end{align}
  We can then compute the difference between $Q_1$ and its estimator as
  \begin{align}
    \norm{\widehat{Q}_1 - Q_1}_{\mathcal{L}} & = \norm{(Q_1 + Q_2P)(I + P^{\ast}P)^{-1/2} - Q_1}_{\mathcal{L}} \nonumber \\
    & = \norm{\left[Q_1 + Q_2P - Q_1(I + P^{\ast}P)^{1/2}\right](I + P^{\ast}P)^{-1/2}}_{\mathcal{L}} \nonumber \\
    & \leq \norm{Q_1\left[I - (I + P^{\ast}P)^{1/2}\right](I + P^{\ast}P)^{-1/2}}_{\mathcal{L}} + \norm{Q_2P(I + P^{\ast}P)^{-1/2}}_{\mathcal{L}} \nonumber \\
    & \leq \norm{[I - (I + P^{\ast}P)^{1/2}](I + P^{\ast}P)^{-1/2}}_{\mathcal{L}} + \norm{P(I + P^{\ast}P)^{-1/2}}_{\mathcal{L}} \nonumber \\
    & \leq \norm{I - (I + P^{\ast}P)^{1/2}}_{\mathcal{L}} + \norm{P}_{\mathcal{L}} \nonumber \\
    & \leq 2\norm{P}_{\mathcal{L}} \nonumber \\
    & = O_P(T^{-2/5}).
    \label{Q_1_diff}
  \end{align}

  Using the linearity and symmetric properties of inner product, and the fact that $C(\phi_{K+j}) = \lambda_{K+j}\phi_{K+j}$, for $j = 1, \cdots k_{max}-K$, we have
  \begin{align}
    \abs{\widehat{\lambda}_{K+j}} & = \abs{\inp{\widehat{\phi}_{K+j}}{\widehat{C}_{h,q}({\widehat{\phi}_{K+j}})}} \nonumber \\
    & = \abs{\inp{\widehat{\phi}_{K+j} - \phi_{K+j} + \phi_{K+j}}{(\widehat{C}_{h,q}-C+C)({\widehat{\phi}_{K+j} - \phi_{K+j} + \phi_{K+j}})}} \nonumber \\
    & = \left|\inp{\widehat{\phi}_{K+j} - \phi_{K+j}}{(\widehat{C}_{h,q}-C)(\widehat{\phi}_{K+j} - \phi_{K+j})} \right. \nonumber \\
    & \qquad + \inp{\widehat{\phi}_{K+j} - \phi_{K+j}}{C(\widehat{\phi}_{K+j} - \phi_{K+j})} + 2\inp{\widehat{\phi}_{K+j} - \phi_{K+j}}{(\widehat{C}_{h,q}-C)(\phi_{K+j})} \nonumber \\
    & \qquad + \left. 2\inp{\widehat{\phi}_{K+j} - \phi_{K+j}}{C(\phi_{K+j})} \right| + \abs{\inp{\phi_{K+j}}{C(\phi_{K+j})}} \nonumber \\
    & \leq \abs{\lambda_{K+j}} + \abs{\inp{\widehat{\phi}_{K+j} - \phi_{K+j}}{(\widehat{C}_{h,q}-C)(\widehat{\phi}_{K+j} - \phi_{K+j})}} \nonumber \\
    & \qquad + \abs{\inp{\widehat{\phi}_{K+j} - \phi_{K+j}}{C(\widehat{\phi}_{K+j} - \phi_{K+j})}} + 2\abs{\inp{\widehat{\phi}_{K+j} - \phi_{K+j}}{(\widehat{C}_{h,q}-C)(\phi_{K+j})}} \nonumber \\
    & \qquad + 2\abs{\inp{\widehat{\phi}_{K+j} - \phi_{K+j}}{C(\phi_{K+j})}} \quad \text{(triangle inequality)} \nonumber \\
    & \leq \abs{\lambda_{K+j}} + \norm{\widehat{\phi}_{K+j} - \phi_{K+j}}^2 \norm{\widehat{C}_{h,q}-C}_{\mathcal{S}} \nonumber \\
    & \qquad + \norm{\widehat{\phi}_{K+j} - \phi_{K+j}}^2\norm{C}_{\mathcal{S}} + 2\norm{\widehat{\phi}_{K+j} - \phi_{K+j}}\norm{\phi_{K+j}}\norm{\widehat{C}_{h,q}-C}_{\mathcal{S}} \nonumber \\
    & \qquad + 2\abs{\lambda_{K+j}} \norm{\phi_{K+j}}\norm{\widehat{\phi}_{K+j} - \phi_{K+j}}  \quad (\text{Cauchy-Schwarz inequality}) \nonumber \\
    & = \abs{\lambda_{K+j}} + O_P(T^{-4/5}),
    \label{lambda_nonspike}
  \end{align}
  where the last inequality follows from \eqref{Q_1_diff} and Lemmas~\ref{lm1} to~\ref{lm4}.

  Denoting $a \asymp b$ if $a = O_P(b)$ and $b = O_P(a)$, conditions in Assumption~\ref{asp4} indicate that $\lambda_{k+1}/\lambda_{k} \asymp 1$ for $k = 1, \cdots, K-1$, and $\lambda_{K+1}/\lambda_{K} \asymp o(1)$, with $\lambda_K > 0$.
  By~\eqref{lambda_diff}, for $k = 1, \cdots, K-1$, we then have
  \begin{equation}
    \frac{\widehat{\lambda}_{k+1}}{\widehat{\lambda}_{k}} = \frac{\lambda_{k+1}+ O_P\left(T^{-2/5}\right)}{\lambda_{k} + O_P\left(T^{-2/5}\right)} \asymp 1 .
    \label{th0_eq8}
  \end{equation}
  Similarly, when $k = K$, by~\eqref{lambda_diff} and~\eqref{lambda_nonspike}, we have
  \begin{equation}
    \frac{\widehat{\lambda}_{K+1}}{\widehat{\lambda}_{K}} = \frac{\lambda_{K+1} + O_P\left(T^{-4/5}\right)}{\lambda_{K}+ O_P\left(T^{-2/5}\right)} \overset{P}{\to} 0,
    \label{th0_eq9}
  \end{equation}
  and
  \begin{equation}
    \frac{\widehat{\lambda}_K}{\widehat{\lambda}_1} = \frac{\lambda_{K}+ O_P\left(T^{-2/5}\right)}{\lambda_{1}+ O_P\left(T^{-2/5}\right)} \asymp 1.
  \end{equation}
  Since $\lambda_{K+1} > \lambda_{K+2} > \cdots > \lambda_{k_{max}}$, by~\eqref{lambda_nonspike} we have, for $k = K+1, \cdots, k_{max}$,
  \begin{equation}
    \frac{\widehat{\lambda}_{k}}{\widehat{\lambda}_{1}} \leq \frac{\lambda_{K+1}  + O_P\left(T^{-4/5}\right)}{\lambda_1+O_P\left(T^{-2/5}\right)} = o_P(1),
    \label{th0_eq11}
  \end{equation}
  which is less than the threshold $\tau$ in~\eqref{eq_10}. With~\eqref{th0_eq8}-\eqref{th0_eq11}, we complete the proof of Proposition~\ref{pr1}.
\end{proof}

\begin{proof}[Proof of Theorem~\ref{thm1}]  
  We prove this theorem firstly assuming that $K$ is known. We then have
  \begin{align*}
  & \norm{\sum_{k=1}^{K}\beta_{t,k}\phi_k(u) - \sum_{k=1}^{K}\widehat{\beta}_{k,t}\widehat{\phi}_k(u)} \\
  & \leq  \sum_{k=1}^{K}\norm{ \beta_{t,k}\phi_k(u) - \widehat{\beta}_{k,t}\widehat{\phi}_k(u) } \\
  &=  \sum_{k=1}^{K}\norm{\beta_{t,k}\phi_k(u) + \beta_{t,k}\widehat{\phi}_k(u) - \beta_{t,k}\widehat{\phi}_k(u) - \widehat{\beta}_{k,t}\widehat{\phi}_k(u)} \\
  & \leq  \sum_{k=1}^{K} \left\lbrace |\beta_{t,k} |\norm{\phi_k(u) - \widehat{\phi}_k(u)} + |\beta_{t,k} -\widehat{\beta}_{t,k} |\norm{\widehat{\phi}_k(u)}   \right\rbrace \quad (\text{triangle inequality}) \\
  & \leq \sum_{k=1}^{K}\sqrt{|\beta_{t,k}|^2 \norm{\phi_k(u) - \widehat{\phi}_k(u)}^2} + \sum_{k=1}^{K}\sqrt{\{\beta_{t,k} -\widehat{\beta}_{t,k} \}^2 \norm{\widehat{\phi}_k(u)}^2} \quad (\text{Cauchy-Schwarz inequality}) \\
  & = O_P\left(T^{-2/5}\right), 
  \end{align*}
  where we have used the fact that the estimated eigenfunctions have unit length due to normalization, i.e., $\norm{\widehat{\phi}_k(u)}^2 = 1$, and results of Lemma~\ref{lm4} and Lemma~\ref{lm5} in the last step.

  By Proposition~\ref{pr1}, $\text{Pr}(\widehat{K} = K) \rightarrow 1$. We readily have the unconditional arguments, completing the proof of Theorem~\ref{thm1}.
\end{proof}

\subsection{Consistency of local feature estimators}\label{app_a2}
\begin{lemma} 
  Define the Besov space ball $B_{P,Q}^{\alpha}(M)$ as
  \begin{equation*}
    B_{P,Q}^{\alpha} = \left\lbrace f \in L^P: \sum_{j\leq0}(2^{j(\alpha+1/2-1/P)}\norm{\bm{D}_{j\cdot}}_P)^Q  < M  \right\rbrace, \qquad 1\leq P,Q \leq \infty
  \end{equation*}
  where $\alpha > 0$ and $\bm{D}_{j\cdot}$ is the vector of wavelet coefficients at the resolution level $j$. 
  Define the NRSI wavelet coefficient estimator as $\widehat{\bm{D}} = \bm{A}\bm{f}_n$, where $\bm{f}_n = \left(f(u_1), \cdots, f(u_n)\right)^{\top}$, and $\bm{A}$ is defined in~\eqref{eq_11}. Under  Assumption~\ref{asp6}, $\widehat{\bm{D}} \in B_{P,Q}^{\alpha}$.
  \label{lm7}
\end{lemma}

\begin{proof}[Proof of Lemma~\ref{lm7}]  
  This lemma is contributes to Theorem 5 and Theorem 6 in \cite{AF01}. Derivation of this lemma is given in the proof to Theorem 6 in \cite{AF01}.
\end{proof}

\begin{remark*}
  Lemma~\ref{lm7} indicates that we interest in performances of wavelet approximation over the Besov space $B_{P,Q}^{\alpha}$. Roughly speaking, the the Besov space $B^{\alpha}_{P,Q}$ contains functions having $\alpha$ bounded derivatives in $L^{P}$ space, the second parameter $Q$ gives a finer gradation of smoothness. Generally, we use $\alpha$ to indicate the degree of smoothness of the underlying signal $f$. See \cite{Meyer92} for definitions and properties of Besov space. Following this lemma, we can apply the results of \cite{HKP99} to derive the consistency of wavelet estimators.
\end{remark*}

\begin{lemma}
  (Proposition 1  in \cite{Cai02})
  Suppose that $u_i \stackrel{ind.}{\sim} N(D_i, \sigma^2_{\ast}), i = 1, \cdots, L$. Let $\widetilde{D}_i = u_i \mathds{1}(S^2 > \lambda L\sigma^2_{\ast})$, where $S^2 = \sum_{i=1}^{L} x_i^2$. Let $\widetilde{\bm{D}}_{L} = [\widetilde{D}_1, \cdots, \widetilde{D}_L]$, and $\bm{D}_L = [D_1, \cdots, D_L]$. In addition, $\lambda = 4.5052$, the root of $\lambda - \log \lambda - 3 = 0$, $L = \log N$ and $\sigma^2_{\ast} = \sigma^2/N$, then
  \begin{equation*}
    E\norm{\widetilde{\bm{D}}_L - \bm{D}_L}_2^2 \leq (2\lambda + 2)(\norm{\bm{D}_L}_2^2 \wedge L\sigma^2)+ 2\lambda \sigma^2N^{-2}\log N.
  \end{equation*}
  \label{lm8}
\end{lemma}

\begin{proof}[Proof of Lemma~\ref{lm8}]  
  This lemma corresponds to Proposition 1 in \cite{Cai02} when the optimal choice of parameters $\lambda$ and $L$ are used. 
\end{proof}

\begin{remark*}
  Lemma~\ref{lm8} gives a risk measure (mean squared error) of the estimated wavelet coefficients after block thresholding. The second term on the right hand side of the risk inequality is negligible, indicating the estimator achieves, within a constant factor, the optimal balance between the variance and the squared bias over the blocks.
\end{remark*}

\begin{definition}
  Let $\mathcal{H} = \mathcal{H}(\alpha_1, \alpha, \gamma, M_1, M_2, M_3, r, v)$, where $0 \leq \alpha_1 \leq \alpha \leq r$, $0 \leq \gamma < \frac{1+2\alpha_1}{1+2\alpha}$, and $M_1, M_2, M_3, v \geq 0$, denote the class of functions $f$ such that for any $j\geq j_0 >0$ there exists a set of integers $A_j$ with cardinality $card(A_j) \leq M_32^{j\gamma}$ for which the following are true:
  \begin{itemize}
    \item for each $p \in A_j$, there exist constants $a_0 = f(2^{-j}p)$, $a_1, \cdots, a_{r-1}$ such that for all $u \in [2^{-j}p,2^{-j}(p+v)]$, $|f(u) - \sum_{m=0}^{r-1}a_m(u-2^{-j}p)^m| \leq M_12^{{-j}\alpha_1}$;
    \item for each $p \notin A_j$, there exist constants $a_0 = f(2^{-j}p)$, $a_1, \cdots, a_{r-1}$ such that for all $u \in [2^{-j}p,2^{-j}(p+v)]$, $|f(u) - \sum_{m=0}^{r-1}a_m(u-2^{-j}p)^m| \leq M_22^{{-j}\alpha}$.
  \end{itemize}
  \label{df2}
\end{definition}

\begin{remark*}
  Broadly speaking, the intervals with indices in $A_j$ are ``bad'' intervals which contain less smooth parts of the function.
  Each function $f \in \mathcal{H}(\alpha_1, \alpha, \gamma, M_1, M_2, M_3, r, v)$ can be approximated by a regular smooth function $f_1$ in the Besov space $B_{\infty,\infty}^{\alpha}(M_2)$ and an irregular perturbation $f_2$: $f = f_1 + f_2$. The perturbation $f_2$ can be, for example, jump discontinuities or high frequency oscillations. Convergence rates are determined by the smooth component $f_1$. The function class $\mathcal{H}(\alpha_1, \alpha, \gamma, M_1, M_2, M_3, r, v)$ contains the Besov class $B_{\infty,\infty}^{\alpha}(M_2)$ as a subset for any given $\alpha_1, \gamma, M_1, M_3, r$ and $v$ \citep[see also Section 3.1 in][]{Cai02}. We hold $\alpha_1, \alpha, \gamma, M_1, M_2, r$ and $v$ fixed, but allow $M_3$ to depend on the sample size. Detailed explanation of each parameter an be found in \cite{HKP99}. 
\end{remark*}

After introducing the function class $\mathcal{H}$, the following lemma shows how the conditions defining $\mathcal{H}$ have direct implications for the wavelet expansion of a function $f \in \mathcal{H}$.

\begin{lemma}
  (Lemma 1 (i) in \cite{Cai02})
  Let $f$ be a function belonging to the function class $\mathcal{H}(\alpha_1, \alpha, \gamma, M_1, M_2, M_3, r, v)$. Assume the wavelets $\{\psi, \Psi\}$ with supports $supp(\psi) = supp(\Psi) \subset [0,v]$. Let $N = 2^J$. For the wavelet coefficients $D'_{J,p}$ and $D_{j,p}$ defined in~\eqref{eq_6}, we have
  \begin{align*}
    |D'_{J,p} - N^{-1/2}f(p/N) | & \leq M_1 \norm{\psi}_1 N^{-(1/2+\alpha_1)} \text{ for all } p \in A_{J}; \\
    |D'_{J,p} - N^{-1/2}f(p/N) | & \leq M_2 \norm{\psi}_1 N^{-(1/2+\alpha)} \text{ for all } p \notin A_{J}; \\
    |D_{j,p}| & \leq M_1 \norm{\Psi}_1 2^{-j(1/2+\alpha_1)} \text{ for all } p \in A_{j}; \\
    |D_{j,p}| & \leq M_2 \norm{\Psi}_1 2^{-j(1/2+\alpha)} \text{ for all } p \notin A_{j}.
  \end{align*}
  \label{lm9}
\end{lemma}

\begin{proof}[Proof of Lemma~\ref{lm9}] 
  This is Lemma 1 (i) in \cite{Cai02}, which follows directly Proposition 1 in \cite{HKP99}. The proof is therefore omitted.
\end{proof}

\begin{lemma}
  Let $N=2^J$ denote the number of realizations on curve $f(u)$. Denote $L^{\ast} = \log(2^J)$ and $\lambda^{\ast} = 4.5052$ as the optimal parameters for the blockwise thresholding. Use the notation $\widetilde{D}_{j,p}$ for the estimated wavelet coefficients after the optimal blockwise thresholding, and $D_{j,p}$ for the true coefficients of $f \in B^{\alpha}_{P,Q}$. We then have
  \begin{equation*}
    E\left[\sum_{j=j_0}^{J-1}\sum_{p}(\widetilde{D}_{j,p} - D_{j,p})^2\right] = O\left(N^{-2\alpha/(1+2\alpha)}\right),
  \end{equation*}
  where $\alpha > 0$ is fixed.
  \label{lm10}
\end{lemma}

\begin{proof}[Proof of Lemma~\ref{lm10}] 
  This lemma is part of global adaptivity results of \cite{Cai02}, with detailed proof provided in the proof to Theorem 4 in Section 8.2 of \cite{Cai02}.
\end{proof}

\begin{proof}{Proof of Theorem~\ref{thm2}} $ $\newline
  The residual function after FPCA is given by
  \begin{align*}
    \widehat{e}_t(u) & = \X_t(u) - \widehat{\mu}(u) - \sum_{k=1}^{\widehat{K}}\widehat{\beta}_{t,k}\widehat{\phi}_k(u) \\
    & = Z_t(u) +  \mu(u) - \widehat{\mu}(u) + \sum_{k=1}^{K}\beta_{t,k}\phi_k(u) - \sum_{k=1}^{\widehat{K}}\widehat{\beta}_{t,k}\widehat{\phi}_k(u) + \varepsilon_t(u)
  \end{align*}
  The NRSI initially interpolates residual observations on grids $\{u_1, \cdots, u_{n_t}\}$ into a vector $\widehat{\bm{e}}_t = [\widehat{e}_t(u_1), \cdots, \widehat{e}_t(u_N)]$ with $N = 2^J > n_t$ equally spaced points. According to \cite{AF01}, approximation errors in this step caused by moving nondyadic points to dyadic points are negligible. Hence, discretized FPCA residuals can be expressed as 
  \begin{equation}
    \widehat{e}_{t}(u_i) = Z_t(u_i) + \mu(u_i) - \widehat{\mu}(u_i) + \sum_{k=1}^{K}\beta_{t,k}\phi_k(u_i) - \sum_{k=1}^{\widehat{K}}\widehat{\beta}_{t,k}\widehat{\phi}_k(u_i) + \varepsilon_t(u_i),
    \label{eti}
  \end{equation} 
  where $i = 1, \cdots, N$, $u_i = i/N$. It follows from Definition 3.1 in \cite{Bosq00} that strong $H$-white noise process (independently and identically distributed sequence of random variables with mean 0 and constant variance taking values in $H$) can be expressed as 
  \begin{equation*}
    \varepsilon_t(u_i) = (\mathscr{W}(u_i) - \mathscr{W}(u_{i-1}))\cdot\sigma, \qquad 0 \leq u_{i-1} < u_i \leq 1, \quad  t \in \mathbb{Z}, 
  \end{equation*}
  where $\mathscr{W}(u)$, for $u \geq 0$ with $\mathscr{W}(0) = 0$ is a measurable bilateral Wiener process, and $\sigma$ is the noise level. By definition, the Wiener process has independent Gaussian increments, i.e., $\mathscr{W}(u) - \mathscr{W}(0) \sim \text{Normal}(0, u)$, and for $0 \leq u_{a} < u_b < u_c < u_d \leq 1$, $\mathscr{W}(u_b) - \mathscr{W}(u_a)$ independent of $\mathscr{W}(u_d) - \mathscr{W}(u_c)$. Since $u_i = i/N$ for $i = 1, \cdots, N$, we have equally sized increments $u_i - u_{i-1} = 1/N$. The sequence $\varepsilon_t(u_1), \cdots, \varepsilon_t(u_N)$ therefore follows an i.i.d. $\text{Normal}(0, \sigma^2/N)$ distribution.

  We consider least asymmetric wavelets $\{\psi, \Psi\}$ constructed by \cite{D92}. Using the ``subband filtering schemes'' discussed by \citet[][Chapter~5]{D92}, the true function $Z_t(u)$ can be approximated by discretized observations as
  \begin{equation*}
    \widetilde{Z}_t(u) = \sum_{i=1}^{N}N^{-1/2}Z_{t}(u_i)\psi_{J,i}(u).
  \end{equation*}
  Let $D'_{j_0,p,t}$ and $D_{j,p,t}$ denote the true wavelet coefficients of $Z_t(u)$, i.e., $D'_{j_0,p,t} = \inp{Z_t}{\psi_{j_0,p}}$ and $D_{j,p,t} = \inp{Z_t}{\Psi_{j,p}}$. Plugging $Z_{t}(u_i)$ from~\eqref{eti} into the last equation, we have 
  \begin{align}
    \widetilde{Z}_t(u) & = \sum_{i=1}^{N}N^{-1/2}\left\lbrace \widehat{e}_{t}(u_i) - [\mu(u_i) - \widehat{\mu}(u_i)] - \left[\sum_{k=1}^{K}\beta_{t,k}\phi_k(u_i) - \sum_{k=1}^{\widehat{K}}\widehat{\beta}_{t,k}\widehat{\phi}_k(u_i)\right] - \varepsilon_t(u_i) \right\rbrace \psi_{J,i}(u) \nonumber \\
    & = \sum_{i=1}^{N}\left\lbrace D'_{J,i,t}  + [N^{-1/2}\widehat{e}_{t}(u_i) - D'_{J,i,t}] -  N^{-1/2}[\epsilon_{i} N^{-1/2}\sigma] \right\rbrace \psi_{J,i}(u) \nonumber \\
    & \qquad  + \sum_{i=1}^{N} \left[\widehat{\mu}(u_i) - \mu(u_i) + \sum_{k=1}^{\widehat{K}}\widehat{\beta}_{t,k}\widehat{\phi}_k(u_i) - \sum_{k=1}^{K}\beta_{t,k}\phi_k(u_i) \right]  \psi_{J,i}(u) \nonumber \\
    & = \sum_{p=1}^{2^{j_0}}\left\lbrace D'_{j_0,p,t} + a'_{j_0,p,t} +  \sigma\epsilon_{j_0,p,t}/N \right\rbrace\psi_{j_0,p}(u) 
    +  \sum_{j={j_0}}^{J-1} \sum_{p=1}^{2^j} \left\lbrace D_{j,p,t} + a_{j,p,t} + \sigma\epsilon_{j,p,t}/N \right\rbrace\Psi_{j,p}(u) \nonumber \\
    & \qquad + \sum_{i=1}^{N} \left[\widehat{\mu}(u_i) - \mu(u_i) + \sum_{k=1}^{\widehat{K}}\widehat{\beta}_{t,k}\widehat{\phi}_k(u_i) - \sum_{k=1}^{K}\beta_{t,k}\phi_k(u_i) \right]\psi_{J,i}(u),
    \label{ztilde}
  \end{align}
  where $\epsilon_i$'s are i.i.d. $\text{Normal}(0, 1)$ such that $var(\varepsilon_t(u_i)) = var(\epsilon_{i} N^{-1/2}\sigma) = \sigma^2/N$. In~\eqref{ztilde}, $D'_{j0,p,t}$ and $D_{j,p,t}$ are the orthogonal transform of $\{D'_{J,i,t}\}_{i=1}^{N}$ via the DWT base matrix $\bm{W}$, likewise $a'_{j_0,p,t}$ and $a_{j,p,t}$ the transform of $\{N^{-1/2}\widehat{e}_{t}(u_i) - D'_{J,i,t}\}_{i=1}^{N}$, and $\epsilon_{j_0,p,t}$ and $\epsilon_{j,p,t}$ the transform of $\{\epsilon_i\}_{i=1}^N$. The $\epsilon_{j_0,p,t}$ and $\epsilon_{j,p,t}$ are i.i.d. $\text{Normal}(0, 1)$ since $\epsilon_i$'s are i.i.d. $\text{Normal}(0, 1)$. By Lemma~\ref{lm9}, the approximation errors satisfy
  \begin{equation}
    \sum_{p=1}^{2^{j_0} } (a'_{j_0,p,t})^2 + \sum_{j={j_0}}^{J-1} \sum_{p=1}^{2^j} a_{j,p,t}^2 = \sum_{i=1}^{N}[N^{-1/2}\widehat{e}_{t}(u_i) - D'_{J,i,t}]^2 = o(N^{-2\alpha/(1+2\alpha)}).
    \label{sum_a}
  \end{equation}
  More details about the derivation of this result can be found in Page 43 of \cite{HKP99}. 

  Let $\widehat{D}'_{j_0,p,t} = D'_{j_0,p,t} + a'_{j_0,p,t} +  \sigma\epsilon_{j_0,p,t}/N$ and $\widehat{D}_{j,p,t} = D_{j,p,t} + a_{j,p,t} + \sigma\epsilon_{j,p,t}/N$ denote the NRSI wavelet coefficients. By Lemma~\ref{lm7}, $\widehat{\bm{D}}_t \in B_{P,Q}^{\alpha}$. According to Definition~\ref{df2}, the Besov space $B_{P,Q}^{\alpha}$ is a subset of the function class $\mathcal{H}(\alpha_1, \alpha, \gamma, M_1, M_2, M_3, r, v)$ (more details see Example 3.1 in \cite{HKP99}). Thus, we can apply Lemma~\ref{lm9} and Lemma~\ref{lm10} in the following derivations involving NRSI estimator $\widehat{\bm{D}}_t$. 
  Denoting the wavelet coefficients after blockwise thresholding as $\widetilde{\bm{D}}_t$ as in Section~\ref{sec:3.2}, according to \eqref{eq_8} we have
  $\widetilde{D}_{j_0,p,t}' = \widehat{D}_{j_0,p,t}'$ and $\widetilde{D}_{j,p,t} = \widehat{D}_{j,p,t}\mathds{1}(S^2_{j_a} > \lambda^{\ast}L^{\ast}\sigma^2/N)$ for $(j,p) \in j_a$.
  The orthonormal wavelet functions satisfy $\norm{\psi} = \norm{\Psi} = 1$. By the isometry of the function norm and the sequence norm, then by triangle inequality we have
  \begin{align}
    E\norm{Z_t(u) - \widetilde{Z}_t(u)}^2 & \leq \mathfrak{c}_0 \left\lbrace \sum_{p=1}^{2^{j_0}}E(D_{j_0,p,t}'-\widetilde{D}_{j_0,p,t}')^2 + \sum_{j={j_0}}^{J-1} \sum_{p=1}^{2^j}E(D_{j,p,t} - \widetilde{D}_{j,p,t})^2 + \sum_{j=J}^{\infty}\sum_{p=1}^{2^j}D_{j,p,t}^2 \right. \nonumber \\
    & \qquad + \left. \sum_{i=1}^{N}E[\widehat{\mu}(u_i) - \mu(u_i)]^2 + \sum_{i=1}^{N}E\left[\sum_{k=1}^{\widehat{K}}\widehat{\beta}_{t,k}\widehat{\phi}_k(u_i) - \sum_{k=1}^{K}\beta_{t,k}\phi_k(u_i) \right]^2 \right\rbrace,
    \label{zdif}
  \end{align}
  where $\mathfrak{c}_0$ is a constant.

  We need to show that $E\norm{Z_t(u) - \widetilde{Z}_t(u)}^2 \rightarrow 0$ as $T,N \rightarrow \infty$. The result of convergence can be easily confirmed since each summand in~\eqref{zdif} converges to zero:
  \begin{itemize}
    \item By Lemma~\ref{lm9} and~\eqref{sum_a},
      \begin{equation*}
        \sum_{p=1}^{2^{j_0}}E(D_{j_0,p,t}'-\widetilde{D}_{j_0,p,t}')^2 + \sum_{j=J}^{\infty}\sum_{p=1}^{2^j}D_{j,p,t}^2 = o(N^{-2\alpha/(1+2\alpha)}).
      \end{equation*}
    \item By Lemma~\ref{lm10} and~\eqref{sum_a},
      \begin{equation*}
        \sum_{j={j_0}}^{J-1} \sum_{p=1}^{2^j}E(D_{j,p,t} - \widetilde{D}_{j,p,t})^2 =  O( N^{-2\alpha/(1+2\alpha)}).
      \end{equation*}
    \item By Proposition~\ref{lemma1}~(\ref{lemma1.2}),
      \begin{equation*}
        \sum_{i=1}^{N}E[\widehat{\mu}(u_i) - \mu(u_i)]^2 <  E\norm{\widehat{\mu}(u)-\mu(u)}^2 = O(1/T).
      \end{equation*}
    \item By Theorem ~\ref{thm1}, 
      \begin{equation*}
        \sum_{i=1}^{N}E\left[\sum_{k=1}^{\widehat{K}}\widehat{\beta}_{t,k}\widehat{\phi}_k(u_i) - \sum_{k=1}^{K}\beta_{t,k}\phi_k(u_i) \right]^2 < E\norm{\sum_{k=1}^{\widehat{K}}\widehat{\beta}_{t,k}\widehat{\phi}_k(u) - \sum_{k=1}^{K}\beta_{t,k}\phi_k(u)}^2 = O(T^{-4/5}).
      \end{equation*}
  \end{itemize}
  Hence, the MSE of the estimator $\widetilde{Z}_t(u)$ satisfies
  \begin{equation*}
    E\norm{Z_t(u) - \widetilde{Z}_t(u)}^2 =  O( N^{-2\alpha/(1+2\alpha)} + T^{-4/5}).
  \end{equation*}
  The Chebyshev's inequality then implies that,
  \begin{equation*}
    \norm{Z_t(u) - \widetilde{Z}_t(u)} = O_P(N^{-\alpha/(1+2\alpha)} +  T^{-2/5}) = o_P(1).
  \end{equation*}
  
\end{proof}

\subsection{Consistency of functional time series estimators}\label{app_a3}

\begin{proof}{Proof of Theorem~\ref{thm3}} $ $\newline
  This theorem can be easily proved with results of Theorems~\ref{thm1} and~\ref{thm2}. By triangle inequality, we have
  \begin{align*}
  E\norm{\X_t(u) - \widehat{\X}_t(u)}^2 & = \norm{\sum_{k=1}^{K}\beta_{t,k}\phi_k(u) + Z_t(u)  + \varepsilon_t(u) - \left(\sum_{k=1}^{\widehat{K}}\widehat{\beta}_{t,k}\widehat{\phi}_k + \widehat{Z}_t(u) \right)}^2 \\
  & \leq E\norm{\sum_{k=1}^{K}\beta_{t,k}\phi_k(u) - \sum_{k=1}^{\widehat{K}}\widehat{\beta}_{t,k}\widehat{\phi}_k}^2 + \norm{ Z_t(u) - \widehat{Z}_t(u)}^2 \\
  & = O_P(T^{-4/5}) +  O( N^{-2\alpha/(1+2\alpha)} + T^{-4/5}).
  \end{align*}
  The  Chebyshev's inequality then implies that
  \begin{equation*}
    \norm{\X_t(u) - \widehat{\X}_t(u)} =  O_P(T^{-2/5}) +  O_P( N^{-\alpha/(1+2\alpha)} + T^{-2/5}).
  \end{equation*}
  Since $N$ is a positive integer, we have $ N^{-\alpha/(1+2\alpha)} > 0$ for $\alpha >0$. Thus,
  \begin{equation*}
    \norm{\X_t(u) - \widehat{\X}_t(u)} = O_P( N^{-\alpha/(1+2\alpha)} + T^{-2/5}).
  \end{equation*}
\end{proof}

\section{}\label{appendixB}

Appendix~\ref{app_b1} first provides additional details about estimating the empirical long-run covariance function of~\eqref{eq_9}. We then present technical details of applying a static version of the FPCA-BTW method in estimating functional time series and its covariance structure in Appendix~\ref{app_b2}. Appendix~\ref{app_b3} presents procedures of making point and interval forecasts using global and local features extracted by the FPCA-BTW method.

\subsection{Estimation of long-run covariance}\label{app_b1}

The optimal bandwidth parameter $\widehat{h}_{\text{opt}}$ in~\eqref{eq_9} is estimated via the ``plug-in'' algorithm of \cite{RS17} as follows. Initially, a pilot estimate of the long-run covariance function is computed utilizing the flat-top weight function $W_{\text{FT}}$ \citep{PR96} of the form 
\begin{equation}
W_{\text{FT}}\left(  x \right) = \begin{cases}
1, & 0 \leq  |x| \leq 0.5 \\
2-2|x|,& 0.5 < | x| < 1 \\
0, & | x| \geq 1
\end{cases},
\label{w_ft}
\end{equation}
with an initial bandwidth $h_1 = T^{1/5}$, and $p=0$, as:
\begin{equation*}
\widehat{C}^{(p)}_{h_1, \text{FT}} (u,s) = \sum_{\ell = -T}^{T} W_{\text{FT}} \left(\frac{\ell}{h_1} \right) |\ell|^p \widehat{c}_{\ell}(u,s).
\end{equation*}
Then, the bandwidth parameter can be estimated by
\begin{equation}
\hat{h}_{\text{opt}} = \widehat{\mathcal{C}}_0(h_1)  T^{1/5},
\label{hopt}
\end{equation}
where 
\begin{equation*}
\widehat{\mathcal{C}}_0(h_1) = \left(4 \norm{\omega \widehat{C}^{(2)}_{h_1, \text{FT}}}^2 \right)^{1/5} \left( \left( \norm{\widehat{C}^{(0)}_{h_1, \text{FT}}}^2 + \left( \int_0^1\widehat{C}^{(0)}_{h_1, \text{FT}}(u,u)\diff u \right)^2  \right)  \int_{-\infty}^{\infty} W_{\text{QS}}(x)\diff x \right) ^{-1/5},
\end{equation*}
and $\omega = 18\pi^2/125$; the quadratic spectral (QS) kernel function $W_{\text{QS}}$ is defined as
\begin{equation}
W_{\text{QS}}(x) = \frac{25}{12\pi^2x^2} \left( \frac{\sin(6\pi x/5)}{6\pi x/5} - \cos(6\pi x/5) \right).
\label{w_qs}
\end{equation}
Finally, plugging in the estimated bandwidth parameter into~\eqref{eq_9} yields
\begin{equation*}
\widehat{C}_{\widehat{h}_{\text{opt}}}(u,s) = \sum_{\ell = -T}^{T} W_{\text{QS}} \left(  \frac{\ell}{\widehat{h}_{\text{opt}}} \right)  \widehat{c}_{\ell}(u,s).
\end{equation*}

\subsection{Static FPCA-BTW method}\label{app_b2}

Global and local features extracted by the proposed FPCA-BTW method has been proved to make improved estimation of 
functional time series in the main document. When functional data possess weak serial dependence, static FPCA would adequately extract global features of the data. We illustrate that the static FPCA-BTW also contributes to more accurate estimation of the functional process and its covariance structure. This appendix serves as supplementary to Appendix~\ref{appendixA} that presents main proofs of theoretical results involving the dynamic FPCA-BTW method.

We illustrate that local features contribute to improved estimation of the functional process via an example involving functional time series $\{\X_t(u)\}_{t=1}^T$ as defined in~\eqref{eq_1}. For simplicity, we assume a zero mean function and weak serial dependence in the data.Non-zero mean functions in practice are handled by centralizing the functional observations.

Consider functional time series $\X_t(u) = \sum_{k=1}^{K}\beta_{t,k}\phi_k(u)  + Z_t(u) + \varepsilon_t(u)$ as in~\eqref{eq_1} with a fixed integer $K$. The sample covariance function for the observed $\{\X_t(u)\}_{t=1}^T$ is computed as
\begin{equation*}
  \widehat{c}_0(u,s) = \frac{1}{T}\sum_{t=1}^{T}[\X_t(u)-\widehat{\mu}(u)][\X_t(s)-\widehat{\mu}(s)], \qquad u,s \in [0,1],
\end{equation*}
where $\widehat{\mu}(u) = \frac{1}{T}\sum_{t=1}^{T}\X_t(u)$ is the empirical mean function.
Decomposing $\widehat{c}_0(u,s)$ yields global features $\sum_{k=1}^{\widehat{K}} \widehat{\beta}_{t,k}\widehat{\phi}_k(u)$ of the considered functional time series. Applying regularized wavelet approximations to FPCA residuals recover local features $\widehat{Z}_t(u)$. With the extracted global and local features, we obtain FPCA estimators $\X_t^{\text{FPCA}}(u)$ and FPCA-BTW estimators $\X_t^{\text{FW}}(u)$ given by
\begin{align*}
  \widehat{\X}_t^{\text{FPCA}}(u) & = \sum_{k=1}^{\widehat{K}} \widehat{\beta}_{t,k}\widehat{\phi}_k(u), \qquad \text{and} \\
  \widehat{\X}_t^{\text{FW}}(u) & = \sum_{k=1}^{\widehat{K}} \widehat{\beta}_{t,k}\widehat{\phi}_k(u) + \widehat{Z}_t(u).
\end{align*}
Computing the mean squared error (MSE) for both estimators then yields that
\begin{align*}
  & \frac{1}{T}\sum_{t=1}^{T}\norm{\X_t(u) - \widehat{\X}^{\text{FW}}_t(u)}^2 \\
  & = \frac{1}{T} \sum_{t=1}^{T}\norm{\sum_{k=1}^{K}\beta_{t,k}\phi_k(u)  + Z_t(u) + \varepsilon_t(u) - \sum_{k=1}^{\widehat{K}} \widehat{\beta}_{t,k}\widehat{\phi}_k(u) - \widetilde{Z}_t(u)}^2 \\
  & \leq \frac{1}{T}\sum_{t=1}^{T}\norm{\sum_{k=1}^{K}\beta_{t,k}\phi_k(u) - \sum_{k=1}^{\widehat{K}} \widehat{\beta}_{t,k}\widehat{\phi}_k(u)}^2 + \frac{1}{T}\sum_{t=1}^{T}\norm{Z_t(u) - \widetilde{Z}_t(u)}^2 \\
  & \leq \frac{1}{T}\sum_{t=1}^{T}\norm{\sum_{k=1}^{K}\beta_{t,k}\phi_k(u) - \sum_{k=1}^{\widehat{K}} \widehat{\beta}_{t,k}\widehat{\phi}_k(u)}^2 + \frac{1}{T}\sum_{t=1}^{T}\norm{Z_t(u)}^2 \\
  & \leq \frac{1}{T} \sum_{t=1}^{T}\norm{\sum_{k=1}^{K}\beta_{t,k}\phi_k(u)  + Z_t(u) + \varepsilon_t(u) - \sum_{k=1}^{\widehat{K}} \widehat{\beta}_{t,k}\widehat{\phi}_k(u)}^2 \\
  & = \frac{1}{T}\sum_{t=1}^{T}\norm{\X_t(u) - \widehat{\X}^{\text{FPCA}}_t(u)}^2.
\end{align*}
Thus, incorporating the extracted local features $\widetilde{Z}_t(u)$ produces a more efficient estimator $\widehat{\X}^{\text{FW}}_t(u)$ than the FPCA estimator $\widehat{\X}^{\text{FPCA}}_t(u)$. 

The true covariance function for the process $\X_t(u)$ can be expressed as
\begin{align}
  c_0(u,s) & = E\left[\left\lbrace \sum_{k=1}^{K}\beta_{t,k}\phi_k(u)  + Z_t(u) + \varepsilon_t(u) \right\rbrace \left\lbrace \sum_{k=1}^{K}\beta_{t,k}\phi_k(s)  + Z_t(s) + \varepsilon_t(s) \right\rbrace \right] \\
  & = E\left[ \sum_{k=1}^{K}\beta_{t,k}^2 \phi_k(u)\phi_k(s) \right] + E\left[Z_t(u)Z_t(s)\right].
  \label{c0_hat}
\end{align}
Derivation of~\eqref{c0_hat} follows that $\{\phi_k(u)\}_{k=1}^K$ are real-valued orthogonal functions with $K$ a fixed positive integer; a set of real numbers $\{\bm{\beta}_{t,k}\}_{k=1}^K = \{\beta_{t,1}, \cdots, \beta_{t,K}\}$ satisfy that $var(\bm{\beta}_j\widehat{D}_{j,p,t}\bm{\beta}_k) = 0$ for any $j \neq k$; $\{Z_t(u)\}_{t \in \mathbb{Z}}$ is a set of functions pairwise orthogonal with $\{\phi_k(u)\}_{k=1}^K$; $\{\varepsilon_t(u)\}_{t \in \mathbb{Z}}$ is Gaussian $H$-white noise with $E\left\{\varepsilon_t(u)\right\}=0$. Using only the extracted global features, we have an FPCA estimator for the covariance function given by
\begin{equation*}
  \widehat{c}^{\text{FPCA}}_0(u,s) = \frac{1}{T}\sum_{t=1}^{T} \sum_{k=1}^{\widehat{K}} \widehat{\beta}^2_{t,k}\widehat{\phi}_k(u)\widehat{\phi}_k(s).
\end{equation*}
Another estimator constructed with the extracted global and local features is given by
\begin{equation*}
  \widehat{c}^{\text{FW}}_0(u,s)  = \frac{1}{T}\sum_{t=1}^{T} \left\lbrace \sum_{k=1}^{\widehat{K}} \widehat{\beta}^2_{t,k}\widehat{\phi}_k(u)\widehat{\phi}_k(s) + \widetilde{Z}_t(u)\widetilde{Z}_t(s) \right\rbrace.
\end{equation*}
It can be easily seen that 
\begin{align*}
  & \norm{c_0 - \widehat{c}^{\textbf{Fw}}_0} \\
   & = \norm{E\left[ \sum_{k=1}^{K}\beta_{t,k}^2 \phi_k(u)\phi_k(s) \right] + E\left[Z_t(u)Z_t(s)\right] - \frac{1}{T}\sum_{t=1}^{T}\sum_{k=1}^{\widehat{K}} \widehat{\beta}^2_{t,k}\widehat{\phi}_k(u)\widehat{\phi}_k(s)  - \frac{1}{T}\sum_{t=1}^{T}\widetilde{Z}_t(u)\widetilde{Z}_t(s)} \\
  & \leq \norm{E\left[ \sum_{k=1}^{K}\beta_{t,k}^2 \phi_k(u)\phi_k(s) \right] + E\left[Z_t(u)Z_t(s)\right] - \frac{1}{T}\sum_{t=1}^{T}\sum_{k=1}^{\widehat{K}} \widehat{\beta}^2_{t,k}\widehat{\phi}_k(u)\widehat{\phi}_k(s)} \\
  & = \norm{c_0 - \widehat{c}^{\textbf{F}}_0},
\end{align*}
since $\widetilde{Z}_t(u)$ has sparse and highly localized spikes over $u \in [0,1]$. 
Hence, incorporating the extracted local features $\widetilde{Z}_t(u)$ gives improved estimator for covariance of the considered process.

The static FPCA-BTW can be applied to extract features of generated data described in Monte Carlo experiments in Section~\ref{sec:5.1} of the main article. Table~\ref{table_app_1} reports mean RSEs and running time of combinations of static FPCA and various local feature extraction methods. It can be seen that the BTW method outperforms both competing methods at recovering local features when applied to static FPCA residuals. Given that all RSEs reported are smaller than 1, extracted local features are tested to improve static FPCA performance.

\begin{table}[!htbp]
  \centering
  \small
    \tabcolsep 0.3in  
\caption{\small Mean RSE and running time of various local feature extraction methods (standard errors in parentheses). The bold entries highlighting the best performing method for each setting.}
{\renewcommand{\arraystretch}{1}
 \begin{tabular}{@{\extracolsep{5pt}}L{2cm}C{1cm}C{2.5cm}C{2.5cm}C{2.5cm}@{}}
 \toprule
     Sample size & & SFPCA & TWFPCA & BTW \\
     \midrule
    \multirow{2}{*}{$T=25$} & RSE & 0.665 (0.073) & 0.732 (0.056) & \textBF{0.644} (0.070) \\ 
     & Time & 15.286 (0.537) & 20.692 (2.946) & \textBF{0.092} (0.012) \\[5pt]
       \multirow{2}{*}{$T=50$} & RSE & 0.636 (0.057) & 0.720 (0.040) & \textBF{0.620} (0.053) \\ 
     & Time & 33.890 (1.222) & 19.494 (3.601) & \textBF{0.135} (0.016) \\[5pt] 
     \multirow{2}{*}{$T=100$} & RSE & 0.620 (0.040) & 0.710 (0.029) & \textBF{0.604} (0.039) \\ 
     & Time & 87.710 (3.008) &  19.927 (3.075) & \textBF{0.261} (0.096) \\
     \bottomrule
  \end{tabular}} \label{table_app_1}
\end{table}

\begin{figure}[!ht]
  \centering
  \includegraphics[width = 30pc]{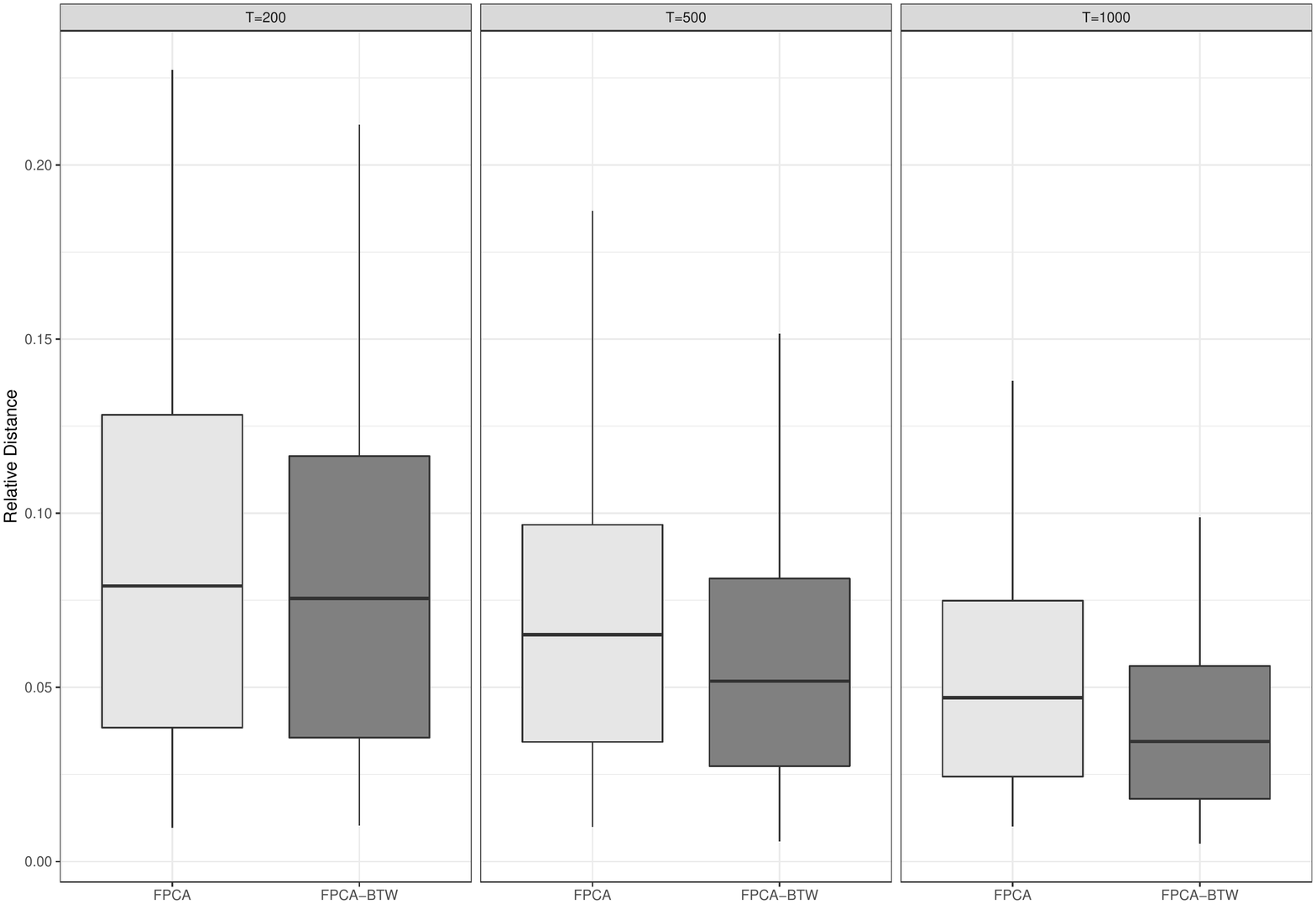}
  \caption{Relative errors of covariance estimators.}\label{fig:app_3}
\end{figure}

\begin{figure}[!ht]
\centering
\subfloat[\label{fig:4a}]{\includegraphics[width = 11pc]{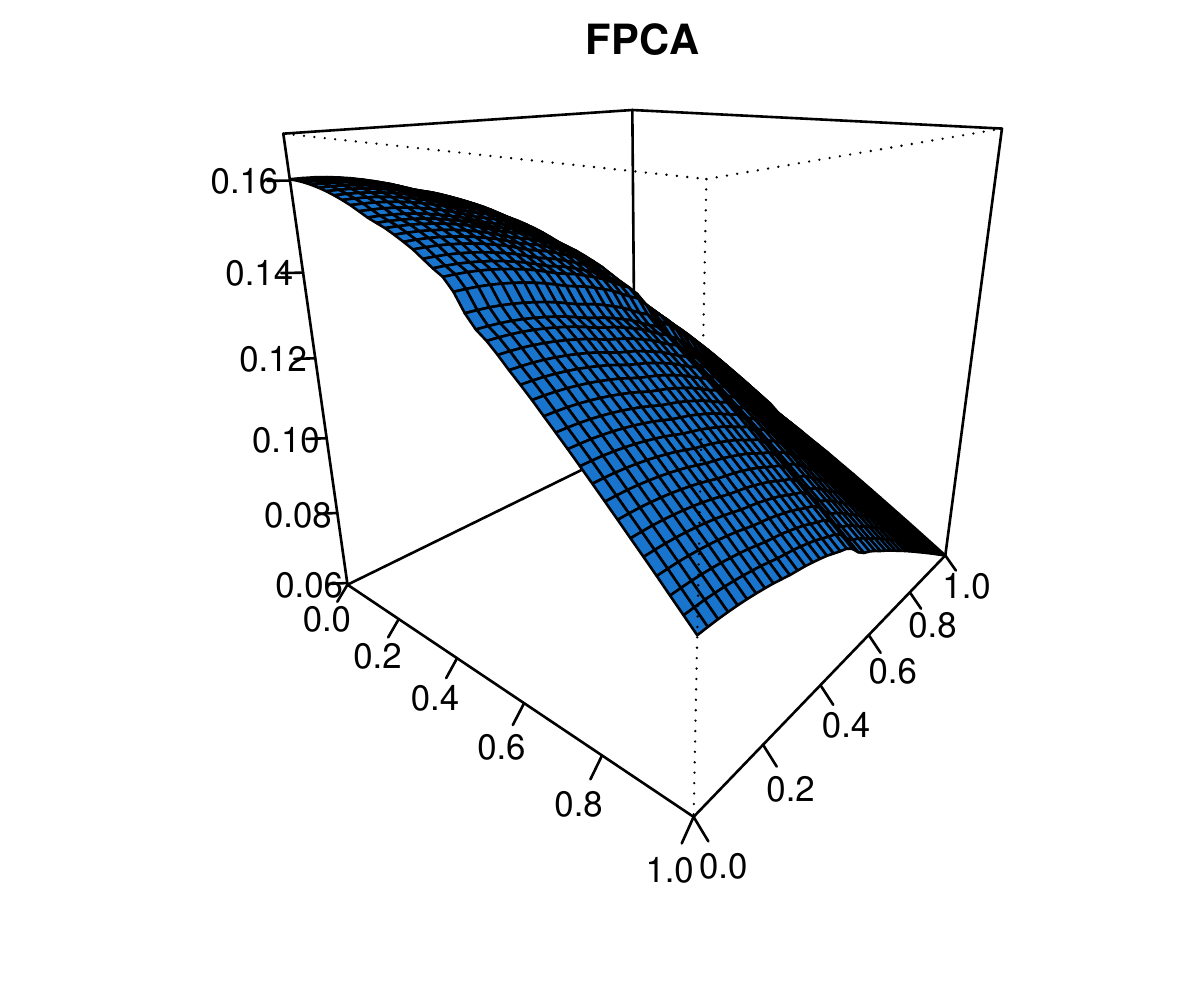}}
\subfloat[\label{fig:4b}]{\includegraphics[width = 11pc]{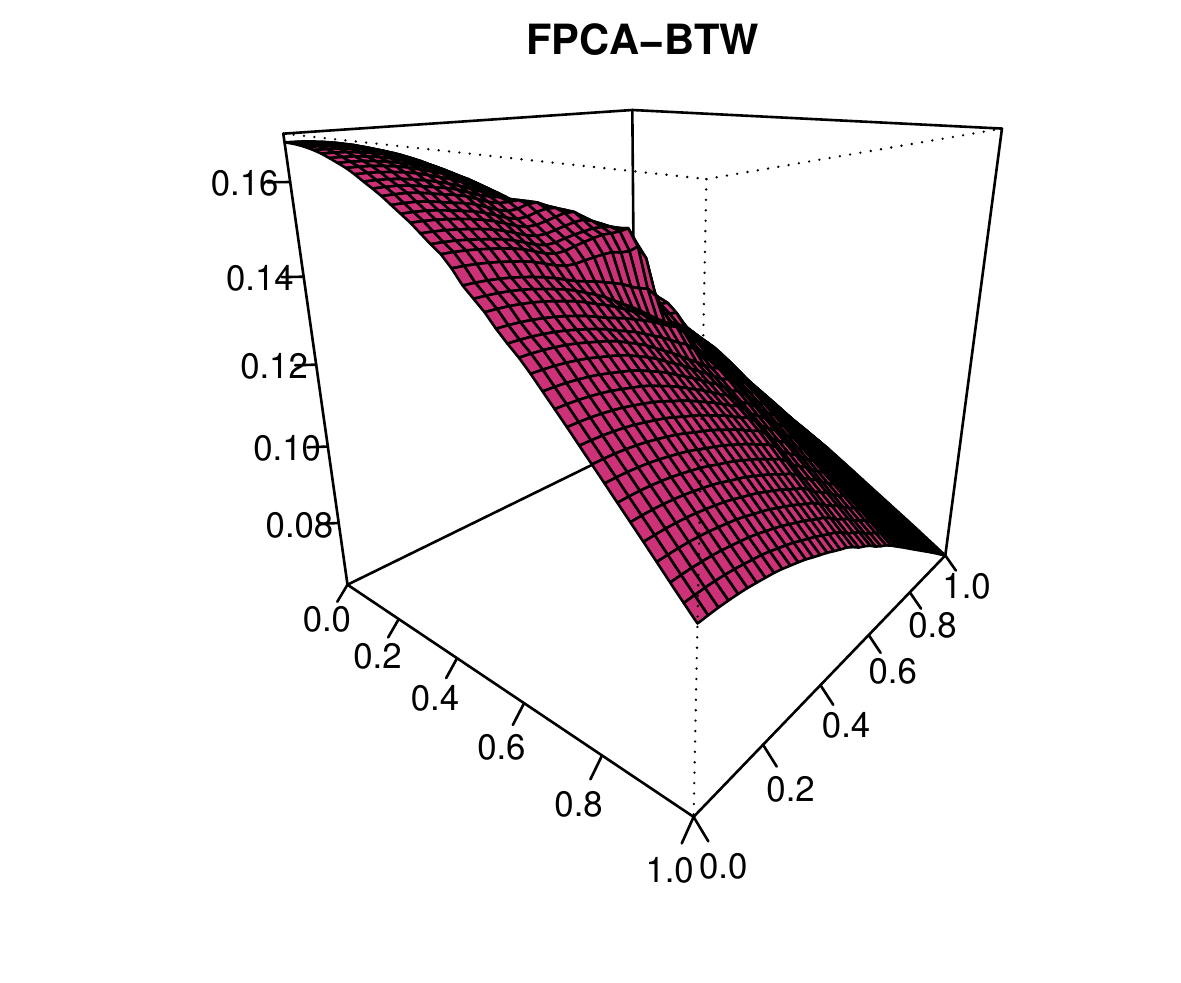}}
\subfloat[\label{fig:4c}]{\includegraphics[width = 11pc]{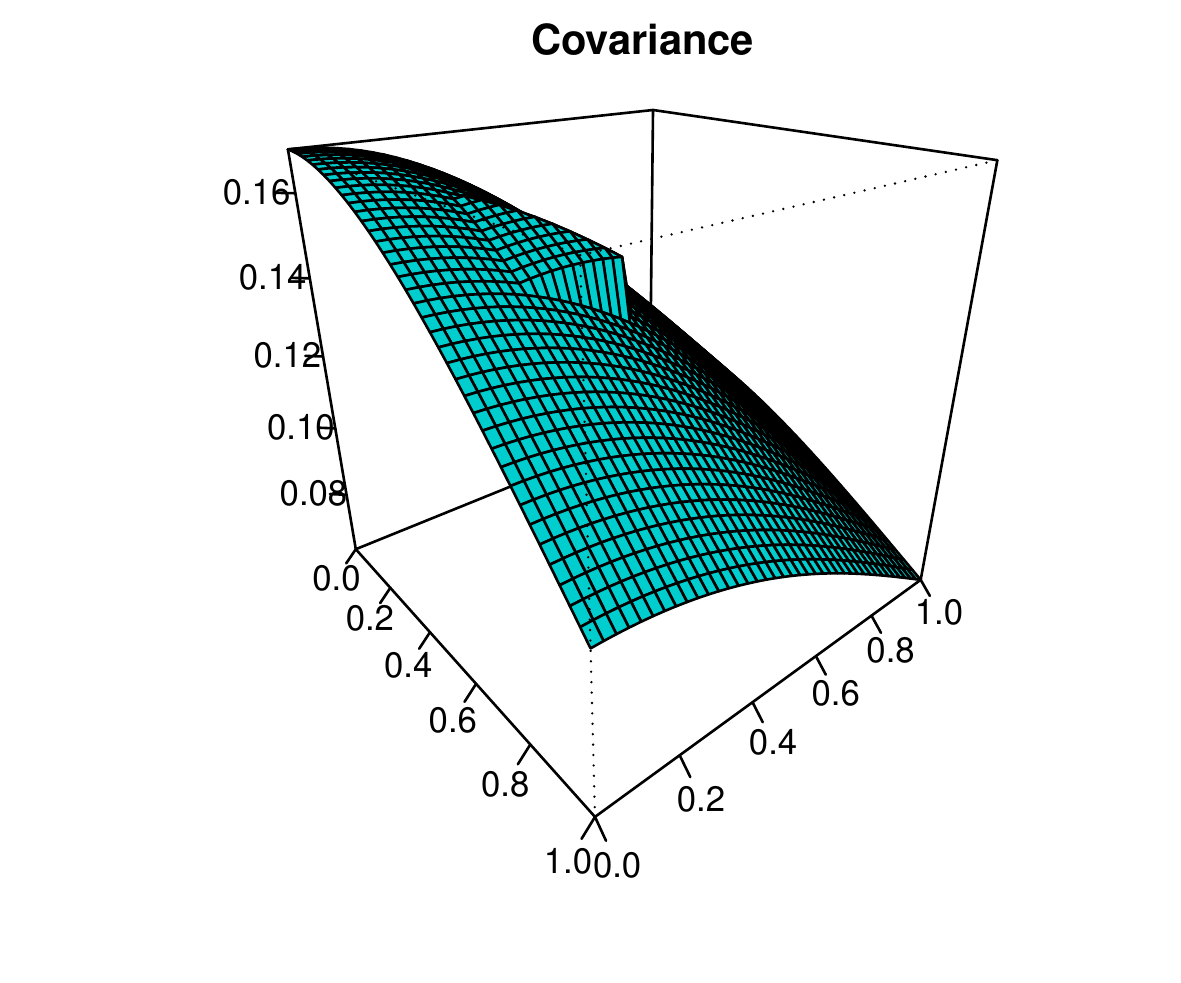}} 
\caption{Surface plots of the mean covariance estimators over 100 simulations obtained by FPCA (blue) and FPCA-BTW (red) for $T = 200$, along with the true theoretical functions (cyan).}\label{fig:app_4}
\end{figure}

Using the static FPCA-BTW method in Experiment 2 in Section~\ref{sec:5.2} of the main article produces smaller reconstruction errors than the static FPCA method, as shown in Fig.~\ref{fig:app_3}. Theoretical covariance functions of generated data have ``pyramid-shaped bumps" corresponding to local features assumed in data generating processes. Figure~\ref{fig:app_4}\subref{fig:4c} visualize the theoretical covariance function for sample size $T = 200$. As shown in Fig.~\ref{fig:app_4}\subref{fig:4a} and~\ref{fig:app_4}\subref{fig:4c}, extracted local features significantly improve covariance function estimation accuracy. 

We have demonstrated that BTW can be used to improve feature extraction performance of static FPCA. In the next section, we show that FPCA-BTW can be used to obtain more accurate forecasts Section~\ref{app_b3}.

\subsection{Forecasting functional time series} \label{app_b3}
We can produce an $h$-step-ahead point forecast for functional time series $\{\X_t(u)\}_{t=1}^T$ utilizing the extracted global and local features. The empirical principal component scores $\{\widehat{\beta}_{t,k}\}_{k=1}^{\widehat{K}}$ are uncorrelated satisfying that $var(\widehat{\beta}_{t,j}\widehat{\beta}_{t,k}) = 0$ for any $j \neq k$. The estimated wavelet coefficients at different resolution levels are also uncorrelated such that $var(\widehat{D}_{j_1,p,t}\widehat{D}_{j_2,p,t}) =0 $ for any $j_1 \neq j_2$. Within a particular resolution level $j$, non-zero wavelet coefficients $\widetilde{D}_{j,p,t}$ after the blockwise thresholding of~\eqref{eq_8} show very weak correlations since sparse local features rarely overlap. For these reasons, we use univariate time series models, such as autoregressive integrated moving average (ARIMA) models, to make forecasts for $\widehat{\beta}_{t,k}$ and $\widetilde{D}_t$ \citep[see, e.g.,][]{HU07,HS09}. 

Conditioning on the estimated $\widehat{\bm{\Phi}} = \{\widehat{\phi}_1(u), \cdots, \widehat{\phi}_K(u)\}$, and $\widetilde{\bm{D}} = \{\widetilde{\bm{D}}_{1}, \cdots, \widetilde{\bm{D}}_{T} \} $, an out-of-sample $h$-step-ahead point forecast can be obtained as 
\begin{align}
  \widehat{\X}_{T+h}(u) & = E[\X_{T+h}|\widehat{\bm{\Phi}}, \widetilde{\bm{D}}] \\
  & = \widehat{\mu}(u) + \sum_{k=1}^{\widehat{K}} \widehat{\beta}_{T+h|T,k}\widehat{\phi}_k(u) + \bm{A}^{\top} \widetilde{\bm{D}}_{T+h|T},
  \label{forecast}
\end{align}
where $\widehat{\beta}_{T+h|T,k}$ and $\widetilde{D}_{T+h|T}$ represents the time series forecasts of the principal component scores and wavelet coefficients, respectively.

We now present a Monte Carlo experiment to demonstrate that extracted local features help to improve point forecast accuracy. The FPCA extracted global features have been commonly used in forecasting functional time series \citep[see, e.g.,][]{HU07,Shang19}. We investigate if the empirically extracted local features contribute to more accurate point forecasts. Applying FPCA to smoothed near-infrared spectroscopy spectra of wood panels data as illustrated in Fig.~\ref{fig:1}\subref{fig:1d} in the main article yields two FPCs for global features, with serially correlated residuals following an autoregressive integrated moving average process of order (0,2,0) (\textsc{ARIMA}(0,2,0)). The data generating process for this experiment is then calibrated using the analysis results of the wood near-infrared spectroscopy data as follows. Choose $\phi_1(u) = \sin(\pi u)$ and $\phi_2(u) = \sin(2\pi u)$ as basis functions for global features, with their associated coefficients generated independently from AR(1) models of the form $\beta_{t,k} = \theta_k \beta_{t-1,k} + \omega_{t,k}$. Select $\theta_1 = 0.2$ and $\omega_{t,1} \sim N(0,10)$ for $\{\beta_{t,1}\}_{t=1}^T$, while choosing $\theta_2 = 0.8$ and $\omega_{t,2} \sim N(0,4)$ for $\{\beta_{t,2}\}_{t=1}^T$. Construct the basis function for local features as
\begin{equation*}
  \phi_3(u) =\begin{cases}
    \sin(\frac{u-a_1}{b_1-a_1} \pi), &  a_1 \leq u < b_1 \\
    2\sin(\frac{u-a_2}{b_2-a_2} \pi), &   a_2 \leq u < b_2 \\
  \end{cases},
\end{equation*}
where $a_1 \sim U(0.05,0.4)$ and $a_2 \sim U(0.55,0.8)$, and $b_1 - a_1 = b_2 - a_2 = 0.1$. Generate $\{\beta_{t,3}\}_{t=1}^T$ from an \textsc{ARIMA}(0,2,0) model with i.i.d. $\text{Normal}(0, 1)$ innovations.  Generate independent noise $\varepsilon_t(u) = \sqrt{0.1}B_t(u)$ with $B_t(u)$ i.i.d. standard Brownian motion $\{B_t(u), u \in [0,1]\}_{t=1}^T$. Finally, orthonormalize three basis functions and calculate simulated functional time series as $\X_t(u) = \beta_{t,1}\phi_{1}(u) + \beta_{t,2}\phi_{2}(u) + \beta_{t,3}\phi_{3}(u) + \varepsilon_t(u)$ for $u \in [0,1]$.

For each $T\in\{25,45,85\}$, apply the FPCA-BTW method to extract global and local features of $\{\X_t(u)\}_{t=1}^T$. $h$-step-ahead point forecasts can be computed using the obtained features as
\begin{equation*}
  \widehat{\X}_{T+h}(u) = \widehat{\mu}(u) + \sum_{k=1}^{\widehat{K}} \widehat{\beta}_{T+h|T,k}\widehat{\phi}_k(u) + \bm{A}^{\top} \widetilde{\bm{D}}_{T+h|T},
\end{equation*}
where $\bm{A}$ is defined in~\eqref{eq_11}, and $\{\widehat{\phi}_k\}_{k=1}^{\widehat{K}}$ are the empirical functional components; $\widehat{\beta}_{T+h|T,k}$ and $\widetilde{D}_{T+h|T}$ are the forecasts obtained via univariate time series methods. We consider one- to five-step-ahead point forecasts, i.e., $h = 1, \cdots 5$, and adopt the expanding window method of \cite{ZW06} to increase the training size by 1 in each iteration. Forecasts obtained in 100 replications for each sample size are then evaluated by the mean absolute forecast error (MAFE), and the root mean squared forecast error (RMSFE). For a particular forecast horizon $h$, the evaluation measures are given by
\begin{align*}
\text{MAFE}(h) &= \frac{1}{(6-h)\times 100}\sum^{5}_{\varsigma = h}\sum^{100}_{i=1}|\X_{T-5+\varsigma}(u_i) - \widehat{\X}_{T-5+\varsigma|T-5+\varsigma-h}(u_i)|, \\
\text{RMSFE}(h) &= \sqrt{\frac{1}{(6 -h)\times 100}\sum^{5}_{\varsigma = h}\sum^{100}_{i=1}\left[\X_{T-5+\varsigma}(u_i) - \widehat{\X}_{T-5+\varsigma|T-5+\varsigma-h}(u_i)\right]^2},
\end{align*}
where $\widehat{\X}_{T-5+\varsigma|T-5+\varsigma-h}$ represents the $h$-step-ahead forecast obtained based on a training set $\{\X_t\}_{t=1}^{T-5+\varsigma-h}$, and $\X_{T-5+\varsigma}$ is the corresponding actual observation; $i = \{1,\cdots, 100\}$ denote equally spaced grid points over $[0,1]$
 
Table~\ref{table_app_2} presents point forecast evaluation results for the FPCA-BTW method and the benchmark FPCA method under different settings. Because the local features extracted from the sample (long-run) covariance function inherit serial dependence of the original data, the FPCA-BTW extraction method produces more accurate point forecasts at each forecast horizon $h$ than the FPCA method. Note that the data generating process for this experiment uses a small AR coefficient $\theta_1 = 0.2$ in generating $\{\beta_{t,1}\}_{t=1}^T$. The resulted functional time series possess only mild serial dependence, for which data the static version of FPCA performs well. In practice, NIR spectra recorded by spectrometers over time often have moderate-to-strong serial dependence, such as the wood panel NIR data illustrated in Figure~\ref{fig:1}\subref{fig:1a} in the main article. 

\begin{table}[!htbp]
  \centering
  \small
  \tabcolsep 0.13in
  \caption{\small Mean MAFEs and RMSFEs of point forecasts averaged over 100 replications. The bold entries highlight the feature extraction method with higher forecast accuracy.} 
  \label{table_app_2}
  {\renewcommand{\arraystretch}{0.9}%
  \begin{tabular}{@{\extracolsep{5pt}}L{2.5cm}C{2cm}cccC{0.8cm}cc@{}}
    \toprule
    & & & \multicolumn{2}{c}{MAFE} & &\multicolumn{2}{c}{RMSFE} \\\cline{4-5}\cline{7-8}
     & Sample size & $h$ & FPCA & FPCA-BTW && FPCA & FPCA-BTW \\
    \midrule 
    \multirow{15}{*}{\textBF{Covariance}} &  & 1 & 0.420 & \textBF{0.404} && 0.531 &\textBF{0.505} \\ 
      & & 2 & 0.446 & \textBF{0.427} && 0.559 & \textBF{0.529} \\ 
      &   $T=25$  & 3 & 0.473 & \textBF{0.452} && 0.588 & \textBF{0.558} \\ 
      &   & 4 & 0.498 & \textBF{0.479} && 0.610 & \textBF{0.579} \\ 
      &   & 5 & 0.496 & \textBF{0.473} && 0.599 & \textBF{0.559} \\ 
    \cline{2-8}
      &  & 1 & 0.404 & \textBF{0.380} && 0.415 & \textBF{0.479} \\ 
      &   & 2 & 0.431 & \textBF{0.406} && 0.547 & \textBF{0.510} \\ 
    &   $T=45$  & 3 & 0.448 & \textBF{0.420} && 0.568 & \textBF{0.526} \\ 
      &   & 4 & 0.464 & \textBF{0.432} && 0.581 & \textBF{0.534} \\ 
      &   & 5 & 0.476 & \textBF{0.442} && 0.579 & \textBF{0.525} \\ 
    \cline{2-8}
      & & 1 & 0.390 & \textBF{0.362} && 0.507 & \textBF{0.465} \\ 
      &   & 2 & 0.427 & \textBF{0.399} && 0.545 & \textBF{0.501} \\ 
      & $T=85$  & 3 & 0.439 & \textBF{0.411} && 0.554 & \textBF{0.509} \\ 
      &   & 4 & 0.446 & \textBF{0.417} && 0.557 & \textBF{0.507} \\ 
      &   & 5 & 0.467 & \textBF{0.431} && 0.569 & \textBF{0.509} \\ 
    \midrule
    \multirow{15}{*}{\textBF{Long-run}}  & & 1 & 0.415 & \textBF{0.410} && 0.525 & \textBF{0.513} \\
    \multirow{15}{*}{\textBF{Covariance}} &   & 2 & 0.427 & \textBF{0.391} && 0.554 & \textBF{0.535} \\
      & $T=25$  & 3 & 0.470 & \textBF{0.461} && 0.585 & \textBF{0.570} \\
      &   & 4 & 0.488 & \textBF{0.479} && 0.600 & \textBF{0.584} \\
      &   & 5 & 0.488 & \textBF{0.476} && 0.590 & \textBF{0.567} \\
    \cline{2-8} 
      & & 1 & 0.405 & \textBF{0.390} && 0.514 & \textBF{0.492} \\
      &   & 2 & 0.424 & \textBF{0.407} && 0.537 & \textBF{0.513} \\
      & $T=45$  & 3 & 0.438 & \textBF{0.420} && 0.552 & \textBF{0.524} \\
      &   & 4 & 0.458 & \textBF{0.437} && 0.570 & \textBF{0.539} \\
      &   & 5 & 0.478 & \textBF{0.454} && 0.580 & \textBF{0.543} \\
    \cline{2-8}   
      &   & 1 & 0.383 & \textBF{0.369} && 0.494 & \textBF{0.472} \\
      &   & 2 & 0.424 & \textBF{0.415} && 0.538 & \textBF{0.518} \\
      & $T=85$  & 3 & 0.439 & \textBF{0.429} && 0.550 & \textBF{0.529} \\
      &   & 4 & 0.448 & \textBF{0.435} && 0.553 & \textBF{0.528} \\
      &   & 5 & 0.470 & \textBF{0.453} && 0.571 & \textBF{0.537} \\
    \bottomrule
  \end{tabular}
  }
\end{table}

\newpage
Interval forecasts can be used to assess forecast uncertainty of models involving the extracted global and local features. Supplement to point forecasts, we adopt the method of~\cite{ANH15} and construct pointwise prediction intervals as follows.
\begin{enumerate}[1)]
\item Using all observed data, compute the empirical FPCs $\{\widehat{\phi}_1(u), \cdots, \widehat{\phi}_K(u)\}$ with their associated estimated principal component scores $\{\widehat{\bm{\beta}}_{1}, \cdots, \widehat{\bm{\beta}}_{\widehat{K}}\}$, where $\widehat{\bm{\beta}}_k = [\widehat{\beta}_{1,k}, \cdots, \widehat{\beta}_{T,k}]$. Compute the regularized wavelet coefficients $\widetilde{\bm{D}} = \{\widetilde{\bm{D}}_{1}, \cdots, \widetilde{\bm{D}}_{T} \}$. In-sample forecasts are then constructed as
\begin{equation*}
\widehat{\X}_{\xi+h}(u) = \widehat{\beta}_{\xi+h,1}\widehat{\phi}_1(u) +\cdots + \widehat{\beta}_{\xi+h,\widehat{K}}\widehat{\phi}_{\widehat{K}}(u) + A^{\top} \widetilde{\bm{D}}_{\xi+h|\xi}, \qquad \xi \in \{\widehat{K}, \cdots,T-h\},
\end{equation*}
where $\{\widehat{\beta}_{\xi+h,1},\cdots,\widehat{\beta}_{\xi+h,K}\}$ and $\widetilde{\bm{D}}_{\xi+h|\xi}$ are $h$-step-ahead forecasts produced by univariate time series models based on $\{\widehat{\bm{\beta}}_{k}\}_{k=1}^{\widehat{K}}$ and $\{\widetilde{\bm{D}}_{t}\}_{t=1}^{\xi}$, respectively.
\item With the in-sample point forecasts, we calculate the in-sample point forecasting errors
\begin{equation*}
\widehat{\epsilon}_{\zeta}(u) = \X_{\xi+h}(u) - \widehat{\X}_{\xi+h}(u),
\end{equation*}
where $\zeta \in \{1,2,\cdots, M \} $ and $M = T-h-\widehat{K}+1$.
\item Based on these in-sample forecasting errors, we sample with replacement to obtain a series of bootstrapped forecasting errors. Denote upper bounds and lower bounds for point forecasts by $\gamma_{\text{lb}}(u)$ and $\gamma_{\text{ub}}(u)$, respectively. We seek a tuning parameter $\psi_{\alpha}$ such that $\alpha \times 100\%$ of in-sample forecasting errors satisfy
\begin{equation*}
\psi_{\alpha} \times \gamma_{\text{lb}}(u) \leq \widehat{\epsilon}_{\zeta}(u) \leq \psi_{\alpha} \times \gamma_{\text{ub}}(u).
\end{equation*}
In-sample forecasting errors $\left\lbrace \widehat{\epsilon}_{1}(u), \cdots, \widehat{\epsilon}_{M}(u)\right\rbrace $ are expected to be approximately stationary, and by the law of large numbers, to satisfy
\begin{multline*}
\frac{1}{M} \sum_{\zeta=1}^{M} \mathds{1} \Big( \psi_{\alpha} \times \gamma_{\text{lb}}(u) \leq \widehat{\epsilon}_{\zeta}(x) \leq \psi_{\alpha} \times \gamma_{\text{ub}}(u) \Big) \\
\approx \text{Pr} \Big[ \psi_{\alpha} \times \gamma_{\text{lb}}(u) \leq \X_{T+h}(u) - \widehat{\X}_{T+h|n}(u) \leq \psi_{\alpha} \times \gamma_{\text{ub}}(u)  \Big].
\end{multline*}
\end{enumerate}

Instead of computing the standard deviation of $\{\widehat{\epsilon}_{1}(u), \cdots, \widehat{\epsilon}_{M}(u)\}$ as done in \cite{ANH15}, we follow~\cite{Shang17} and use the nonparametric bootstrap approach to calculate pointwise prediction intervals. Specifically, we determine a $\pi_{\alpha}$ such that $\alpha \times 100\%$ of in-sample forecasting errors satisfy
\begin{equation*}
\pi_{\alpha} \times \gamma_{\text{lb}}(u_i) \leq \widehat{\epsilon}_{\zeta}(u_i) \leq \pi_{\alpha} \times \gamma_{\text{ub}}(u_i), \qquad i = 1,\dots,n.
\end{equation*}
Then, the $h$-step-ahead pointwise prediction intervals are given as 
\begin{equation*}
\pi_{\alpha} \times \gamma_{\text{lb}}(u_i) \leq \X_{T+h}(u_i) - \widehat{\X}_{T+h|T}(u_i) \leq \pi_{\alpha} \times \gamma_{\text{ub}}(u_i),
\end{equation*}
where $i$ symbolizes the discretized data points. 

\end{appendices}

\clearpage 

\newpage
\bibliographystyle{agsm}
\bibliography{datadriven}